\documentclass[12pt, draftcls,onecolumn]{IEEEtran}

\usepackage{cite}
\ifCLASSINFOpdf
   \usepackage[pdftex]{graphicx}
\else
   \graphicspath{{../eps/}}
\fi
\usepackage{amsmath}
\usepackage{bm}
\usepackage{amsfonts}

\usepackage{algorithmic}
\usepackage{array}
\ifCLASSOPTIONcompsoc
  \usepackage[caption=false,font=normalsize,labelfont=sf,textfont=sf]{subfig}
\else
  \usepackage[caption=false,font=footnotesize]{subfig}
\fi

\usepackage{color}
\usepackage{tikz}
\usepackage{pgfplotstable}
\usetikzlibrary{%
	calc,%
	shapes%
}
\usepackage{bigints}
\usepackage{setspace}
\usepackage{pgfkeys}
\def\addlegendimage{\csname pgfplots@addlegendimage\endcsname}

\usepackage{algorithm,algorithmic}
\usepackage{amsthm}

\usepackage{subfloat}

\usepackage{accents}

\usepackage{tablefootnote}

\usepackage{amssymb}
\usepackage{hyperref}

\DeclareSymbolFont{euler}{U}{eur}{m}{n}
\DeclareMathSymbol{\sdelta}{\mathord}{euler}{"0E}

\newtheorem{lemma}{Lemma}

\newtheorem{definition}{Definition}
\newtheorem{remark}{Remark}

\allowdisplaybreaks[1]

\newcommand{\modi}[1]{\textcolor{black}{#1}}
\newcommand{\cc}[1]{\textcolor{black}{#1}}
\newcommand{\ac}[1]{\textcolor{black}{#1}}
\newcommand{\ccc}[1]{\textcolor{black}{#1}}

\begin{document}
%
% paper title
% Titles are generally capitalized except for words such as a, an, and, as,
% at, but, by, for, in, nor, of, on, or, the, to and up, which are usually
% not capitalized unless they are the first or last word of the title.
% Linebreaks \\ can be used within to get better formatting as desired.
% Do not put math or special symbols in the title.
%\title{Residual Clipping Noise in ACO-OFDM-based Multilayer Optical OFDM schemes}
\title{Residual Clipping Noise in Multi-layer Optical OFDM: Modeling, Analysis, and Application}
%
%
% author names and IEEE memberships
% note positions of commas and nonbreaking spaces ( ~ ) LaTeX will not break
% a structure at a ~ so this keeps an author's name from being broken across
% two lines.
% use \thanks{} to gain access to the first footnote area
% a separate \thanks must be used for each paragraph as LaTeX2e's \thanks
% was not built to handle multiple paragraphs
%

\author{Zhenyu~Zhang,
        Anas~Chaaban,~%\IEEEmembership{Senior~Member,~IEEE}
        and~Mohamed-Slim~Alouini,~%\IEEEmembership{Fellow,~IEEE}% <-this % stops a space
\thanks{
Related preliminary results were presented in the 1st Global LiFi Congress \cite{laco-perf2-exp/bitLoading}.%

Zhenyu Zhang and Anas Chaaban are with the School of Engineering, University of British Columbia, Kelowna, BC V1V 1V7, Canada, e-mail: zhenyu.zhang@alumni.ubc.ca; anas.chaaban@ubc.ca.%}% <-this % stops a space
%\thanks{{

Mohamed-Slim Alouini is with the Computer, Electrical, and Mathematical Sciences and Engineering (CEMSE) division, King Abdullah University of Science and Technology (KAUST), Thuwal 23955-6900, Kingdom of Saudi Arabia, e-mail: slim.alouini@kaust.edu.sa.%}% <-this % stops a space
}}

\maketitle

% As a general rule, do not put math, special symbols or citations
% in the abstract or keywords.
\begin{abstract}
Optical orthogonal frequency division multiplexing (O-OFDM) schemes are variations of OFDM schemes which produce non-negative signals. \ac{Asymmetrically-clipped O-OFDM (ACO-OFDM) is a single-layer O-OFDM scheme, whose spectral efficiency can be enhanced by adopting multiple ACO-OFDM layers or a combination of ACO-OFDM and other O-OFDM schemes. However, since symbol detection in such enhanced ACO-OFDM (eACO-OFDM) is done iteratively, erroneous detection} leads to residual clipping noise (RCN) which can degrade performance in practice. \ac{Thus, it is important to develop an} accurate model for RCN \ac{which can be used to design} RCN-aware eACO-OFDM schemes. To this end, this paper provides a mathematical analysis of RCN leading to an accurate model of RCN power. The obtained model is used to analyze the performance \ac{of various eACO-OFDM schemes. It is shown that the model provides} an accurate evaluation of symbol error rate (SER), which would be underestimated if RCN is ignored. Moreover, the model is shown to be useful for designing an RCN-aware resource allocation that increases the robustness of the system \ac{in terms of meeting a target SER}, compared to an RCN-unaware design.
%Optical orthogonal frequency division multiplexing (O-OFDM) schemes are variations of OFDM schemes which produce non-negative time-domain signals, and be constructed using a single or multiple layers.
%Multi-layer O-OFDM can outperform single layer O-OFDM in energy and spectral efficiency. 
%There are various multi-layer O-OFDM schemes, which can enhance the spectral efficiency of the signal layer asymmetrically-clipped O-OFDM (ACO-OFDM), known as eACO-OFDM.
%In the demodulation of eACO-OFDM, successive interference cancellation is used to remove the clipping noise of ACO-OFDM, which introduces residual clipping noise (RCN) which in turn degrades the system performance. 
%The common approach to deal with RCN is to assume that it can be eliminated by coding. However, for a more practical performance evaluation, we need an accurate RCN model and an RCN-aware eACO-OFDM design, which are missing in the current literature. 
%To this end, the purpose of this paper is to provide a mathematical analysis of RCN leading to  an accurate RCN power model. 
%To demonstrate its quality, the obtained model is used to analyze the system performance, and is shown to provide an accurate \slimc{symbol error rate (SER)} evaluation, which would otherwise be underestimated if the RCN is ignored. Moreover, the relevance of the model is demonstrated by showing how an RCN-aware resource allocation can increase the robustness of the system by meeting the target SER performance, contrary to an RCN-unaware design.
\end{abstract}

% Note that keywords are not normally used for peerreview papers.

% For peer review papers, you can put extra information on the cover
% page as needed:
% \ifCLASSOPTIONpeerreview
% \begin{center} \bfseries EDICS Category: 3-BBND \end{center}
% \fi
%
% For peerreview papers, this IEEEtran command inserts a page break and
% creates the second title. It will be ignored for other modes.
\IEEEpeerreviewmaketitle

\section{Introduction} \label{Sec:Intro}
% In this section, 
% 1) I will go through the motivation and development of optical wireless communication;
% 2) go through the modulation schemes and their features;
% 3) point out the importance of how hybrid O-OFDM;
% 4) point out the lack of RCN model in current literature. 

\IEEEPARstart{O}{ptical} wireless communication (OWC) has received significant research focus in recent years \cite{owc1,owc4}, since it is expected to complement radio wireless transmission in providing ultra-fast communication. Intensity modulation and direct detection (IM/DD) is widely used to realize OWC because of its simplicity \cite{imdd,book1}. \ac{Among other schemes,} orthogonal frequency division multiplexing (OFDM), a core modulation technique adopted in 4G \cc{Long-Term Evolution (LTE)} and 5G \cc{New Radio (NR)}, has proved its potential in OWC \cite{ofdm1}. %OFDM has advantages such as high spectral efficiency, simple channel equalization, and robustness to multipath interference, which are desired properties in wireless communication.

%The recent development xxx and open challenges. 

IM/DD OWC uses solid-state lighting devices as transmitters, which constrains the modulating current to be non-negative. Therefore, to apply OFDM in IM/DD OWC, the traditional complex-valued OFDM design should be revisited. Several techniques have been designed to address this issue. For instance, DC-biased optical OFDM (DCO-OFDM) uses Hermitian symmetry to construct a real-valued signal, and a direct current (DC) bias to ensure nonnegativity. On the other hand, asymmetrically-clipped optical OFDM (ACO-OFDM) \cite{aco} only loads symbols onto odd subcarriers (DC subcarrier is indexed by $0$), uses Hermitian symmetry, and clips (sets to zero) the negative samples of the time-domain signal. \cc{This clipping only introduces noise in the unused even subcarriers}, \ac{which does not affect detection}. \ac{Compared to DCO-OFDM, this saves energy by avoiding an extra DC bias, at the expense of lower} spectral efficiency \ac{since fewer subcarriers are used}. Pulse-amplitude-modulation discrete multi-tone (PAM-DMT) \cite{pam-dmt} is another scheme which avoids {an extra DC bias}, by loading purely imaginary PAM symbols onto all subcarriers while ensuring Hermitian symmetry, and \ac{then clipping negative samples of the time-domain signal}. \cc{This clipping} introduces purely real-valued noise in the frequency domain, which \ac{does not affect the detection of the purely imaginary symbols \cite{pam-dmt}.} PAM-DMT has the same spectral efficiency as ACO-OFDM, because only the imaginary parts of the subcarriers are modulated. 

Improving the spectral efficiecy while maintaining the energy efficiency in a clipping-based O-OFDM scheme is the main motivation behind multi-layer O-OFDM schemes. Consider an extra signal which only uses even subcarriers combined with an ACO-OFDM signal. \ac{The extra signal does not interfere with the ACO-OFDM signal, but the ACO-OFDM signal's clipping noise interferes with the extra signal.} Thus, detection can be applied first on the ACO-OFDM signal (odd subcarriers) interference-free. Then, \ac{the clipping noise can be reconstructed and subtracted from the received signal, and detection can be applied on the extra signal (even subcarriers).} This is the basic idea of enhanced ACO-OFDM (eACO-OFDM) schemes. A similar idea can be used for enhancing PAM-DMT \cite{pam-dmt-layered} and {a} digital-Hartley-transform-based scheme \cite{dht} which are not the focus of this paper.

The extra signal occupying only even subcarriers can be a DCO-OFDM signal leading to the asymmetrically clipped DC-biased optical OFDM (ADO-OFDM) scheme \cite{ado}, a PAM-DMT signal leading to the hybrid ACO-OFDM (HACO-OFDM) scheme \cite{haco}, or another set of ACO-OFDM layers leading to the layered ACO-OFDM (LACO-OFDM) scheme \cite{laco} (also known as spectral-and-energy-efficient OFDM (SEE-OFDM) \cite{see2}). %The difference between SEE-OFDM and LACO-OFDM is that SEE-OFDM {also has} a time-domain process as an alternative way of removing clipping noise and separating {the layers}. A similar idea is adopted in \cite{haco-receiver2} to provide a fast HACO-OFDM receiver. 
\ac{While this enhances spectral efficiency (all subcarriers are used), some problems arise.} The addition of multiple OFDM signals leads to a more severe peak-to-average power ratio (PAPR) resulting in more peak clipping distortion\cite{haco-papr,dht,laco-perf3/papr-reduction, wang2018analysis}. Moreover, erroneous detection at the receiver leads to erreneous reconstruction of clipping noise, which introduces distortion to the extra layers \cite{lowery2016comparisons,laco-perf2-exp/bitLoading, laco-perf3/papr-reduction, cmp5/powerAlloc}. This distortion is called inter-layer interference in \cite{laco-perf3/papr-reduction} and  residual clipping noise (RCN) in \cite{laco-perf2-exp/bitLoading}.

RCN in eACO-OFDM is dealt with in two different ways in the literature. Some works assume that it can be ignored under a high signal-to-noise ratio (SNR) given a constellation size, \cc{such that detection errors are negligible} \cite{cmp2/eaco,cmp5/powerAlloc}. However, \ac{without} an explicit relation between RCN, SNR, constellation size, and error rate, this assumption only applies to examined cases. %To provide a general system design reference, a study on the relation between RCN and other transmission parameters is needed.}
%a high SNR assumption does not necessarily hold under practical considerations, \cc{especially when certain error rates are allowed in the transmission.} Thus, a study on the relation between RCN and SNR is needed.
Other studies ignore RCN by arguing that a perfect coding scheme can eliminate \ac{detection errors, and hence also} RCN \cite{cmp3-informationRate,laco-rx3-multiClassCoding}. 
A recent coding scheme for LACO-OFDM suggests using a dedicated codebook for each layer (multi-class coding) \cite{laco-rx3-multiClassCoding}. Theoretically, this can eliminate detection errors in each layer and thus eliminate RCN. However, \ac{the use of} multiple codebooks \ac{requires} rate-matching and \ac{has} high complexity at both the transmitter and the receiver. This is especially true for the last few layers of LACO-OFDM, which only have a few subcarriers, making the use of a new codebook for each layer inefficient. %One possible solution is to encode the last several layers jointly, where the total noise power (including RCN) in those layers is needed for choosing the codebook.
%Although multi-class coding is {shown to provide good performance, the resource allocation process for it has challenges such as rate-matching and high complexity, because of adopting multiple codebooks. As each layer uses one dedicated codebook in the multi-class coding scheme, it is more suitable for separating layers to convey independent traffic streams at different rates. For high-speed service which {requires using} all the layers at once, multi-class coding become unsuitable and there is no existed study to show how to realize it using multi-class coding scheme.}
Alternatively, if we want to encode all layers of LACO-OFDM together in the presence of RCN, performance can only be maintained if the distortion caused by RCN is quantified. Thus, an RCN power model is needed for quantifying and maintaining the performance of eACO-OFDM schemes in practice.

RCN power models \ac{have been studied in} \cite{laco-perf2-exp/bitLoading, laco-perf3/papr-reduction}. RCN power is modeled as a portion of noise power in \cite{laco-perf2-exp/bitLoading}, which shows using experiments that the model can decrease the error-rate at a given bit-rate, and balance the error-rate across layers of LACO-OFDM. However, this RCN power model in \cite{laco-perf2-exp/bitLoading} is not accurate . \cc{In \cite[(13)]{laco-perf3/papr-reduction}, RCN power is \ac{assumed} to be equal to the power of detection error \ac{calculated as the product of} the error probability and the square of the distance of detection errors.} Therefore, \ac{deriving an accurate} RCN power model for eACO-OFDM is still an open problem.

% Shortage of the current resource allocation in eACO-OFDM
Optimized bit and power allocation is \ccc{an aspect} in eACO-OFDM and requires \ccc{to know RCN power}. There are two categories bit-loading problems in multicarrier systems \cite{bitloading1,bitloading3}: the bit rate maximization problem (BRMP) which aims to maximize the overall bit rate under a total power constraint, and the margin maximization problem (MMP) which aims at minimize the overall power consumption for a target bit rate. \ac{Both require an RCN power model. Power allocation and layer assignment for LACO-OFDM was studied in} \cite{laco-powerAlloc}, which ignores RCN and assumes the same total noise power in each layer. \modi{However, %the noise power that affects detection in each layer of LACO-OFDM is related to the number of effective subcarriers in that layer and is not likely to be equal. Thus 
these assumptions lead to an overestimation of the noise power.}

%Optimizing the bit and power allocation is another {research question in} eACO-OFDM. There are basically two categories of resource allocation problems in multicarrier system \cite{bitloading1,bitloading3}, firstly, the bit rate maximization problem (BRMP)  which aims to maximize the overall bit rate under {a total power constraint}, and secondly, the margin maximization problem (MMP) which aims at minimize the overall power consumption for a target bit rate. Instead of {formulating} BRMP and MMP problems for eACO-OFDM, {\color{red}the current literature mostly focuses on `layer-based' optimization, such as layer number assignment and power and bits allocation by layer \cite{ado-reciver/powerAlloc, laco-powerAlloc, see2}, where ignoring RCN makes the conclusions doubtful.}
%Optimal power allocation and layer assignment for LACO-OFDM are studied in \cite{laco-powerAlloc}, which assumes the same {total} noise power in each layer. However, for LACO-OFDM, the effective noise which distorts the detection process in each layer is related to the number of effective subcarriers and not likely to be equal. 
%On the other hand, {from a resource allocation's point of view}, `layer-based' resource allocation {is} generally suboptimal. {To} improve the system's throughput and reliability, it is better to use more freedom by {involving} all possible subcarriers in the resource optimizating process for finding the globally optimal. To perform resource allocation in each subcarrier, RCN {is} a part of the distortion {that must be accounted for}. 

In this paper, we provide a careful study of the RCN of eACO-OFDM schemes. We investigate and model \ac{the process which generates RCN, and we propose a worst-case RCN power model which proves useful for analyzing and optimizing eACO-OFDM schemes.} The contributions of this paper \ac{can be summarized as follows:}
\begin{enumerate}
	\item Three \ac{asymptotic} statistical properties of RCN are demonstrated: RCN is independent and identically distributed in the time domain, RCN is circularly symmetric complex Gaussian with zero mean in all effective subcarriers of the affected layers, and \ccc{the correlation among RCN signals from different layers are negligible}. 
	
	\item Based on these properties, a worst-case RCN power model is proposed which is accurate for a wide range of SNR.
	
	\item Accurate SER evaluations for eACO-OFDM schemes \ac{are provided using the} RCN power model.  
	
	\item A globally optimized RCN-aware resource allocation scheme for eACO-OFDM is given, which shifts complexity to the transmitter side and controls SER to be below a target SER. This leads to a reliable eACO-OFDM scheme which is relevant in practice. 
%	{This leads to a} robust eACO-OFDM without multi-class coding and is more friendly to {the current market-use resource scheduling process defined in both 4G-LTE and 5G-NR protocols.}
\end{enumerate}

The rest of this paper is structured as follows. In Sec. \ref{Sec:Back}, we review the ADO-OFDM, HACO-OFDM, and LACO-OFDM schemes and their components. In Sec. \ref{Sec:ana-rcn}, we explore the statistics of RCN and propose methods for estimating RCN power and total noise power. In Sec. \ref{Sec:serEst}, we derive a {theoretical expression of the SER} of eACO-OFDM with RCN {taken into} consideration. In Sec. \ref{sec:optimization}, an {RCN-aware} SER-controlled LACO-OFDM is demonstrated. Finally, Sec. \ref{Sec:sim} shows simulation results and Sec. \ref{Sec:conclusion} concludes the paper. 

\ac{The following notations are used throughout the paper.} Bold letters represent vectors, where a lower case ($\mathbf{x}$) is used to denote a discrete-time signal and an upper case ($\bf{X}$) \ac{is used to denote} the frequency-domain counterpart of $\mathbf{x}$. %A time-domain signal $\ut{\mathbf{x}} \overset{\Delta}{=} \mathbf{x} - \mathbb{E}[\mathbf{x}]$ 
FFT/IFFT$(\cdot)$ denote the fast Fourier transform and its inverse, i.e., for $\mathbf{x} = [x(n)]\cc{_{n=0}^{N-1}}$, \ac{$\mathbf{X} = [X(k)]\cc{_{k=0}^{N-1}} = \text{FFT} \{\mathbf{x}\} = \sum_{n=0}^{N-1} x(n) e^{ -\mathsf{j} \frac{2\pi}{N}kn }$ and} $\mathbf{x} = \text{IFFT} \{\mathbf{X}\} = \frac{1}{N}\sum_{k=0}^{N-1} X(k) e^{ \mathsf{j} \frac{2\pi}{N}kn }$, \cc{where $\mathsf{j} \triangleq \sqrt{-1}$}. 
\ac{The operators $\mathbb{E}[\cdot]$ and $\mathbb{V}[\cdot]$ represent the expectation and variance (element-wise),
%a time-domain signal $\mathbf{x}$ is always consider to be constructed by an alternating-current (AC) denoted by $\ut{\mathbf{x}} = \mathbf{x} - \mathbb{E}[\mathbf{x}]$ and a direct-current (DC) part $d = \mathbb{E}[\mathbf{x}]$, i.e., $\mathbf{x} = \ut{\mathbf{x}} + d$, where $\mathbb{E}\{\cdot\}$ and $\mathbb{V}\{\cdot\}$ represent the statistical expectation and variance, respectively;
$|\cdot|$ denotes the absolute value (element-wise) of a real number/vector or the cardinality of a set, $(\cdot)^*$ denotes the conjugate, $(\cdot)^+$ denotes the nonnegative part of a signal so that $(\mathbf{x})^+ = \frac{\mathbf{x}+|\mathbf{x}|}{2}$, $\|\cdot\|$ denotes the $l_2$-norm, $\lfloor\cdot\rfloor $ denotes the floor operator, and $\otimes$ denotes convolution.} The set $\mathbb{R}$ is the set of real numbers. For a random vector ${\mathbf{X}}=[X_i]\cc{_{i=0}^{N-1}}$, $\mathcal{P}\{ {\mathbf{X}} \} = \frac{1}{N}\sum\cc{_{i=0}^{N-1}}\mathbb{E} [|X_i|^2] $ denotes \ac{the average power of $\bf{X}$. All logarithms are base 2, and $\mathcal{N}(\mu,\sigma^2)$ denotes a Gaussian distribution with mean \modi{$\mu$} and variance $\sigma^2$.} %We write $\modi{X}\sim P_X(x)$ to indicate that \modi{$X$} is distributed according to some probability density function (pdf) $P_X(x)$.}

\section[back]{System Model} \label{Sec:Back}
% 1. Here I introduce all eACO-OFDM schemes: including xxx

\cc{Three eACO-OFDM schemes are studied in this paper} (ADO-OFDM, HACO-OFDM, LACO-OFDM). In what follows, we first \cc{introduce the IM/DD channel model}, then we introduce the construction of some single-layer O-OFDM schemes (ACO-OFDM, DCO-OFDM and PAM-DMT), and finally introduce the three eACO-OFDM schemes under a \cc{unified} framework. 

\subsection{Channel Model}
\ac{In IM/DD OWC,} a real positive signal $x(n)$ is transmitted by an LED, \ac{and received by a photodiode (PD)} through a channel with impulse response $h(n)$ and real\ac{-valued} additive white Gaussian noise (AWGN) $v_0(n)$. \ac{The received signal $y(n)$ is given by}
\begin{align}
	y(n) &=  h(n) \otimes x(n) + v_0(n).
\end{align}
%Let the linear dynamic range\footnote{The linear dynamic range of an LED is the current range where the LED has a linear optical power response to the injected current.} of the LED be $d_0$ to $\mathcal{A}$, where $d_0$ is the turn-on voltage and $\mathcal{A}$ is the peak limit, which requires $x(n) = (\ut{x}(n) + d) \in [d_0,\mathcal{A}]$. As peak-clipping distortion will not be considered in our analysis, the turn-on voltage and peak limit are not critical so that we set $d_0 = 0$ and $\mathcal{A} = \infty$ for convenience throughout the rest discussions.
Note that this requires $x(n) \geq d_0$ where $d_0$ is the turn-on current of the LED. Since the value of $d_0$ does not affect the analysis (can be absorbed into the DC component of $x(n)$), we set $d_0 = 0$ for convenience. 
\ccc{In the frequency domain, we have $Y(k)=H(k)X(k)+V_0(k)$, where $Y(k)$, $H(k)$, $X(k)$ and $V_0(k)$ are the FFT counterpart of $y(n)$, $h(n)$, $x(n)$ and $v_0(n)$, respectively. For a flat channel, $H(k)=1$ for all $k$. For a frequency-selective channel, we adopt channel-inversion equalization in the frequency domain, which gives $Y'(k)=H^{-1}(k)Y(k)=X(k)+V(k)$, where $V(k)=H^{-1}(k)V_0(k)$ and $H(k)$ can be obtained by channel estimation. Therefore, the equivalent time-domain signal after channel equalization, $y'(n) = x(n) + v(n)$, has a zero-mean colored Gaussian noise $v(n)$.}

%For a flat channel, $h(n)={\sdelta}(n)$ is a Dirac impulse, and $y(n)=x(n)+v_0(n)$. For a frequency-selective channel, channel-inversion equalization gives $y'(n) = h^{-1}(n) \otimes y(n) = x(n) + v(n)$, where $h^{-1}(n) \otimes h(n) = {\sdelta}(n)$ and $v(n) = h^{-1}(n) \otimes v_0(n)$ is zero-mean colored Gaussian noise.

The power of $x(n)$ can be classified into optical power $P_\text{opt} \triangleq \eta_{\rm eo} \lim_{T\to\infty}\frac{1}{T}\sum_{n=0}^{T-1}x(n)$ \ccc{(the electrical-to-optical conversion efficiency of the LED times the DC value of $x(n)$), where we assume a unified electrical-to-optical conversion efficiency, i.e., $\eta_{\rm eo}=1 (\rm lumens\;per\;ampere)$, without loss of generality},
%\footnote{\ac{The actual optical power is $\eta_{\rm eo}P_{\rm opt}$, where $\eta_{\rm eo}$ is the electrical-to-optical conversion efficiency of the LED (assumed $1$ herein).}} 
and electrical power $P_\text{elec} \triangleq \lim_{T\to\infty}\frac{1}{T}\sum_{n=0}^{T-1}x^2(n) R$, \ccc{where we assume a unified LED resistance, i.e., $R=1(\rm ohm)$, without loss of generality}. 
\modi{Additionally, we define the effective power $P_\text{eff}$ as the power of the `useful part' of the signal on which performance depends, which is specified for each O-OFDM scheme in the next subsections.}
%Additionally, we define the effective power $P_\text{eff}$ as the power of the `useful part' of the signal on which performance depends (since part of the signal is not used in the detection process as we shall see next). This power $P_\text{eff}$ depends on the O-OFDM scheme and is defined in the next section. 
%We assume that the system is constrained by an electrical power constraint \ccc{in some of the simulations in this paper}. However, for generalizing our simulation results, we provide a table (Table \ref{tab:power}) which relates $P_\text{opt}$, $P_\text{elec}$, and the effective power \modi{$P_\text{eff}$}, and is proved in Appendix \ref{Sec:Appen2}.}
\ccc{Table I relates $P_\text{opt}$, $P_\text{elec}$, and the effective power $P_\text{eff}$, where the relations are proved in Appendix \ref{Sec:Appen2}.}

\begin{table} 
	\renewcommand{\arraystretch}{1.2}
	\caption{ \ccc{Quantity relations among $P_{\rm elec}$, $P_{\rm opt}$, and $P_{\rm eff}$. ($\eta_{\rm eo}=1,R=1$.)}}
	\label{tab:power}
	\centering
	\begin{tabular}{ l |c c c }
		\hline
		& $P_\text{elec}$ & $P_\text{opt}$\tablefootnote{\ccc{Here, the DC levels for all schemes are chosen to be large enough to avoid zero-clipping distortion, as specified in Sec. II-B/C and also in Appendix A.}}  \\
		\hline
		ACO-OFDM & $2 P_\text{eff}$ & $\sqrt{\frac{2P_\text{eff}}{\pi}}$\\% & $1.4P_\text{eff}$ \\
		DCO-OFDM & $10 P_\text{eff}$ & $3 \sqrt{P_\text{eff}}$\\% & $P_\text{eff}$ \\
		PAM-DMT & $2 P_\text{eff}$ & $\sqrt{\frac{2P_\text{eff}}{\pi}}$\\% & $1.4P_\text{eff}$ \\
		ADO-OFDM\tablefootnote{The power relations for ADO-OFDM, HACO-OFDM, and LACO-OFDM are derived under the assumption that the effective power is distributed equally over all effective subcarriers.} & $(6+\frac{6}{\sqrt{2\pi}})P_\text{eff}$  & $(\frac{1}{\sqrt{\pi}}+\frac{3}{\sqrt{2}}) \sqrt{P_\text{eff}}$\\% & $1.2 P_\text{eff}$  \\
		HACO-OFDM$^\textsc{\arabic{footnote}}$ & $(2+\frac{2}{\pi}) P_\text{eff}$ & $\frac{2}{\sqrt{\pi}}\sqrt{P_\text{eff}}$ \\%& $1.4 P_\text{eff}$ \\
		LACO-OFDM$^\textsc{\arabic{footnote}}$ & \modi{$(2- \frac{2}{\pi} + \frac{2}{(3-2\sqrt{2})\pi} \frac{\sqrt{2}^J-1}{\sqrt{2}^J+1})P_\text{eff}$} & \modi{$\sqrt{ \frac{2}{(3-2\sqrt{2})\pi} \frac{\sqrt{2}^J-1}{\sqrt{2}^J+1}P_\text{eff} }$}\\% & $1.4 P_\text{eff}$
	\end{tabular}
	\renewcommand{\arraystretch}{1}
\end{table}

\subsection{Single-layer O-OFDM Schemes} \label{sec:back-sl-oofdm}
\ac{A unipolar signal $x(n)\geq0$ can be constructed using OFDM in different ways, starting from a frequency domain signal ${ \bf{S} } = [S(k)]_{k=0}^{N-1}$, where $N$ is the number of subcarriers and $S(k)$ is the symbol (PAM or QAM) loaded onto the $k$th subcarrier.} %Define $\mathcal{K}$ as the set of the index of all effective subcarriers that can be non-zero. 
To obtain a real-valued time-domain signal, $\bf{S}$ is constrained to be Hermitian symmetric, i.e., $S(k) = S^{*}(N-k)$ \cc{for $k\in\{0,\ldots,\frac{N}{2}$\}}. To guarantee nonnegativity, several approaches can be used as explained next. 

\subsubsection{ACO-OFDM}
\ac{ACO-OFDM only loads symbols onto odd subcarriers of the frequency-domain signal denoted by $\mathbf{S}_{\rm aco}$ (Hermitian symmetric). Thus,} $S_{\rm aco}(k)=0$ for all $k\notin\mathcal{K}_{\rm aco}=\{k|k = 2i-1, i = 1,\ldots, \frac{N}{2} \}$, \ac{and $S_{\rm aco}(k)\in\mathbb{C}$ for effective subcarriers $k\in\mathcal{K}_{\rm aco}$.}
%$S_{\rm aco}(k)\neq0$ for $k\in\mathcal{K}_{\rm aco} = \{k|k = 2i-1, i = 1,\ldots, \frac{N}{2} \}$ \ac{(effective subcarriers of ACO-OFDM)} and $S_{\rm aco}(k)=0$ for all $k\notin\mathcal{K}_{\rm aco}$. 
%and \[{\mathbf{S}}_{\text{aco}} = [0, S_{\text{aco}} (1), 0, S_{\text{aco}} (3), \dots, S_\text{aco} ( \frac{N}{2}-1 ), 0, S_\text{aco}^* ( \frac{N}{2}-1 ), \dots, S_\text{aco}^* (3), 0, S_\text{aco}^* (1) ].\] 
\ac{Due to this construction,} ${\mathbf{s}}_\text{aco} = \text{IFFT}\{ {\mathbf{S}}_\text{aco} \}$ satisfies $s_\text{aco}(n)=-s_\text{aco}(n+\frac{N}{2})$ $\forall n<\frac{N}{2}$, and clipping at zero will not cause information loss \cite{aco}. Using this property, an ACO-OFDM modulator clips at zero and obtains a non-negative signal ${\mathbf{x}}_\text{aco}$,
\begin{equation*}
	\mathbf{x}_\text{aco} = ( {\mathbf{s}}_\text{aco} ) ^+ = \frac{ {\mathbf{s}}_\text{aco} + |{\mathbf{s}}_\text{aco}| }{2}.
\end{equation*}

Note that the clipping operation \modi{introduces \textit{clipping noise} $\frac{|{\mathbf{s}}_\text{aco}|}{2}$ which only occupies even subcarriers \cite{aco}. Thus,} in ACO-OFDM, the receiver simply ignores the even subcarriers of the received signal and recovers information from the \ac{{\it useful signal} $\frac{{\mathbf{s}}_\text{aco}}{2}$} \modi{in odd subcarriers}. 
%Note that the clipping operation halves the amplitude of ${\mathbf{s}}_\text{aco}$ and introduces \textit{clipping noise} $\frac{|{\mathbf{s}}_\text{aco}|}{2}$ \cite{aco}. However, the clipping noise only occupies even subcarriers \cite{aco}, and thus, the ACO-OFDM demodulator simply ignores the even subcarriers of the received signal and recovers information from the \ac{{\it useful signal} $\frac{{\mathbf{s}}_\text{aco}}{2}$} (odd subcarriers). 
 
\modi{For ACO-OFDM, we have $P_\text{eff} = \mathcal{P}\{\frac{ {\mathbf{s}}_\text{aco} }{2}\}$. The relations among $P_\text{elec}$, $P_\text{opt}$, and $P_\text{eff}$ are listed in Table \ref{tab:power}, and are proved in Appendix \ref{Sec:Appen2}.}
%\cc{To analyze the transmit power of ACO-OFDM, let $\varepsilon_{s,avg}$ represent the average symbol power \ac{(same for all $k$)} and $P_\text{eff} = \mathcal{P}\{\frac{ {\mathbf{s}}_\text{aco} }{2}\}$ represent the power of the useful ACO-OFDM signal $\frac{\mathbf{s}_\text{aco}}{2}$. The relation between $P_\text{elec}$, $P_\text{opt}$, and $P_\text{eff}$ is listed in Table \ref{tab:power}, which is proved in Appendix \ref{Sec:Appen2}.}

%\cc{To analyze the transmitting power, we have equivalently $\mathbf{x}_\text{aco} = \ut{\mathbf{x}}_\text{aco} + d \mathbf{1}$, where $\mathbf{1}$ is an all-one vector of dimension $N$. Then, we have $P_\text{elec} = \mathcal{P}\{\mathbf{x}_\text{aco}\}$, $P_\text{opt} = \mathbb{E}[x_\text{dco}(n)]=d$, $P_\text{AC} = \mathcal{P}\{\ut{\mathbf{x}}_\text{aco}\}$, and $P_\text{DC}=d^2$. Let $\varepsilon_{s,avg}$ represents the average symbol power in a transmission and $P_\text{eff} = \mathcal{P}\{\frac{ {\mathbf{s}}_\text{aco} }{2}\}$ represent the information part of power of $\mathbf{x}_\text{aco}$, termed the effective power. We have $\varepsilon_{s,avg} = \frac{P_\text{eff}}{|\mathcal{K}|}$. The power relations are listed in Table \ref{tab:power}, which are proved in Appendix \ref{Sec:Appen2}.}

\subsubsection{DCO-OFDM} 
\ac{In DCO-OFDM, symbols are loaded onto all subcarriers of the signal ${\mathbf{S}}_\text{dco}$ while satisfying Hermitian symmetry, and $S_\text{dco}(0) = S_{\rm dco}(\frac{N}{2})=0$. Thus, the effective subcarriers of DCO-OFDM are subcarriers $k \in \mathcal{K}_\text{dco} = \{k|k \ne \frac{N}{2}, k \in\{ 1,\ldots, N-1\} \}$.} The DCO-OFDM modulator then applies a DC bias $d_{\rm dco}$ to \ac{the real-valued} ${\mathbf{s}}_\text{dco} = \text{IFFT}\{ {\mathbf{S}}_\text{dco} \}$ and clips at zero, i.e.,
\begin{equation}
	\mathbf{x}_\text{dco} = (\mathbf{s}_\text{dco} + d_{\rm dco}\mathbf{1})^+ ,
\end{equation}
where $\mathbf{1}$ is an all-one vector with size $N$. We fix $d_{\rm dco} = 3 \sqrt{ \mathbb{V}[s_\text{dco}(n)] }$ to \ccc{avoid} clipping distortion, and hence $\mathbf{x}_\text{dco} \approx \mathbf{s}_\text{dco} + d_{\rm dco}\mathbf{1}$. \ac{The receiver uses an FFT operation and then detects the symbols from the effective subcarriers.} The relations between $P_\text{elec}$, $P_\text{opt}$, and \ac{$P_\text{eff}=\mathcal{P}\{\mathbf{s}_{\rm dco}\}$} are listed in Table \ref{tab:power}, and are proved in Appendix \ref{Sec:Appen2}.

\subsubsection{PAM-DMT}
\ac{A PAM-DMT signal is constructed from ${\mathbf{S}}_\text{pam}$, where $S_{\rm pam}(k)$ is a purely imaginary PAM symbol for $k \in \mathcal{K}_\text{pam} = \{k|k \ne \frac{N}{2}, k \in\{ 1,\ldots, N-1\} \}$ (effective subcarriers of PAM-DMT) and \modi{$S_\text{pam}(0) = S_{\rm pam}(\frac{N}{2})=0$}.} Due to this construction, ${\mathbf{s}}_\text{pam} = \text{IFFT}\{ {\mathbf{S}}_\text{pam} \}$ has $s_\text{pam}(n) = -s_\text{pam}(N-n-1)$ $\forall n<\frac{N}{2}$, and clipping at zero will not cause information loss \cite{pam-dmt}. Similar to ACO-OFDM, a PAM-DMT modulator clips at zero and obtains a non-negative signal ${\mathbf{x}}_\text{pam}$,
\begin{equation*}
\mathbf{x}_\text{pam} = ( {\mathbf{s}}_\text{pam} ) ^+ = \frac{ {\mathbf{s}}_\text{pam} + |{\mathbf{s}}_\text{pam}| }{2}.
\end{equation*}  
The clipping noise $\frac{|{\mathbf{s}}_\text{pam}|}{2}$ \ac{has a frequency-domain counterpart which is nonzero in all effective subcarriers,} but is real valued \cite{pam-dmt}, and thus it does not interfere with the useful signal $\frac{{\mathbf{s}}_\text{pam}}{2}$ which is imaginary. A PAM-DMT receiver can simply ignore the real part of each subcarrier and recover the information from the imaginary part.

\modi{For PAM-DMT, we have $P_\text{eff} = \mathcal{P}\{\frac{\mathbf{s}_\text{pam}}{2}\}$. The relations among $P_\text{elec}$, $P_\text{opt}$, and $P_\text{eff}$ are listed in Table \ref{tab:power}, and are proved in Appendix \ref{Sec:Appen2}.}

\ac{Next, we discuss multi-layer schemes constructed as a superposition of an ACO-OFDM layer and additional ACO-OFDM, DCO-OFDM, or PAM-DMT layers.}

\subsection{Multi-layer O-OFDM schemes \modi{(eACO-OFDM)}} \label{sec:back-ml-oofdm}

\begin{figure}
	\centering
	\subfloat[Transmitter.]{
		\includegraphics[width=0.5\textwidth ]{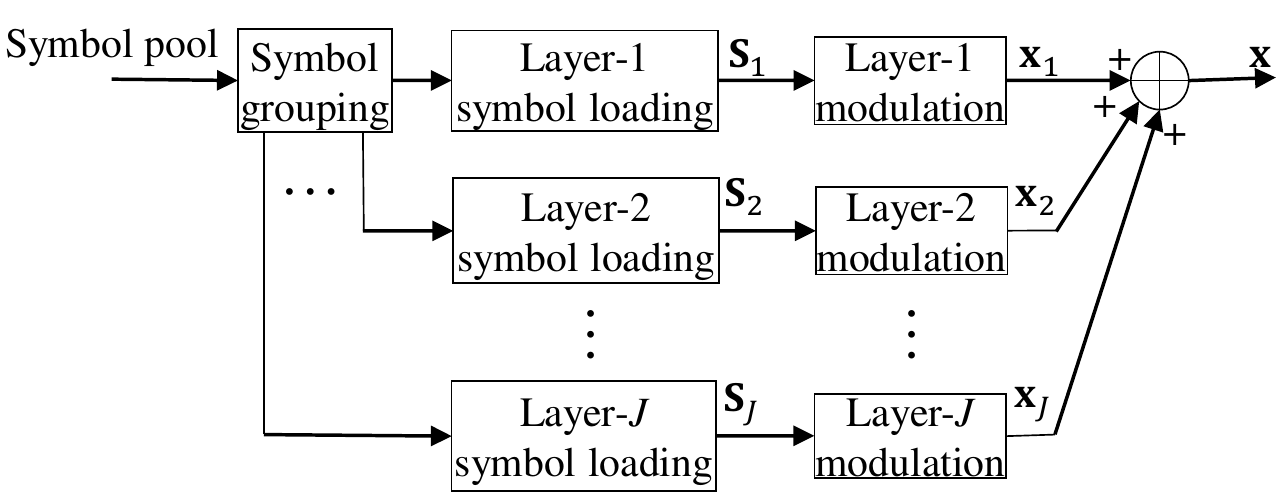}
		\label{fig:modulator}
	}
	\subfloat[Receiver.]{
		\includegraphics[width=0.4\textwidth ]{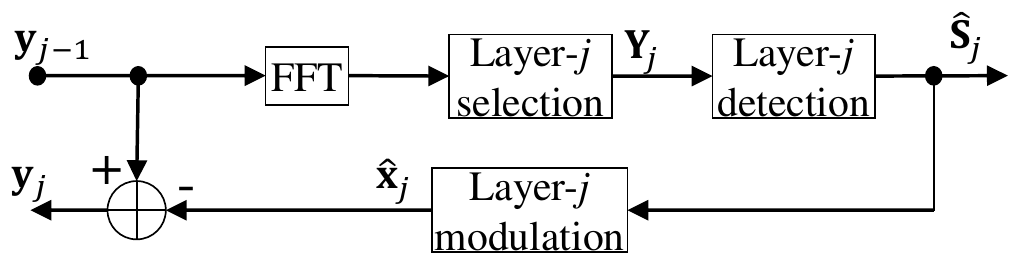}
		\label{fig:demodulator}
	}
	\caption{Flowchart of eACO-OFDM transmitter and receiver. Here, $J=2$ in ADO-OFDM and HACO-OFDM, and $J\le \log \frac{N}{2}$ for LACO-OFDM.}
\label{fig:flowchar}
\end{figure}

Fig. \ref{fig:flowchar} describes the transmitter and receiver \ac{of multi-layer O-OFDM schemes, which is applicable to  ADO-, HACO- and LACO-OFDM. In what follows, we first describe the transmitter in its three variants, and then describe a unified receiver.}

\subsubsection{\modi{eACO}-OFDM Transmitter}
\ac{At the transmitter side, the (PAM or QAM) constellation symbols are first divided into $J$ groups, where each group is transmitted over one of the $J$ layers (symbol grouping). We denote the set of effective subcarriers of layer $j$ by $\mathcal{K}_j$. Then, symbols in each group are loaded onto the effective subcarriers of each layer, and the remaining subcarriers $k\notin\mathcal{K}_j$ are set to zero (layer-$j$ symbol loading). This leads to $\mathbf{S}_{j}$, which depends on the single-layer scheme used in layer $j$. Finally, $\mathbf{S}_{j}$ is modulated to $\mathbf{x}_{j}$ using the O-OFDM modulation scheme of layer $j$ (layer-$j$ modulation), and the sum of all $\mathbf{x}_{j}$ is transmitted.}

\ac{In ADO-, HACO-, and LACO-OFDM, the first layer is an ACO-OFDM layer, i.e., \cc{$\mathcal{K}_1 = \{ k | k=2i-1, i \in\{1, \ldots, \frac{N}{2}\} \}$, $\mathbf{S}_{1} = {\mathbf{S}}_\text{aco}$, and thus \modi{${\mathbf{x}}_{1} = {\mathbf{x}}_\text{aco} = \frac{ {\mathbf{s}}_\text{aco} + |{\mathbf{s}}_\text{aco}| }{2}$}.} The subsequent layers can be DCO-OFDM, PAM-DMT, or ACO-OFDM layers, as described next.}

%\cc{The differences between the three eACO-OFDM schemes considered in this work are the construction of the remaining layers, \cc{which is described in the following subsections. To analyze the transmitting power, we have equivalently $\mathbf{x} = \ut{\mathbf{x}} + d\mathbf{1}$, where $\ut{\mathbf{x}} = \sum_{j=1}^{J} { \ut{\mathbf{x}}_{L_j} }$. Then for eACO-OFDm schemes, $P_\text{elec} = \mathcal{P}\{\mathbf{x}\}$, $P_\text{opt} = \mathbb{E}[x(n)]=d$, $P_\text{AC} = \mathcal{P}\{\ut{\mathbf{x}}\}$, $P_\text{DC}=P_\text{opt}^2$. 
%It is also needed to define $\varepsilon_{s,avg}^{(j)}$ as the average symbol power of layer $j$.} 
%where we use $P_t$ and $P_{L_j}$ ($j=1,\dots, J$) to denote the average power of $\bf{x}$ and ${\mathbf{x}}_{L_j}$, respectively, i.e., $P_t = \mathcal{P}\{ {\mathbf{x}} \}$, $P_{L_j} = \mathcal{P} \{ {\mathbf{x}}_{L_j} \}$, and we use $P_t'$ and $P_{L_j}'$ ($j=1,\dots, J$) to denote the \emph{effective power} of $\bf{x}$ and ${\mathbf{x}}_{L_j}$, respectively.

\paragraph{ADO-OFDM}

\modi{In this scheme, $J=2$ and the second layer employs DCO-OFDM with $\mathcal{K}_2 = \{ k | k \ne \frac{N}{2}, k=2(i-1), i \in\{2, \ldots, \frac{N}{2}\} \}$ (even subcarriers).}  Let ${\mathbf{S}}_{2} = {\mathbf{S}}_\text{dco}^{(2)}$, where $S_\text{dco}^{(2)}(k)$ is a QAM symbol when $k \in \mathcal{K}_2$ \modi{and $S_\text{dco}^{(2)}(k)=0$ \ccc{otherwise}.}
%, i.e., 
%\[{\mathbf{S}}_\text{dco}^{(2)} = [d, 0, S_\text{dco}(2),0, \dots, S_\text{dco}(\frac{N}{2}-2), 0, 0, 0,  S_\text{dco}^{*}(\frac{N}{2}-2), \dots, 0,S_\text{dco}^{*}(2),0],\] {where $d$ is chosen to ensure $\min\{ {\mathbf{x}}_\text{dco}^{(2)} \} \ge 0$.} 

\ac{Then, ${\mathbf{x}}_\text{dco}^{(2)}$ is constructed from ${\mathbf{S}}_\text{dco}^{(2)}$ using DCO-OFDM to yield} ${\mathbf{x}}_\text{dco}^{(2)} = ({\mathbf{s}}_\text{dco}^{(2)} + d_{\rm dco}^{(2)}\mathbf{1})^+$, where ${\mathbf{s}}_\text{dco}^{(2)} = \text{IFFT}\{ {\mathbf{S}}_\text{dco}^{(2)} \}$ and $d_{\rm dco}^{(2)}$ is sufficient large to avoid zero-clipping. Then we have ${\mathbf{x}}_{2} = {\mathbf{x}}_\text{dco}^{(2)}$ and the ADO-OFDM signal is
\begin{equation*}
\modi{{\mathbf{x}}_{\rm ado} = \sum_{j=1}^{2} {\mathbf{x}}_{j} = \mathbf{x}_\text{aco} + \mathbf{x}_\text{dco}^{(2)}.}
\end{equation*}

\modi{For ADO-OFDM, we have $P_\text{eff} = \mathcal{P}\{\frac{\mathbf{s}_\text{aco}}{2} + \mathbf{s}_\text{dco}^{(2)}\}$. \ac{Assuming that $P_\text{eff}$ is distributed equally in all effective subcarriers,} i.e., $\mathbb{E}[|\frac{S_\text{aco}(k_1)}{2}|^2] = \mathbb{E}[|S_\text{dco}^{\ccc{(2)}}(k_2)|^2]$, $\forall k_1 \in \mathcal{K}_1, \forall k_2\in \mathcal{K}_2$, then we have the relations among $P_\text{elec}$, $P_\text{opt}$, and $P_\text{eff}$ as listed in Table \ref{tab:power}, which are proved in Appendix \ref{Sec:Appen2}.}

%\cc{When $\varepsilon_{s,avg}^{(1)} = \varepsilon_{s,avg}^{(2)}$, we have the following relation as proved in Appendix:
%\begin{align}
%	\varepsilon_{s,avg} &= \frac{P_\text{eff}}{|\mathcal{K}_1| + |\mathcal{K}_2|}  \notag \\
%	P_\text{elec} & \approx 5.7 P_\text{eff}, \notag \\
%	P_\text{DC} & = P_\text{opt}^2 \approx 4.5 P_\text{eff},  \notag \\
%	P_\text{AC} & \approx 1.2 P_\text{eff}.
%\end{align}}

%\cc{Recalling the effective power of ACO-OFDM is a half of the total power,} we have the following relation between {the} power and \emph{effective power} {of ADO-OFDM}: 
%\begin{align} \label{eq:ado-P-Peff}
%P_t &= P_{L_1} + P_{L_2} = 2P_{L_1}' + P_{L_2}' + d^2,\\
%P_t' &= P_{L_1}' + P_{L_2}'.
%\end{align}

%\cc{Note that layer 2 signal ${\mathbf{x}}_{L_2}$ is disturbed by the clipping noise from layer 1. }

\paragraph{HACO-OFDM Transmitter}

\modi{In this scheme, $J=2$ and PAM-DMT is employed in the second layer with $\mathcal{K}_2 = \{ k | k \ne \frac{N}{2}, k=2(i-1), i \in\{ 2, \ldots, \frac{N}{2}\}  \}$.} Let ${\mathbf{S}}_{2}={\mathbf{S}}_\text{pam}^{(2)}$, where $S_\text{pam}^{(2)}(k)$ \ac{is a PAM symbol} when $k \in \mathcal{K}_2$ \modi{and $S_\text{pam}^{(2)}(k)=0$ otherwise}. Let ${\mathbf{s}}_\text{pam}^{(2)} =  { \text{IFFT} \{ {\mathbf{S}}_\text{pam}^{(2)} \} } $ and ${\mathbf{x}}_\text{pam}^{(2)} = ( { {\mathbf{s}}_\text{pam}^{(2)} } ) ^+$. Then, we have ${\mathbf{x}}_{2} = {\mathbf{x}}_\text{pam}^{(2)}$ and the HACO-OFDM modulation output is
\begin{equation*}
	\modi{{\mathbf{x}}_{\rm haco} = \sum_{j=1}^{2} {\mathbf{x}}_{j} = \mathbf{x}_\text{aco} + \mathbf{x}_\text{pam}^{(2)}.}
\end{equation*}

\modi{For HACO-OFDM, we have $P_\text{eff} = \mathcal{P}\{\frac{\mathbf{s}_\text{aco}}{2} + \frac{\mathbf{s}_\text{pam}^{(2)}}{2}\}$. \ac{Assuming that $P_\text{eff}$ is distributed equally in all effective subcarriers}, i.e., $\mathbb{E}[|\frac{S_\text{aco}(k_1)}{2}|^2] = \mathbb{E}[|\frac{S_\text{pam}^{(2)}(k_2)}{2}|^2]$, $\forall k_1 \in \mathcal{K}_1, \forall k_2\in \mathcal{K}_2$, then we have the relations between $P_\text{elec}$, $P_\text{opt}$, and $P_\text{eff}$ as listed in Table \ref{tab:power}, which are proved in Appendix \ref{Sec:Appen2}.}

%\cc{When $\varepsilon_{s,avg}^{(1)} = \varepsilon_{s,avg}^{(2)}$, the relations between $P_\text{elec}$, $P_\text{opt}$, $P_\text{DC}$, $P_\text{AC}$ and $P_\text{eff}$ are listed in Table \ref{tab:power}, which are proved in Appendix \ref{Sec:Appen2}. }

%\cc{When $\varepsilon_{s,avg}^{(1)} = \varepsilon_{s,avg}^{(2)}$, we have the following relation as proved in Appendix:
%\begin{align}
%\varepsilon_{s,avg} &= \frac{P_\text{eff}}{|\mathcal{K}_1| + |\mathcal{K}_2|}  \notag \\
%P_\text{elec} &= (2- \frac{1}{\pi}) P_\text{eff} \approx 1.7 P_\text{eff}, \notag \\
%P_\text{DC} &= \frac{1}{\pi} P_\text{eff} \approx P_\text{opt}^2 \approx 0.3 P_\text{eff},  \notag \\
%P_\text{AC} &= (2 - \frac{2}{\pi}) P_\text{eff} \approx 1.4 P_\text{eff},
%\end{align}
%where the last relation $P_\text{AC} = (2 - \frac{2}{\pi}) P_\text{eff} \approx 1.4 P_\text{eff}$ always holds even without the condition. }

%\cc{Note that layer 2 signal ${\mathbf{x}}_{L_2}$ is affected by the imaginary-part clipping noise from layer 1. }

\paragraph{LACO-OFDM Transmitter}

In contrast to ADO-OFDM and HACO-OFDM which \ac{combine} all the muted subcarriers of the first-layer (ACO-OFDM) \ac{into} a single extra layer \ac{(DCO-OFDM layer or PAM-DMT layer, respectivley)}, LACO-OFDM adopts another ACO-OFDM \ac{layer} as the second layer, which again leaves some subcarriers \cc{unloaded}. Then, additional ACO-OFDM layers \ac{can be included using the remaining subcarriers} \cite{laco,see2}. \cc{In detail,} let $\mathbf{S}_{j} ={\mathbf{S}}_\text{aco}^{(j)}$, $j = 1,\ldots,J$, be the QAM symbols of layer $j$, where ${S}_\text{aco}^{(j)}(k)$ is non-zero for \cc{$k \in \mathcal{K}_j = \{ k | k=2^{j-1}(2i-1), i \in\{1, \ldots, \frac{N}{2^j}\} \}$} and zero otherwise. This leads to a maximum number of layers $J=\log{\frac{N}{2}}$ for an $N$-subcarrier system.
%Therefore, layer $j$ uses $\frac{N}{2^j}$ subcarriers and contains $\frac{N}{2^{(j+1)}}$ symbols because of Hermitian symmetry, and the last layer only uses two subcarriers and contains 1 symbol for the same reason. The maximum number of layers available for an $N$-subcarrier system is $J=\log{\frac{N}{2}}$. 
%{\color{red}The following gives a detailed expression of LACO-OFDM construction.} Let the symbols of layer 1 be ${\mathbf{s}}_\text{aco}^{(1)} = {\mathbf{s}}_\text{aco}$, and the symbols of layer $j$ be ${\mathbf{s}}_\text{aco}^{(j)} = [S_\text{aco}^{(j)}(k)]_{k=0}^{N-1}$, where $S_\text{aco}^{(j)}(0) = S_\text{aco}^{(j)}(\frac{N}{2}) = 0$, and $S_\text{aco}^{(j)}(k) =0 $, if $k = 2^j (i-1),\;\; i = 1,\dots, \frac{N}{2^{j}} $. 
Let $ {\mathbf{s}}_\text{aco}^{(j)} = \text{IFFT}\{ {\mathbf{S}}_\text{aco}^{(j)} \} $ and ${\mathbf{x}}_\text{aco}^{(j)}= ( { {\mathbf{s}}_\text{aco}^{(j)} } ) ^+$. Then, $ {\mathbf{x}}_{j} = {\mathbf{x}}_\text{aco}^{(j)}$ and the \ac{overall} LACO-OFDM \ac{signal} is
\begin{equation*}
{\mathbf{x}}_{\rm laco}
= \sum_{j=1}^{J} { {\mathbf{x}}_{j} } 
= \sum_{j=1}^{J} { {\mathbf{x}}^{(j)}_\text{aco} }.
\end{equation*}

\modi{For LACO-OFDM, we have $P_\text{eff} = \mathcal{P}\{\sum_j \frac{\mathbf{s}_\text{aco}^{(j)}}{2}\}$. \ac{Assuming that $P_\text{eff}$ is distributed equally in all effective subcarriers,} i.e., $\mathbb{E}[|\frac{S_\text{aco}^{(j)}(k)}{2}|^2] = \varepsilon$, $\forall j\in\{1,\ldots,J\}$ and $\forall k\in\mathcal{K}_{j}$, for some $\varepsilon$, then we have the relations between $P_\text{elec}$, $P_\text{opt}$, and $P_\text{eff}$ as listed in Table \ref{tab:power}, which are proved in Appendix~\ref{Sec:Appen2}. }

%\cc{When $\varepsilon_{s,avg}^{(j)} = \varepsilon_{s,avg}^{(t)}$, $\forall j,t$, the relations between $P_\text{elec}$, $P_\text{opt}$, $P_\text{DC}$, $P_\text{AC}$ and $P_\text{eff}$ are listed in Table \ref{tab:power}, which are proved in Appendix \ref{Sec:Appen2}. }

%\cc{When $\varepsilon_{s,avg}^{(j)} = \varepsilon_{s,avg}^{(t)}$, $\forall j,t$, we have the following relation as proved in Appendix:
%\begin{align}
%	\varepsilon_{s,avg} &= \frac{P_\text{eff}}{\sum_j |\mathcal{K}_j|}  \notag \\
%	P_\text{elec} & = 2 P_\text{eff}, \notag \\
%	P_\text{DC} &= P_\text{opt}^2 = \frac{2}{\pi} P_\text{eff},  \notag \\
%	P_\text{AC} &= (2 - \frac{2}{\pi}) P_\text{eff}
%\end{align}
%where the last relation $P_\text{AC} = (2 - \frac{2}{\pi}) P_\text{eff} \approx 1.4 P_\text{eff}$ always holds even without the condition. }

\subsubsection{\cc{eACO-OFDM Receiver}}

A unified receiver for eACO-OFDM is introduced in this part, which applies for ADO-, HACO-, and LACO-OFDM. \ac{For the specific receiver of each scheme, the reader is referred to \cite{ado,haco,laco}.} For convenience, we start the analysis after channel-inversion equalization, and write the received signal as
%For convenience, we start the analysis of demodulation process after sampling and channel-inverse equalization, and simply write the received signal as
\begin{equation} \label{eq:y}
	{\bf{y}} = {\mathbf{x}} + {\mathbf{v}},
\end{equation}
where ${\bf{y}}=[y(n)]_{n=1}^{N}$, ${\mathbf{x}}=[x(n)]_{n=1}^{N}$, \ccc{and ${\mathbf{v}}=[v(n)]_{n=1}^{N}$ is the noise after channel-inversion equalization.}
%where ${\bf{y}}=[y(n)]_{n=1}^{N}$, ${\mathbf{x}}=[x(n)]_{n=1}^{N}$ and ${\mathbf{v}}=[v(n)]_{n=1}^{N}$ with $v(n) = h^{-1}(t) \otimes v_0(n)$.

%\cc{Note that layer $j$ signal ${\mathbf{x}}_{L_j}$ is affected by the clipping noise from all fromer layers, layer $t<j$.}

\ac{As shown in Fig. \ref{fig:demodulator}, detection is done iteratively, starting with layer $1$ ($j=1$ in Fig. \ref{fig:demodulator}) followed by layers $j=2,3,\ldots$, respectively, until symbols of all layers are detected.}

\ac{The signal used for detection of symbols in layer $j$ is given by}
\begin{equation}\label{eq:yj}
	{\bf{y}}_{j-1} = {\bf{y}} - \sum_{t=1}^{j-1} { \hat{\mathbf{x}}_{t} },
\end{equation}
for $j>1$, and by ${\bf{y}}_0 = {\bf{y}}$ for $j=1$. Thus, $\mathbf{y}_{j-1}$ can be understood as the remaining signal after detecting \ac{and subtracting layers $1,\ldots,j-1$} and will be used to detect layer $j$.

\ac{The signal $\mathbf{y}_{j-1}$ is first transformed to the frequency domain (FFT). Then, the effective subcarriers of layer $j$ are selected (layer-$j$ selection) leading to $\mathbf{Y}_{j}$, and the symbols loaded onto these subcarriers are detected (layer-$j$ detection) leading to $\hat{\mathbf{S}}_{j}$. To remove the contribution of this layer from the subsequent layers, $\hat{\mathbf{S}}_{j}$ is remodulated (layer-$j$ modulation) leading to $\hat{\mathbf{x}}_{j}$, which is then subtracted from $\mathbf{y}_{j-1}$ leading to $\mathbf{y}_j$.}

%After FFT, symbols in layer $j$, ${\bf{Y}}_{L_j}$, are selected, and maximum likelihood detection is applied to obtain $\hat{\mathbf{S}}_{L_j}$. Then $\hat{\mathbf{x}}_{L_j}$, which is a reconstruction of ${\mathbf{x}}_{L_j}$, can be obtained by re-modulating $\hat{\mathbf{S}}_{L_j}$ using the specific modulation scheme of layer $j$. 

\ac{This receiver process induces distortion due to residual clipping noise (RCN) as follows.} The first ACO-OFDM layer has clipping noise in even subcarriers. This will interfere with the layers $2,\ldots,J$. If $\hat{\mathbf{x}}_{1}$ is not a perfect reconstruction of ${\mathbf{x}}_{1}$ \ac{(due to detection errors in layer $1$)}, then $\mathbf{y}_1$ in \eqref{eq:yj} will contain RCN from the first layer. This will only affect layer 2 in ADO-OFDM and HACO-OFDM. Since LACO-OFDM may have $J>2$ layers of ACO-OFDM, RCN is accumulated layer by layer. 

\ac{To understand the effects of RCN on the performance of all eACO-OFDM schemes, the next section analyses RCN in detail.}

\section[RCNana]{Analysis on Residual Clipping Noise} \label{Sec:ana-rcn}
% 1. In this section RCN's statistic characters, power estimation model are given. 
% 2. RCN is derived in two-layer system first then multi-layer system. 
% 3. The accuracy of estimation is dependent on how large the neighboring region of a QAM point are considered in detection error estimation. 

\ac{In this section, we analyse RCN by deriving its mathematical expression, proving some statistical properties, and then using \modi{these} to model the power of RCN in each layer, \modi{which enables the estimation of RCN power and total noise power in each subcarrier.}}

\subsection{Mathematical Expression of RCN and Total Noise} \label{Sec:RCN math}

In the eACO-OFDM receiver, the output of {the detection} process in Fig. \ref{fig:demodulator} can be written as 
\cc{\begin{equation} \label{eq:hat-s-L-j}
	\hat{\mathbf{S}}_{j} = {\mathbf{S}}_{j} + {\mathbf{E}}_{j},
\end{equation}
where ${\mathbf{E}}_{j} = [E_{j}(k)]_{k=0}^{N-1}$ \ac{and $E_{j}(k)$ is the detection-error in subcarrier $k$ of layer $j$}. In the time domain, we have $\hat{\mathbf{s}}_{j} = {\mathbf{s}}_{j} + {\mathbf{e}}_{j}$, where $\hat{\mathbf{s}}_{j} = \text{IFFT}\{\hat{\mathbf{s}}_{j}\}$ and ${\mathbf{e}}_{j} = \text{IFFT}\{{\mathbf{E}}_{j}\}$.}  Since the first layer is an ACO-OFDM layer, {then $\hat{\mathbf{x}}_{1}$ is given by}
\begin{align}
\hat{\mathbf{x}}_{1} 
= ( \hat{\mathbf{s}}_{1} ) ^+  
= \frac{1}{2} ( {\mathbf{s}}_{1} + {\mathbf{e}}_{1} + | {\mathbf{s}}_{1} + {\mathbf{e}}_{1} | ).
\end{align}
Then, from \eqref{eq:yj}, the second iteration at the receiver starts from 
\begin{align} \label{eq:y1}
	{\bf{y}}_1 &= \sum_{t=1}^{J} { \mathbf{x}}_{t} + {\mathbf{v}} - \hat{\mathbf{x}}_{1} 
	\nonumber \\
	&= \sum_{t=2}^{J} { \mathbf{x}}_{t} + {\mathbf{v}} + \frac{1}{2} ( {\mathbf{s}}_{1} + | {\mathbf{s}}_{1} | ) - \frac{1}{2} ( {\mathbf{s}}_{1}  + {\mathbf{e}}_{1} + | {\mathbf{s}}_{1} + {\mathbf{e}}_{1} | )
	\nonumber \\
&= \sum_{t=2}^{J} { \mathbf{x}}_{t} + {\mathbf{v}} - \frac{1}{2}{\mathbf{e}}_{1}  + \frac{1}{2} ( |{\mathbf{s}}_{1}| - | {\mathbf{s}}_{1} + {\mathbf{e}}_{1} | ).
\end{align}

In \eqref{eq:y1}, $\bf{v}$ interferes with all layers; \ac{$\frac{1}{2}{\bf{e}}_{1}$ has a frequency-domain counterpart $\frac{1}{2}\mathbf{E}_{1}$ which is only nonzero in the effective subcarriers of layer 1 \ac{(odd sybcarriers)}, and thus will not interfere with layers higher than 1}. The term $\frac{1}{2} ( |{\mathbf{s}}_{1}| - | {\mathbf{s}}_{1} + {\mathbf{e}}_{1} | )$ is the RCN which interferes {with layers} $j>1$. 
%Thus, the total noise after removing the first layer can be denoted by
%\begin{equation} \label{eq:z1}
%	{\bf{z}}_1 = {\mathbf{v}} + {\boldsymbol{\delta}}_{1}.
%\end{equation}

\ac{To generalize this, consider iteration $j+1$ at the receiver. From \eqref{eq:yj}, this iteration starts from}
\begin{align} \label{eq:yj_full}
{\bf{y}}_j 
	&= \sum_{t=1}^{J} { {\mathbf{x}}_{t} }+ {\mathbf{v}} - \sum_{t=1}^{j} { \hat{\mathbf{x}}_{t} } 
	\nonumber \\
	&= \sum_{t>j}^{J} { {\mathbf{x}}_{t} }  + {\mathbf{v}} - \frac{1}{2}\sum_{t=1}^{j} { {\mathbf{e}}_t } + \sum_{t=1}^{j} { \frac{1}{2} ( |{\mathbf{s}}_{t}| - | {\mathbf{s}}_{t} + {\mathbf{e}}_t | ) } .
\end{align}
In \eqref{eq:yj_full}, $\frac{1}{2}\sum_{t=1}^{j} { {\mathbf{e}}_t }$ is \ac{the time-domain error signal induced by the} detection errors in {layers} $1,\ldots,j$, whose \ac{frequency-domain counterpart is nonzero only in the effective subcarriers of layers $1,\ldots,j$,} and thus does not interfere with layers $j+1,\ldots,J$. The term $\sum_{t=1}^{j} { \frac{1}{2} ( |{\mathbf{s}}_{t}| - | {\mathbf{s}}_{t} + {\mathbf{e}}_t | ) }$ in \eqref{eq:yj_full} is the RCN caused by layers $1,\ldots,j$, which interferes with all layers $j+1,\ldots,J$. For convenience, we denote the RCN from layer $t$ as 
\begin{equation}\label{eq:rcn}
	{\boldsymbol{\delta}}_{t} =  { \frac{1}{2} ( |{\mathbf{s}}_{t}| - | {\mathbf{s}}_{t} + {\mathbf{e}}_t | ) }. 
\end{equation}
\emph{Note that $|\delta_t(n)| \le \frac{1}{2}|e_t(n)|$ which can be shown using the reverse triangle inequality}.\footnote{\ccc{Reverse triangle inequality states that,} $\forall a,b \in \mathbb{R}$, $\big||a|-|b|\big| \le |a-b|$.}

\ac{Using this, we can write the total noise after removing layer $j$ as}
\begin{equation}\label{eq:z}
	{\bf{z}}_j = {\mathbf{v}} + \sum_{t=1}^{j} { {\boldsymbol{\delta}}_{t} },
\end{equation}
for $j >0$, and ${\bf{z}}_0 = {\mathbf{v}}$ for $j=0$. 

Note that \ac{the total noise expression \eqref{eq:z} applies for} the three eACO-OFDM {schemes}, with { $j\in \{0,1\}$} for ADO-OFDM and HACO-OFDM; and {$j\in \{0, \ldots, J-1\}$} for LACO-OFDM. \ac{Next, we study some statistical properties of RCN, which are useful for \ccc{modeling the RCN power}.}

\subsection{Statistical Properties of RCN} \label{sec:statistics_analysis}

RCN \modi{$\boldsymbol{\delta}_t$} and noise $\mathbf{v}$ result in detection errors. The error rate can be evaluated analytically for {a given QAM constellation} if the statistics of total noise is known \cite{book-digitalComm5th}. \ac{This necessitates studying the statistics of RCN. In this section,} we \cc{provide} three statistical properties of RCN which will facilitate error rate evaluation of eACO-OFDM \ac{(discussed next section)}. 

First, we state two lemmas which are required to prove statistical properties of RCN, and can be proved using the Lyapunov central limit theorem \cite[Example 27.4]{book-Partrick}.

\begin{lemma}\label{Lem1}
	Consider $N$ complex-valued random variables $X(k)$, $k=0,1,\ldots, N-1$, that satisfy $\mathbb{E}[X(k)]=0$ and $|X(k)| \le K <\infty$ $\forall k$, where $K$ is a constant, and Hermitian symmetry, i.e., \cc{$X(k) = X^*(N-k)$} for \cc{$0< k < \frac{N}{2}$}. Let $\mathbb{V}[ X(k) ] = \sigma_X^2(k)$ and $x(n) =  \frac{1}{N} \sum_{k=0}^{N-1} X(k) e^{ \mathsf{j} \frac{2\pi}{N}kn }$. Then $x(n) \xrightarrow{d} \mathcal{N}\big(0, \frac{1}{N^2} \sum_{k=0}^{N-1} \sigma_X^2(k)\big)$ as $N \rightarrow \infty$.\footnote{$\xrightarrow{d}$ represents convergence in distribution.}
\end{lemma}

\begin{lemma}\label{Lem2}	
 	Consider $N$ \textit{i.i.d.} real-valued  random variables $x(n)$, $n=0,1,\dots, N-1$, that have mean $\mu_x$, variance $\sigma_x^2$, and satisfy $|x(n)| \le A < \infty$ $\forall n$, where $A$ is a  constant. Let $X(k) =  \sum_{n=0}^{N-1} x(n) e^{ -\mathsf{j} \frac{2\pi}{N}kn }$. Then, $X(k) \xrightarrow{d} \mathcal{CN}(0,  N \sigma_x^2)$ when $N \rightarrow \infty$ \cc{$\forall k\notin\{0,\frac{N}{2}\}$,} and \cc{$X(\frac{N}{2}) \xrightarrow{d} \mathcal{N}(0,  N \sigma_x^2)$}.
\end{lemma}

Now let ${\boldsymbol{\Delta}}_{t} = [\Delta_t(k)]_{k=0}^{N-1}= \text{FFT} \{ \boldsymbol{\delta}_t \}$ and ${\mathbf{E}}_t = [E_t(k)]_{k=0}^{N-1}= \text{FFT} \{ \mathbf{e}_t \}$ \cc{be the IFFT of $\boldsymbol{\delta}_t$ and $\mathbf{e}_t$, respectively}, \ac{and let $\mathcal{B}_t \triangleq \{ k | k=2^{t-1}2(i-1), i \in\{ 1, \ldots, \frac{N}{2^t}\}, k\neq 0, k\neq \frac{N}{2} \}$ denote the set of subcarriers affected by RCN from layer $t$.} Then for an eACO-OFDM system with Gaussian noise, \ccc{channel-inversion} equalization, QAM and ML-detection, we have the following \textit{three properties} of RCN:
\begin{enumerate}
	\item  When $|\mathcal{K}_t|$ is large enough, \cc{the time-domain RCN} $\delta_t(n)$ {is} independent and identical distributed (\emph{i.i.d.}) with respect to $n$;
	\item  When $|\mathcal{K}_t|$ is large enough, \cc{the frequency-domain RCN} \modi{$\Delta_t(k) \xrightarrow{d} \mathcal{CN}(0, N \ac{\sigma_{\delta_t}^2})$, for all $k \in \mathcal{B}_t$, \ac{where $\sigma_{\delta_t}^2=\mathbb{V}[\delta_t(n)]$ which is independent of $n$ by property 1};	}
	%is identically distributed \cc{for all $k\in\mathcal{B}_t$} and converges in distribution to a {circularly symmetric complex Gaussian \cc{with zero-mean} \ac{and variance $N \mathbb{V}[\delta_t(n)]$};}
	\item \ac{The covariance of the frequency-domain RCN from layers $t_1$ and $t_2$ is negligible for any $k$ and $t_2\neq t_1$. Since $\mathbb{E}\{\Delta_{t}(k)\}\ccc{=} 0$ by property 2, this implies that $\mathbb{E}\{\Delta_{t_1}(k)\Delta_{t_2}^*(k)\}\approx 0$.}
\end{enumerate}
The first two properties can be proved using Lemmas \ref{Lem1} and \ref{Lem2}. The third property \ac{will not be proved, but is supported by simulations}. 

\cc{\textit{Proof of Property 1 and 2:} First recall from the definition of the time-domain RCN in \eqref{eq:rcn} that ${\boldsymbol{\delta}}_{t} =  { \frac{1}{2} ( |{\mathbf{s}}_{t}| - | {\mathbf{s}}_{t} + {\mathbf{e}}_t | ) }$.} %, where $s_{L_t}(n) =\frac{1}{N} \sum_{k=0}^{N-1} S_{L_t}(k) e^{ \mathsf{j} \frac{2\pi}{N}kn }$ and $e_t(n) =\frac{1}{N} \sum_{k=0}^{N-1} E_t(k) e^{ \mathsf{j} \frac{2\pi}{N}kn }$. 
The proof \ac{follows five steps which show that (i) $s_{t}(n)$ is \emph{i.i.d.} Gaussian, (ii) $e_1(n)$ is \emph{i.i.d.} Gaussian, (iii) $\delta_1(n)$ is \emph{i.i.d.}, and (iv) $\Delta_1(k)$ is zero-mean circularly-symmetric complex Gaussian identically $\forall k\in \mathcal{B}_1$. Then, we use these steps to prove properties 1 and 2 by induction.}

Step (i): As QAM symbols are transmitted on $S_{t}(k)$ equiprobably \ac{for $k\in\mathcal{K}_t$ and $S_{t}(k)=0$ $\forall k\notin\mathcal{K}_t$,} it follows that $S_{t}(k)$ has zero-mean and is bounded. Moreover, $\mathbf{S}_{t}$ has Hermitian-symmetry. Then it follows from Lemma \ref{Lem1} that $s_{t}(n) \xrightarrow{d} \mathcal{N}\big(0, \frac{1}{N^2} \sum_{k=0}^{N-1} \mathbb{V}[ S_{t}(k) ] \big)$ as $|\mathcal{K}_t| \rightarrow \infty$ \ac{for all $n$. Moreover, due to the} approximate orthogonality of the vectors $[ e^{ \mathsf{j}\frac{2\pi}{N}k n_1 } ]_{k=0}^{N-1}$ and $[ e^{ \mathsf{j}\frac{2\pi}{N}k n_2 } ]_{k=0}^{N-1}$ when $N \rightarrow \infty$ (which holds if $|\mathcal{K}_t|\to\infty$), it follows that $s_{t}(n_1)$ and $s_{t}(n_2)$ are independent, $\forall$ $n_1 \ne n_2$ \ac{when $|\mathcal{K}_t|$ is large.}

Step (ii): From \eqref{eq:hat-s-L-j}, \ac{we have that} $E_1(k) = \hat{S}_{1}(k) - S_{1}(k)$ for $k\in\mathcal{K}_1$ and $E_1(k)=0$ otherwise, where $\hat{S}_{1}(k)$ is the detection outcome of $S_{1}(k)$. \ac{Since $\mathbf{S}_{1}$ is Hermitian-symmetric, and the real-valued time-domain} noise has a Hermitian-symmetric frequency-domain counterpart, it follows that $\hat{\mathbf{S}}_{1}$ is Hermitian-symmetric, \ac{and hence} $\mathbf{E}_1$ is also Hermitian-symmetric. \ac{Note also that since $\hat{S}_{1}(k)$ and $S_{1}(k)$ are bounded (bounded QAM constellation), then $E_1(k)$ is also bounded.} Now since $Z_0(k) = V(k)$ \ac{(cf. \eqref{eq:z})} has zero-mean and a symmetric probability density function, then $E_1(k)$ has zero-mean. \ac{Using Lemma 1, we conclude that} $e_1(n) \xrightarrow{d} \mathcal{N} \big( 0, \frac{1}{N^2} \sum_{k=0}^{N-1} \mathbb{V}[ E_1(k) ] \big)$ as $|\mathcal{K}_1| \rightarrow \infty$ \ac{for any $n$.} The independence between $e_1(n_1)$ and $e_1(n_2)$, $\forall$ $n_1 \ne n_2$ when $N \rightarrow \infty$ can be proved similar to $s_{t}(n)$ in the previous paragraph.

Step (iii): Note that $\delta_1(n) =  \frac{1}{2} ( |s_{1}(n)| - | s_{1}(n) + e_1(n) | ) $, where $s_{1}(n)$ and $e_1(n)$ are respectively \emph{i.i.d.} \ac{(from (i) and (ii)). Moreover, the independence of $s_{1}(n_1)$ and $e_1(n_2)$ for any $n_1\neq n_2$ when $N \rightarrow \infty$ can be proved similar to $s_{t}(n)$ in step (i). Thus,} $\delta_1(n)$ is \emph{i.i.d.}. \ac{Denote its variance by $\sigma_{\delta_1}^2$.}

Step (iv): %Due to the ACO-OFDM \ac{construction}, the frequency-domain RCN $\Delta_t(k)$ is non-zero  $k \in \mathcal{G}_t \triangleq \{ k | k=2^{t-1}2(i-1), i = 1, \dots, \frac{N}{2^t} \}$, where $\Delta_t(k)$ is complex-valued. 
For $k \in \mathcal{B}_1$, $\Delta_1(k) = \sum_{n=0}^{N-1} {\delta_1(n) e^{ -\mathsf{j}\frac{2\pi}{N}kn } }$, where $\delta_1(n)$ is \emph{i.i.d.} (cf. step (iii)). Since $|\delta_1(n)| \le \frac{1}{2}|e_1(n)|$ \ac{(cf. \eqref{eq:rcn})} and $e_1(n)$ is a Gaussian random variable with small variance for large $|\mathcal{K}_1|$ \ac{(cf. proof of (ii))}, then $|\delta_1(n)|$ is bounded for large $|\mathcal{K}_1|$. Thus, from Lemma \ref{Lem2}, we have $\Delta_1(k) \xrightarrow{d} \mathcal{CN}(0,  N \ac{\sigma_{\delta_1}^2})$ as $|\mathcal{K}_1|\to\infty$ \ac{for all $k\in\mathcal{B}_1$.} 

\ac{Now the above steps can be used to prove properties 1 and 2 by induction.} Since $\Delta_1(k)$ is zero-mean Gaussian for $k\in\mathcal{B}_t$ (iv), then $Z_t(k) = V(k) + \sum_{i=1}^{t} \Delta_i(k)$ with $t=1$ is zero-mean Gaussian for $k\in\mathcal{B}_1$. Following the same lines of steps (ii), (iii), and (iv), it follows that $\delta_2(n)$ is \emph{i.i.d.} and $\Delta_2(k)$ is zero-mean Gaussian for $k\in\mathcal{B}_2$. Repeating these steps for all $t=2,3,\ldots$ with $|\mathcal{K}_t|$ large proves properties 1 and 2.

\textit{Evidence of Property 3:} {We do not provide a formal proof of property 3, but rather an intuitive explanation supported by numerical simulations in Fig. \ref{fig:cor}. As system noise $\mathbf{v}$ is always the major stimulator of detection errors, then ${\Delta}_{t}(k)$ depends more on system noise than on RCN from lower layers. Since system noise is independent across layers, ${\Delta}_{t}(k)$ will be almost independent across layers. Fig. \ref{fig:Prcn_cmp_flat_major} shows that system noise power is much larger than RCN power for a wide range of SNR. This supports property 3.

\begin{remark}
Properties 1-3 do not require equal bit-loading and uniform noise power spectrum. 
\end{remark}

\begin{remark}
Properties 1-3 are only exact asymptotically when $|\mathcal{K}_t|\rightarrow \infty$, and become less precise when $|\mathcal{K}_t|$ is small. Nevertheless, the approximation is still acceptable for small $|\mathcal{K}_t|$ as shown in Fig. \ref{fig:rcn_pdf}, \ref{fig:Prcn_cmp_flat_major}, and \ref{fig:Prcn_cmp_fs_major}.%, and it provides good performance approximations even for small $|\mathcal{K}_t$ such as 4, 8, 16.}
\end{remark}

\begin{remark}\label{Rem3}
Note that we also have $\Delta_t(\frac{N}{2}) \xrightarrow{d} \mathcal{N}(0,  N \sigma_{\delta_t}^2)$ which can be proved by Lemma \ref{Lem2}. Moreover, \modi{as $\delta_t(n)$ is i.i.d.}, it is straightforward to prove that $\Delta_t(0) = \sum_{n=0}^{N-1}\delta_t(n)$ satisfies \modi{$\mathbb{E}[\Delta_t(0)] =N\mathbb{E}[\delta_t(n)]$ and $\mathbb{V}[\Delta_t(0)] =  N \ac{\sigma_{\delta_t}^2}$ which is equal to $\mathbb{V}[\Delta_t(k)]$}, \ac{ $\forall k\in\mathcal{B}_t\cup\{\frac{N}{2}\}$}. This will be needed in Sec. \ref{subsec:Prcn_est}.
\end{remark}

\subsection{Worst-case RCN Power Model} \label{subsec:Prcn_est}

Recall that $|{\delta}_{t}(n)| \le \frac{1}{2}|e_t(n) |$, \modi{which implies
\begin{equation} \label{eq:Prcnle_t}
	\mathcal{P}\{ \boldsymbol{\delta}_{t} \} \le \frac{1}{4}\mathcal{P}\{ \mathbf{e}_t \} = \frac{1}{4N}\mathcal{P}\{\mathbf{E}_t\} = \frac{1}{4N^2}\sum_{k=0}^{N-1}\mathcal{P}\{E_t(k)\},
\end{equation}
\ac{from Parseval's theorem.} This forms a worst-case model for the power of the time-domain RCN $\mathcal{P}\{ \boldsymbol{\delta}_{t} \}$ in terms of $\mathcal{P}\{E_t(k)\}$. \ac{A worst-case model of the power of the frequency-domain RCN can be obtained using  $\mathcal{P}\{\boldsymbol{\Delta}_t\} = N \mathcal{P}\{\boldsymbol{\delta}_t\}$ from Parsevel's theorem. It remains to derive a worst-case model of the power of the frequency-domain RCN per subcarrier, i.e., $\Delta_t(k)$.}} 

\ac{Using Remark \ref{Rem3} \modi{and property 2}, we have $\mathcal{P}\{\Delta_t(0)\} =\mathbb{V}[\Delta_t(0)]+\mathbb{E}^2\{\Delta_t(0)\}\geq\mathbb{V}[\Delta_t(k)] = \mathcal{P}\{\Delta_t(k)\}$ for any $k \in \mathcal{B}_t$. 
%Also, using Remark \ref{Rem3} and Property 2, we have $ \mathcal{P}\{\Delta_t(\frac{N}{2})\} = \mathcal{P}\{\Delta_t(k)\}$ for any $k \in \mathcal{B}_t$. 
Thus, $\mathcal{P}\{\Delta_t(0)\}\geq \mathcal{P}\{\Delta_t(\frac{N}{2})\}= \mathcal{P}\{\Delta_t(k)\}$ for any $k \in \mathcal{B}_t$. Moreover, $\mathcal{P}\{\Delta_t(k)\} = 0$ for $k \notin \mathcal{B}_t \cup \{0, \frac{N}{2}\}$. Using these, and since $\Delta_t(k)$ is i.i.d. for $k\in\mathcal{B}_t$ (property 2), then $\mathcal{P}\{\boldsymbol{\Delta}_t\}=\frac{1}{N}(\mathcal{P}\{\Delta_t(0)\}+(|\mathcal{B}_t|+1)\mathcal{P}\{\Delta_t(k)\})\geq \frac{|\mathcal{B}_t|+2}{N}\mathcal{P}\{\Delta_t(k)\}$ for any $k\in\mathcal{B}_t$. Then, since $\mathcal{P}\{\boldsymbol{\Delta}_t\} = N \mathcal{P}\{\boldsymbol{\delta}_t\}$, we obtain the following for $k \in \mathcal{B}_t$
\begin{align} \label{eq:Prcnle}
	\mathcal{P}\{\Delta_t(k)\} \le \frac{N^2}{|\mathcal{B}_t| + 2} \mathcal{P}\{ \boldsymbol{\delta}_{t} \} \le \frac{N^2}{|\mathcal{B}_t| + 2} \frac{1}{4N^2}\sum_{\tilde{k}=0}^{N-1}\mathcal{P}\{E_t(\tilde{k})\} = \frac{1}{4|\mathcal{K}_t|}\sum_{\tilde{k}\in \mathcal{K}_t} \mathcal{P}\{E_t(\tilde{k})\},
\end{align}
where the last equality follows since $\mathcal{P}\{E_t(k)\} = 0$ for $k \notin \mathcal{K}_t$ and since $|\mathcal{K}_t|=|\mathcal{B}_t| + 2$. This provides a worst-case estimation of $\mathcal{P}\{\Delta_t(k)\}$ in terms of the detection error power $\mathcal{P}\{E_t(k)\}$. Next, we relate $\mathcal{P}\{E_t(k)\}$ with the QAM constellation and noise power using the following definition.}

\begin{definition} \label{definition}
For a complex AWGN channel \ac{$Y=X+Z$ with input $X$ uniformly distributed over an $M$-QAM constellation with a minimum Euclidean distance $d_\text{min}$, Gaussian noise $Z$ with variance $\sigma^2$, and detection outcome $\hat{X}$, let $\mathsf{f}(d_\text{min}, \sigma^2, M) $ be a function which maps $(d_\text{min}, \sigma^2, M)$ to a real number which equals the power of detection errors over this channel, i.e., $\mathsf{f}(d_\text{min}, \sigma^2, M)=\mathbb{E}\{|X-\hat{X}|^2\}$.} 
\end{definition}

A method for approximating $\mathsf{f}(d_\text{min}, \sigma^2, M) $ is described in Appendix \ref{Sec:Appen1} and will be used in the simulations. %\ac{Now, we can write $\mathcal{P}\{E_{t}(k) \} = \cc{\mathsf{f}}(d_\text{min}, \sigma^2, M)$, which can be used to model $\mathcal{P}\{ \Delta_{t}(k) \}$.}

%Come back to the estimation of $\mathcal{P}\{ E_t(k) \}$. 
Denote the QAM constellation size used in subcarrier $k$ of layer $t$ by $M_t(k)$. Suppose that before detection, ${\bf{Y}}_{j}$ is scaled by 2 to {recover} the {original} scale of {the constellation} before clipping\footnote{Recall that the clipping in ACO-OFDM halves the amplitude of the effective subcarriers, as described in Sec. \ref{Sec:Back}.}. Then, the power of noise in subcarrier $k$ of layer $t$ becomes $4\mathcal{P}\{ Z_{t-1}(k) \}$ (to be derived in the next subsection). The minimum Euclidean distance of the constellation in subcarrier $k$ \ac{can be written as} \cite{book-digitalComm5th} 
\begin{equation} \label{eq:de}
	d_t(k) = \sqrt{ \frac{ 6 \mathcal{P}\{S_{t}(k)\} } { M_t(k) - 1 } }.
\end{equation}
Note that \eqref{eq:de} is only accurate for square QAM constellations or rectangular QAM constellations with a large constellation size \cite{book-digitalComm5th}. Using Definition \ref{definition}, we have
\begin{equation} \label{eq:P_Et}
	\mathcal{P}\{ E_t(k) \} = \cc{\mathsf{f}}\big( d_t(k) , \modi{4}\mathcal{P}\{ Z_{t-1}(k) \}, M_t(k) \big) .
\end{equation}

Then from \eqref{eq:Prcnle}, we have
\begin{equation} \label{eq:prcn_le}
	\mathcal{P}\{ {{\Delta}}_{t}(k)\} \le P_{t}^{\rm w} \triangleq \frac{1}{4 \cc{|\mathcal{K}_t|}} \cc{\sum_{\ccc{\tilde{k}} \in \mathcal{K}_t}} {\cc{\mathsf{f}} \big( d_t(\tilde{k}) , \modi{4}\mathcal{P}\{ Z_{t-1}(\tilde{k}) \}, M_t(\tilde{k}) \big)  },
\end{equation}
\ccc{where $P_{t}^{\rm w}$ represents the worst-case RCN power. For each $t$, $P_{t}^{\rm w}$ is the same for $k\in\mathcal{B}_t$ and is always 0 for $k\notin\mathcal{B}_t$.}

%{
%	As a remark, from \eqref{eq:prcn_le}, we have the following relation between worst-case RCN power and $\frac{\mathcal{P}\{S_{L_t}(k)\}}{\mathcal{P}\{ Z_{t-1}(k) \}}$:
%	\begin{equation}
%	\frac{P_{t}^{w}}{\mathcal{P}\{ Z_{t-1}(k) \}} \propto \frac{\mathcal{P}\{S_{L_t}(k)\}}{\mathcal{P}\{ Z_{t-1}(k) \}} \text{Q}\left(\epsilon(k) \frac{\mathcal{P}\{S_{L_t}(k)\}}{\mathcal{P}\{ Z_{t-1}(k) \}} \right) \le ,
%	\end{equation}
%	where 
%}	

\subsection{Worst-Case Total Noise Power}

The frequency-domain counterpart of \eqref{eq:z} is 
\begin{align}\label{eq:z_f}
	Z_j(k) = {V(k)} + \sum_{t=1}^{j} { {\Delta}_{t}(k) }, \;\; j>0,
\end{align}
Thus,
\begin{align} \label{eq:p_z_k_part1}
	\mathcal{P}\{ Z_j(k) \} 
	%= \mathbb{E}\left\{ |{Z_j(k)}|^2 \right\} %\nonumber\\
	= \mathbb{E}[ {Z_j(k)} {Z_j^*(k)} ] %\nonumber \\
	= \mathbb{E} \big[ [{V(k)} + \sum_{t=1}^{j} { {\Delta}_{t}(k) }] [{V(k)} + \sum_{t=1}^{j} { {\Delta}_{t}(k) }]^* \big].
\end{align}
\modi{Since ${\Delta}_{t}(k)$ is RCN from layer $t-1$ and $V(k)$ is noise in layer $t$, }
%\ac{Since $V(k)$ will not introduce RCN in subcarrier $k$ (but to other subcarriers through RCN), then} 
${\Delta}_{t}(k)$ is independent of $V(k)$ for \cc{$k\in \mathcal{B}_t$}. Also, \ac{$\mathbb{E} [\Delta_{t_1}(k)\Delta_{t_2}^*(k)] \approx 0$ for $t_1 \ne t_2$ (Property 3)}. Then, \eqref{eq:p_z_k_part1} \ac{can be written as}
\begin{align} \label{eq:Pzjle}%\label{eq:p_z_k_part2}
	\mathcal{P}\{ Z_j(k) \} = \mathcal{P} \{V(k)\} + \sum_{t=1}^{j} { \mathcal{P} \{ {\Delta}_{t}(k) \} \modi{\le \mathcal{P} \{V(k)\} + \sum_{t=1}^{j} { P_{t}^{\rm w} }} },
\end{align}
\modi{where the last inequality follows from \eqref{eq:prcn_le}.}
%and using \eqref{eq:prcn_le}, we obtain
%\begin{align}\label{eq:Pzjle}
%	\mathcal{P}\{ Z_j(k) \} \le \mathcal{P} \{V(k)\} + \sum_{t=1}^{j} { P_{t}^{\rm w} }.
%\end{align}
%With both \eqref{eq:Prcnle} and \eqref{eq:Pzjle}, $\mathcal{P}\{ Z_j(k) \}$ and $P_{t}^{w}$ can be obtained iteratively, starting from $t=0$, e.g., $P_{0}^{w} = 0$, $\mathcal{P}\{ Z_0(k) \} = \mathcal{P}\{ V(k) \}$; $P_{1}^{w} = \frac{1}{2N} \sum_{k=0}^{N-1} {P_e \big( d_{L_1}(k) , 2\mathcal{P}\{ V(k) \}, M_1(k) \big)  }$, $\mathcal{P}\{ Z_1(k) \} \le \mathcal{P}\{ V(k) \} + P_{1}^{w}$; and so on.

Now, using \eqref{eq:prcn_le} and \eqref{eq:Pzjle}, we can %\modi{estimate}
\ac{upper bound} the RCN power $\mathcal{P}\{ {{\Delta}}_{t}(k)\}$ and the total noise power $\mathcal{P}\{ Z_j(k) \}$. \ac{The bounds can be used to obtain worst-case values of RCN and total noise power iteratively, as follows. For $t=0$, we have $P_{0}^{\rm w} = 0$ and $\mathcal{P}\{ Z_0(k) \} = \mathcal{P}\{ V(k) \}$. Consequently, for $t=1$, we obtain $P_{1}^{\rm w} = \frac{1}{2N} \sum_{k\in\mathcal{K}_1} \mathsf{f} \big( d_{1}(k) , \ac{4}\mathcal{P}\{ V(k) \}, M_1(k) \big)  $ and $\mathcal{P}\{ Z_1(k) \} \le \mathcal{P}\{ V(k) \} + P_{1}^{\rm w}$. Using $P_1^{\rm w}$ and $\mathcal{P}\{ V(k) \} + P_{1}^{\rm w}$ as worst-case RCN and total noise powers, respectively, we obtain $P_{2}^{\rm w} = \frac{1}{N} \sum_{k\in\mathcal{K}_2} \mathsf{f} \big( d_2(k) , 4(\mathcal{P}\{ V(k)\}+P_1^{\rm w}), M_2(k) \big)$ and $\mathcal{P}\{ Z_2(k) \} \le \mathcal{P}\{ V(k) \} + P_{1}^{\rm w}+P_2^{\rm w}$ for $t=2$, and so on.} 

The quality of this worst-case RCN power model will be evaluated in Sec. \ref{sec:sim-rcn-est}. In the following sections, we use the RCN power model for performance analysis and system optimization of eACO-OFDM.

\section{Symbol Error Probability of eACO-OFDM} \label{Sec:serEst}
% -> to be updated with $\varepsilon$ and $\sigma$, and stating something such as, for a M-QAM or M-PAM symbol with power $\varepsilon$ disturbed by AWGN with variance $\sigma^2$, ML detection will result to an error probalibty of: xx, xx, respectively.

In this section, we derive the theoretical symbol error probabilities of the three eACO-OFDM schemes \ac{assuming an} AWGN channel and ML detection, \ac{while taking RCN into account.} 

Given an arbitrary $M$-QAM or $M$-PAM symbol of average power $\varepsilon$ transmitted through an AWGN channel with noise variance $\sigma^2$, and given a receiver which adopts ML detection, the symbol error rate is given by 
%Referring to \cite{book-digitalComm5th}, the error probabilities of $M$-QAM and $M$-PAM are
\cite{book-digitalComm5th} 
\begin{align} 
	\mathsf{p}_{\rm e}^{\rm QAM}(M, \varepsilon, \sigma^2) 
	&= 4 \frac{\sqrt{M}-1}{\sqrt{M}} \mathsf{Q}\left( \frac{\sqrt{3\varepsilon}}{\sqrt{M-1}\sigma}\right)% \nonumber\\
	%&\quad\quad  
	\cdot \left[ 1-\frac{\sqrt{M}-1}{\sqrt{M}} \mathsf{Q}\left(  \frac{\sqrt{3\varepsilon}}{\sqrt{M-1}\sigma} \right) \right], \label{eq:pe_qam}\\
\mathsf{p}_{\rm e}^{\rm PAM}(M,\varepsilon, \sigma^2) &=  2\frac{\sqrt{M}-1}{\sqrt{M}} \mathsf{Q}\left( \frac{\sqrt{6\varepsilon}}{\sqrt{M^2-1}\sigma}\right), \label{eq:pe_pam}
\end{align}
\ac{Using these expressions, and the worst-case RCN power model from last section, we can evaluate the performance of eACO-OFDM schemes as described next.}

%where $\cc{\mathsf{Q}}(x) = \frac{1}{\sqrt{2\pi}}{\bigintssss}_{x}^{+\infty}{e^{ -\frac{t^2}{2} }dt}$.} Note that \eqref{eq:pe_qam} is only accurate for square QAM or rectangular QAM with large constellation size \cite{book-digitalComm5th}.
%
%\cc{In the following, the theoretical average SER over one OFDM block for each eACO-OFDM scheme will be given for evaluating system performance. The calculation requires constellation size, allocated power, and noise power of each subcarrier.} 
%For convenience, we term the non-zero subcarriers of each layer before clipping as effective subcarrier, i.e., the non-zero subcarriers in $\mathbf{S}_{L_j}$ (c.f. Sec. \ref{sec:back-ml-oofdm}).

\subsection{ADO-OFDM}
Both layers of ADO-OFDM use QAM. Denote the constellation size of subcarrier $k$ of layer $j$ by $M_j(k)$. In the first layer, the signal amplitude is halved because of clipping, \ac{and hence $\varepsilon$ in \eqref{eq:pe_qam} is equal to $\frac{\mathcal{P}\{ S_{1}(k) \}}{4}$. Also, the total noise power is $ \mathcal{P}\{ V(k) \}$}. Thus the error probability is
\begin{equation}\label{eq:pe_ado_1}
	p_{\rm e,1}^{\rm ado}(k)=\mathsf{p}_{\rm e}^{\rm QAM} \left( M_1(k), \frac{\mathcal{P}\{ S_{1}(k) \}}{4}, \mathcal{P}\{ V(k) \} \right).
\end{equation}
In the second layer, \ac{$\varepsilon$ in \eqref{eq:pe_qam} is equal to $\mathcal{P}\{ S_{2}(k) \}$ since there is no clipping, and the total noise power is $\mathcal{P}\{ Z_1(k)\}\leq \mathcal{P}\{ V(k) \}+P_1^{\rm w}$ by \eqref{eq:Pzjle}.} Thus the error probability is
\begin{align}\label{eq:pe_ado_2}
	p_{\rm e,2}^{\rm ado}(k)
	&=\mathsf{p}_{\rm e}^{\rm QAM} \big( M_2(k), \mathcal{P}\{ S_{2}(k) \}, \mathcal{P}\{ Z_1(k) \} \big) \notag \\
	&\leq\mathsf{p}_{\rm e}^{\rm QAM} \big( M_2(k), \mathcal{P}\{ S_{2}(k) \}, \mathcal{P}\{ V(k) \}+P_1^{\rm w} \big).
\end{align}
%Here we suppose the DC bias can always ensure non-negative time-domain signal. 
%Note that, when $M_j(k)=0$, $p_{e,L_j}^{ado}(k) = 0$, ($j=1,2$).

\cc{Denote the number of loaded effective subcarriers in all layers by $N'\leq N-2$. Then, the overall error probability is the average of the error probabilities of all effective subcarriers, i.e., }
\begin{align}\label{eq:pe_ado_sys}
	p_{\rm e}^{\rm ado} =  \frac{  \sum_{j=1,2}\sum_{k}{p_{{\rm e},j}^{\rm ado}(k)}  } {N'}.
\end{align}

\subsection{HACO-OFDM}
The two layers of HACO-OFDM use QAM and PAM respectively. Both layers have clipping, which halves the signal amplitude \ac{leading to $\varepsilon=\frac{\mathcal{P}\{ S_{j}(k) \}}{4}$ in \eqref{eq:pe_qam} and \eqref{eq:pe_pam}. The noise power in layers 1 and 2 is $\mathcal{P}\{ V(k) \}$ and $\mathcal{P}\{ Z_1(k) \}\leq\mathcal{P}\{ V(k) \}+P_1^{\rm w}$ (using \eqref{eq:Pzjle}), respectively.} Thus the error probabilities of the two layers are
\begin{align}\label{eq:pe_haco_1}
p_{\rm e,1}^{\rm haco}(k)
&=\mathsf{p}_{\rm e}^{\rm QAM} \left( M_1(k), \frac{\mathcal{P}\{ S_{1}(k) \}}{4}, \mathcal{P}\{ V(k) \} \right) ,\\
\label{eq:pe_haco_2}
p_{\rm e,2}^{\rm haco}(k)
&=\mathsf{p}_{\rm e}^{\rm PAM} \left( M_2(k), \frac{\mathcal{P}\{ S_{2}(k) \}}{4}, \mathcal{P}\{ Z_1(k) \} \right) \notag \\
&\ac{\leq\mathsf{p}_{\rm e}^{\rm PAM} \left( M_2(k), \frac{\mathcal{P}\{ S_{2}(k) \}}{4}, \mathcal{P}\{ V(k) \}+P_1^{\rm w} \right).}
\end{align}
%Note that when $M_j(k)=0$ for $j=1,2$, then $\mathsf{p}_{{\rm e},j}^{\rm haco}(k) = 0$.

The overall error probability is 
\begin{align}\label{eq:pe_haco_sys}
p_{\rm e}^{\rm haco} =  \frac{ \sum_{j=1,2}\sum_{k}{p_{{\rm e},j}^{\rm haco}(k) } } {N'},
\end{align}
where $N'$ is the number of loaded effective subcarriers in all layers.

\subsection{LACO-OFDM}

Each layer uses QAM and has clipping, which halves the signal amplitude. \ac{This leads to $\varepsilon=\frac{\mathcal{P}\{ S_{j}(k) \}}{4}$  in \eqref{eq:pe_qam}. Also, the total noise power is $ \mathcal{P}\{ V(k) \}$ in layer 1 and $\mathcal{P}\{ Z_j(k) \}\leq\mathcal{P}\{ V(k) \}+\sum_{t=1}^{j-1}P_j^{\rm w}$ in layer $j+1$, $j\in\{1,\ldots,J-1\}$ (using \eqref{eq:Pzjle}).} Thus the error probability of layer $j$ is 
\begin{align}\label{eq:pe_laco_j}
	p_{{\rm e},j}^{\rm laco}(k)
	&=\mathsf{p}_{\rm e}^{\rm QAM} \left( M_j(k), \frac{\mathcal{P}\{ S_{j}(k) \}}{4}, \mathcal{P}\{ Z_{j-1}(k) \} \right) \notag \\
    &\leq \mathsf{p}_{\rm e}^{\rm QAM} \left( M_j(k), \frac{\mathcal{P}\{ S_{j}(k) \}}{4}, \mathcal{P}\{ V(k) \}+\sum_{t=1}^{j-1}P_t^{\rm w} \right).
\end{align}
%Note that $p_{{\rm e},j}^{\rm laco}(k) = 0$ when $M_j(k)=0$.

The overall error probability is 
\begin{align}\label{eq:pe_laco_sys}
p_{\rm e}^{\rm laco} = \frac{  \sum_{j}\sum_{k} {p_{{\rm e},j}^{\rm laco}(k)}  } {N'},
\end{align}
where $N'$ is the number of loaded effective subcarriers in all layers.

%Note that if the power of total noise in the last iteration of demodulating, $\mathcal{P}\{ Z_{J-1}(k) \}$, is known, \eqref{eq:pe_laco_sys} is equivalent to 
%\begin{align}\label{eq:pe_laco_sys_alter}
%	p_{e}^{laco} = \frac{  \sum_{k} { p_e^{QAM} \left( \sum_j M_j(k), \frac{\mathcal{P}\{ S_{L_j}(k) \}}{4}, \mathcal{P}\{ Z_{J-1}(k) \} \right) }  } {N-2}.
%\end{align}

Now we are able to evaluate the SER of the eACO-OFDM schemes with higher accuracy since the RCN is included in the expression. The quality of this evaluation is shown numerically in Sec. \ref{sec:sim-serEst}. \ac{Next, we show how these results can be used for RCN-aware system optimization.}

\section[optimization]{RCN-Aware System Optimization} \label{sec:optimization}

In this section, we discuss an \cc{\emph{exemplary} RCN-aware resource-allocation application, that serves to demonstrate \ac{the usefulness of the RCN model for} designing a reliable LACO-OFDM scheme in the presence of RCN without using multi-class coding.} 
  
%and then extend it to a system optimization design framework for meeting different optimization goal for eACO-OFDM. In the proposed optimization algorithm and framework, we suppose layer number is decided in prior regarding the largest allowed computation complexity, global optimal is obtained through optimizing the bit and power allocated in each subcarrier in a whole, and no extra complexity is introduced in the receiver. 

%\subsection{SER-controlled LACO-OFDM}

Since RCN is accumulated layer by layer in LACO-OFDM, subcarriers from different layers are distorted by different levels of RCN power. This affects the optimization of bit loading and power allocation. \ac{If RCN is not taken into account in this optimization, the scheme may fail to deliver the expected performance. Using the proposed RCN power model, we propose an RCN-aware SER-controlled LACO-OFDM design which makes the scheme more reliable.}

\ac{We consider a bit-rate maximization problem (BRMP) in LACO-OFDM}, where the \ccc{allocated power and bits} of each subcarrier are adaptively changed according to the channel condition, to meet power and error rate constraints. %Practice can limit the peak voltage by clipping to avoid harming the light emitting device, which is not considered in this paper. 
\ac{Under these assumptions, we propose an RCN-aware iterative algorithm for bit loading and power allocation.} Denote the number of bits loaded into subcarrier $k$ by $B(k)$, \emph{the effective power} allocated to \cc{subcarrier $k$} by $P_s(k)$, i.e., $\mathcal{P}\{\frac{S_j(k)}{2}\}=P_s(k)$, \ac{and the `effective power' budget by $P_{\rm eff}$, which can be obtained from the electrical power budget $P_\text{elec}$ or optical power budget $P_\text{opt}$ using Table \ref{tab:power}.} Also, denote the worst-case total noise power of each subcarrier by $P_z(k)$, which is initialized to $P_z^{(0)}(k)=\mathcal{P}\{ Z_0(k) \} = \mathcal{P}\{ V(k) \}$ at iteration $i=0$. Then, $B(k)$, $P_s(k)$ and $P_z(k)$ are updated iteratively as follows. 
\begin{enumerate}
\item Set $i=0$, and $\Phi=\{1,\ldots,N-1\}$.
\item Calculate $P_s^{(i)}(k)$, $k\in\Phi$, by solving $\max_{P_s^{(i)}(k)}  \sum_{k\in\Phi}  \log \left( 1+\frac{|H(k)|^2 P_s^{(i)}(k)}{P_z^{(i)}(k)} \right)$ subject to $\sum_{k\in\Phi} P_s^{(i)}(k) \le  N^2P_\text{eff}$, where $H(k)$ is the channel magnitude in subcarrier $k$.
\item Set $B^{(i)}(k)=\lfloor R_{\Gamma} (k)\rfloor$, $k\in\Phi$, where $R_{\Gamma} (k) = \log \left( 1 + \frac{|H(k)|^2 P_s^{(i)}(k)}{\mathsf{\Gamma} (p_{\rm e}) P_z^{{(i)}}(k)} \right)$, $p_{\rm e}$ is a given SER constraint, and $\mathsf{\Gamma} (p_{\rm e}) = \frac{1}{3} \left[ Q^{-1}\left( \frac{1}{4} p_{\rm e} \right)  \right] ^2$ \cite{snrGap}. %controls the SER performance of {the} current bit and power allocation scheme, i.e., $B(k)$ and $P_s(k)$.
\item If $B^{(i)}(k) = 0$ for some $k\in\Phi$, update $\Phi$ to $\Phi\setminus\{k|B^{(i)}(k) = 0\}$ and repeat from step 2.%${\mathbf{P}}_s^{(i)}$ using step 2 while fixing $P_s^{(i)}(k) = 0$ for any $k \in \Phi$, and renew $B^{(i)}(k)$ using step 3.
%\item Update $\Phi = \{k|B^{(i)}(k) = 0\}$. If $|\Phi| > 0$, renew ${\mathbf{P}}_s^{(i)}$ by solving $\max_{{\mathbf{P}}_s^{\modi{(i)}}}  \sum_{k=1}^{N-1} { \log {\left( 1+\frac{|H(k)|^2 P_s^{\modi{(i)}}(k)}{P_z^{\ac{{(i)}}}(k)} \right)}}$ subject to $\sum_{k=1}^{N-1} {P_s^{\modi{(i)}}(k)} \le  N^2P_\text{eff}$ and $P_s^{(i)}(k) = 0$ for any $k \in \Phi$.
\item Calculate $P_z^{(i+1)}(k)$ for all $k$ using \eqref{eq:Pzjle}, i.e., $P_z^{(i+1)}(k)=\mathcal{P}\{V(k)\}+\sum_{t=1}^{j_k-1}P_t^{\rm w}$ where $j_k$ is the index of the layer which contains effective subcarrier $k$, i.e., $k\in\mathcal{K}_{j_k}$. To calculate $P_t^{\rm w}$, use \eqref{eq:prcn_le} with $M_t(k)=2^{B^{(i)}(k)}$ and $\mathcal{P}\{S_t(k)\}=4P_s^{(i)}(k)$.
\item If $\|{\bf{P}}_z^{(i+1)} - {\bf{P}}_z^{(i)}\|^2 > \epsilon$ where $\epsilon  \ll \frac{\| {\bf{P}}_z^{(i)} \|^2}{N}$ and ${\bf{P}}_z^{(i)} = [P_z^{(i)}(k)]_{k=0}^{N-1}$, increment $i$, set $\Phi=\{1,\ldots,N-1\}$, and repeat from step 2, \ccc{otherwise, end the algorithm.}
\end{enumerate}

If RCN power is perfectly estimated, then the bit loading scheme from step 3 will be able to maintain the SER below $p_{\rm e}$. Sec. \ref{sec:sim-optimization} shows simulation results of the proposed SER-controlled LACO-OFDM scheme, and it also shows that the number of iterations required in the optimization is small.

\section{Simulations and Discussions} \label{Sec:sim}
% 1. 1024 FFT
% 2. 10^4 Monte-Carlo runs
% 3. ACO-OFDM and DCO-OFDM use 64-QAM and PAM-DMT use 8-QAM except for the last simulation which uses adaptive resource allocation. 
% 4. SNR region is always set at 0-30dB, while in some simulations, only one or two SNR cases are considerred. 
% 5. 3 rims are used for RCN power estimation.
% 6. flat channel are supposed unless otherwise specified. 

In this section, we use simulations to examine the main conclusions of the former sections. In all the following simulations, $N=1024$ and $\mathcal{P}\{ {\mathbf{v}} \} = 1$. All the available subcarriers are used to transmit information, which means LACO-OFDM has $9$ layers. Monte-Carlo simulation uses \cc{$10^4$} independent runs, and in each run, information symbols and noise are generated randomly. Define $\gamma = 10\log_{10}  \frac{P_\text{elec}}{\mathcal{P}\{ {\mathbf{v}} \}}$(dB) as the signal to noise ratio (SNR), and define $\gamma_\text{eff}=10\log_{10} \frac{P_\text{eff}}{ \mathcal{P}\{ {\mathbf{v}} \} }$(dB) as the \emph{effective SNR}. The relation between $\gamma_\text{eff}$ and $\gamma$ follows the relation between $P_\text{eff}$ and $P_\text{elec}$ \ac{in Table \ref{tab:power}.} \cc{In all simulations except for the last one (which adopts power allocation), \modi{$P_\text{eff}$ is equally distributed in all effective subcarriers.} }
%we fix the \ac{effective (?)} symbol power to \ac{be equal in all subcarriers}.} 
To avoid redundancy, some simulations only test LACO-OFDM, as ADO-OFDM and HACO-OFDM show similar results or patterns. Unless otherwise specified, a flat AWGN channel is considered and ACO-OFDM uses $64$QAM in each subcarrier as a default.

\subsection{Statistics of RCN \cc{and model mismatch}} \label{sec:stat}

This simulation examines RCN property 2, i.e., the Gaussianity of $\Delta_{t}(k)$ (its variance is examined in Sec. \ref{sec:sim-rcn-est}.) We choose $\gamma_\text{eff}=0$ or $20$dB. For each $t$, the real and imaginary parts of $\Delta_{t}(k)$, $k\in\mathcal{B}_t$, are collected into two sample sets, each set is normalized by \modi{dividing the variance of the combined sample sets of the real and imaginary parts of $\Delta_{t}(k)$}, and a cumulative distribution function (CDF) is generated for each, \modi{as shown in Fig. \ref{fig:rcn_pdf}, where subcarrier $k=256$ is taken as an example and similar results hold for other tested values of $k$.} %is generated for each} \cite{book-Degroot}. \cc{Then, we compare the resulting histogram \ac{(converged to a line plot)} with the standard Gaussian distribution $\mathcal{N}(0,1)$.} 
%Fig. \ref{fig:rcn_pdf} compares the cdf of $\Delta_{t}(k)$ for $k=256$ and $t=1,\dots, 8$ from the simulation results. 
%Similar results holds for other tested values of $k$. 
It can be seen that the obtained CDF \modi{of the normalized $\Delta_{t}(k)$ samples} is close to $\ccc{\mathcal{CN}}(0,1)$ for most $t$. A mismatch occurs when $t=6,7,8$, which are layers with $16$, $8$, and $4$ effective subcarriers, respectively. The mismatch is mainly caused by the small number of effective subcarriers, which is not enough to obtain reliable statistics. It is worth to note that irrespective of $N$, model mismatch was observed only when the number of effective subcarriers of a layer is less than $32$. Such a mismatch leads to an underestimated RCN power \ccc{in our tested cases}, but is acceptable as shown in the later simulations.
%It can be seen that the obtained probability density function of $\Delta_t(k)$ is close to $\mathcal{N}(0,1)$ for most $t$, for both values of $\gamma_\text{eff}$. \cc{A mismatch occurs when $t=6,7,8$, \ac{which are layers with} $16$, $8$, and $4$ effective subcarriers, respectively.} The mismatch is mainly \ac{caused by the small} number of effective subcarriers, \ac{which is not enough} to obtain reliable statistics. \cc{It is worth to note that irrespective of $N$, model mismatch occurs only when \ac{the number of} effective subcarriers of a layer is less than $32$. Such a mismatch will lead to an underestimated RCN power, but is acceptable as shown in the later simulations.}

\begin{figure}[!t]
\centering
	\subfloat[Normalized samples of the real part of $\Delta_t(k)$.]{
	\includegraphics[width=0.45\textwidth, height=0.4\textwidth]{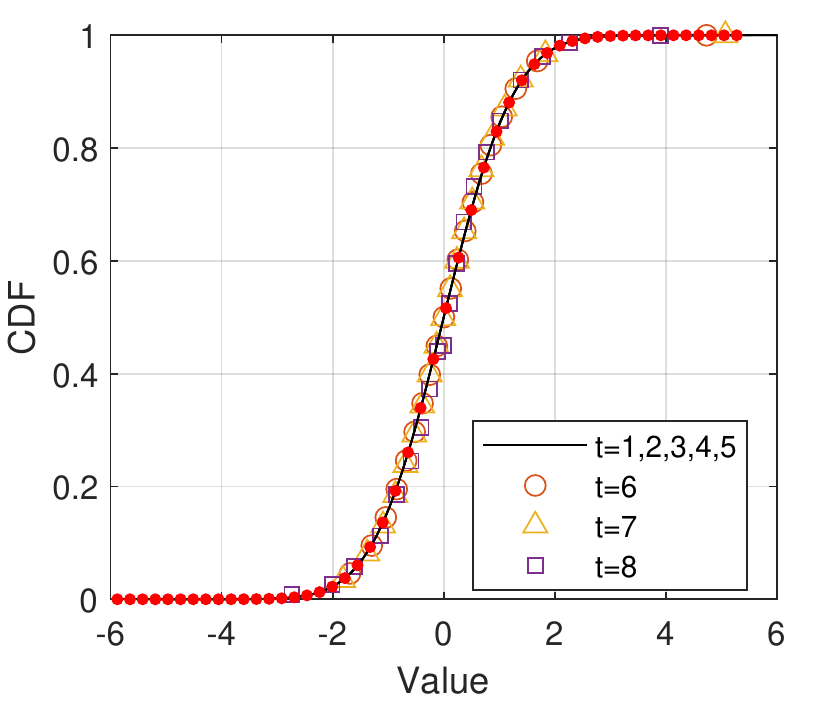}
	}
	\subfloat[Normalized samples of the imaginary part of $\Delta_t(k)$.]{
	\includegraphics[width=0.45\textwidth, height=0.4\textwidth]{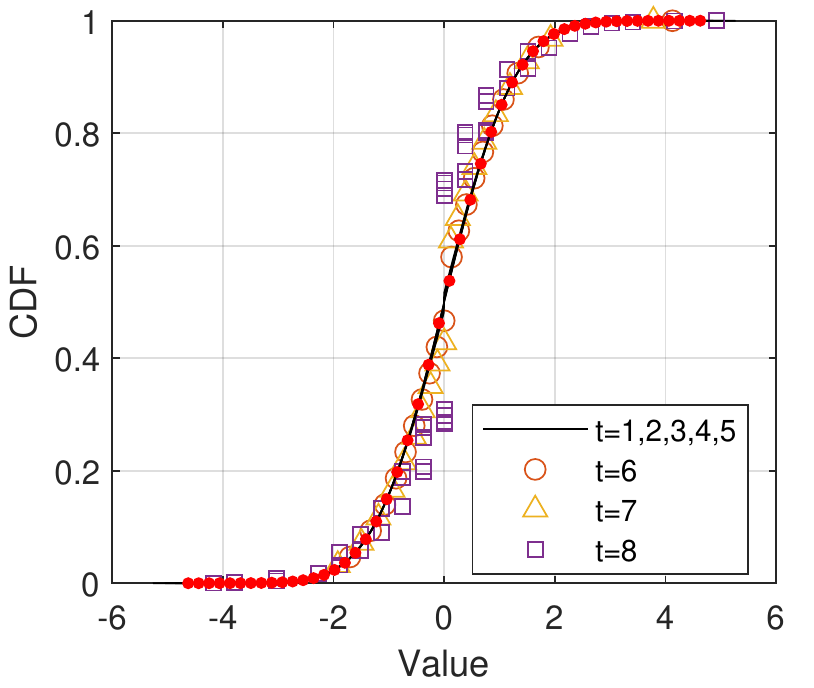}
	}
	\caption{\cc{Simulated CDF of $\Delta_t(k)$, $t=1,\dots,8$, $\modi{k=256}$, with normalized samples. SNR=$0$ and $20$dB \ac{lead to similar results}. \modi{The dots depict the CDF of $\mathcal{N}(0,1)$ as a reference.}}}
	\label{fig:rcn_pdf}
\end{figure}

\subsection{\ac{Covariance of Frequency-Domain RCN}} \label{sec:cor}

This simulation examines \cc{RCN property 3. We evaluate the normalized covariance of $\Delta_{t_1}(k)$ and $\Delta_{t_2}(k)$ for some sample values of $k$, defined as 
\begin{align} \label{eq:corr}
\rho_{t_1,t_2}\modi{(k)}=\frac{ \mathbb{E} \Big\{ \big( \Delta_ {t_1}(k)-\mathbb{E}\{\Delta_ {t_1}(k)\} \big) \big( \Delta_{t_2}^*(k)-\mathbb{E}\{\Delta_{t_2}^*(k)\} \big) \Big\}  }  { \sqrt{\mathbb{V}[\Delta_{t_1}(k)]\mathbb{V}[\Delta_{t_2}(k)]} }. 
\end{align}}
%From our simulation results, $\forall t_1 \ne t_2$, $\rho_{t_1,t_2} < 0.01$ holds for all tested SNR. 
Fig. \ref{fig:cor} shows the result for $\gamma_\text{eff}=0$dB. It can be seen that $\modi{|\rho_{t_1,t_2}(k)|} = 1$ for $t_1=t_2$ and $\modi{|\rho_{t_1,t_2}(k)|} \approx 0$ otherwise. \ac{The same result was observed for $\gamma_\text{eff}=10$, $20$, and $30$dB, and for \modi{randomly} selected values of $k$. Since $\Delta_t(k)$ has zero mean (cf. property 2 and Fig. \ref{fig:rcn_pdf}), this implies that $\mathbb{E}\{\Delta_{t_1}(k)\Delta_{t_2}^*(k)\}\approx 0$, and supports property 3.} 
%\modi{It also worth to mention that the correlation of $\delta_{t_1}(n)$ and $\delta_{t_2}(n)$ for all $t_1 \ne t_2$ is also proved to be negligible through simulation, as its correlation factor (obtained by replacing $\Delta_{t}(k)$ by $\delta_t(n)$ in \eqref{eq:corr}) gives a similar figure as Fig. \ref{fig:cor}, which also supports property 3.}

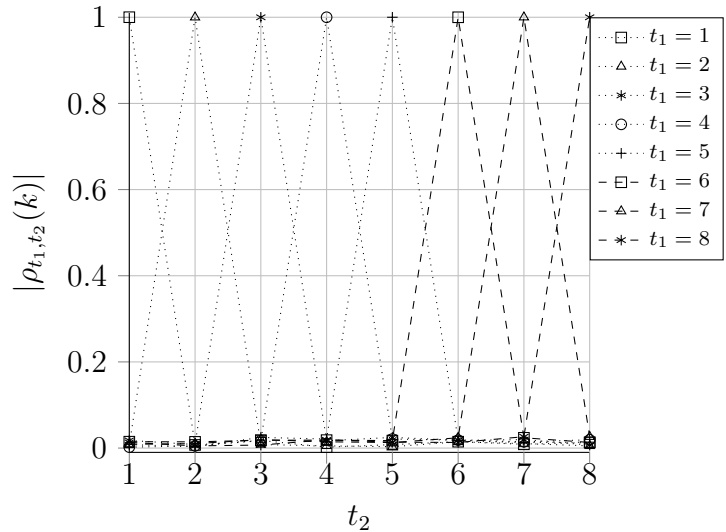
\begin{figure}[!t]
	\centering
	% This file was created by matlab2tikz.
%
%The latest updates can be retrieved from
%  http://www.mathworks.com/matlabcentral/fileexchange/22022-matlab2tikz-matlab2tikz
%where you can also make suggestions and rate matlab2tikz.
%
\definecolor{mycolor1}{rgb}{1.00000,0.00000,1.00000}%
\definecolor{mycolor2}{rgb}{1.00000,1.00000,0.00000}%
\begin{tikzpicture}

\begin{axis}[%
width=2.4in,
height=2.3in,
at={(0,0)},
scale only axis,
xmin=1,
xmax=8,
xtick={1,2,3,4,5,6,7,8},
tick align=outside,
xlabel={t1},
ymin=1,
ymax=8,
ytick={1,2,3,4,5,6,7,8},
ylabel={$t_2$},
zmin=-0.01,
zmax=1.01,
zlabel={\modi{$|\rho_{t_1,t_2}(k)|$}},
%view={-20}{30},
view={90}{0},
axis background/.style={fill=white},
axis x line*=bottom,
axis y line*=left,
axis z line*=left,
xmajorgrids,
ymajorgrids,
zmajorgrids,
legend style={at={(axis cs: 8,8,1)}, anchor=north west, legend cell align=left, align=left}
]
\addplot3 [color=black, dotted, mark=square, mark options={solid, black} ]
 table[row sep=crcr] {%
1	1	1.00012364815903\\
1	2	0.00723419954303165\\
1	3	0.017730601618715\\
1	4	0.00215332135772673\\
1	5	0.00801243932292706\\
1	6	0.0148170171360081\\
1	7	0.00876082975635962\\
1	8	0.0131446815555399\\
};
 \addlegendentry{\scriptsize $t_1 = 1$}

\addplot3 [color=black, dotted, mark=triangle, mark options={solid, black} ]
 table[row sep=crcr] {%
2	1	0.00723419954303165\\
2	2	0.999913663262204\\
2	3	0.00820913160397697\\
2	4	0.00512830174865328\\
2	5	0.00323403447048174\\
2	6	0.0144297857531365\\
2	7	0.0103614054299998\\
2	8	0.00511848250693002\\
};
 \addlegendentry{\scriptsize $t_1 = 2$}

\addplot3 [color=black, dotted, mark=asterisk, mark options={solid, black} ]
 table[row sep=crcr] {%
3	1	0.017730601618715\\
3	2	0.00820913160397697\\
3	3	0.999932624981091\\
3	4	0.0170719820058953\\
3	5	0.0247666748369509\\
3	6	0.0181724644288342\\
3	7	0.0195912948315161\\
3	8	0.00698404371587171\\
};
 \addlegendentry{\scriptsize $t_1 = 3$}

\addplot3 [color=black, dotted, mark=o, mark options={solid, black} ]
 table[row sep=crcr] {%
4	1	0.00215332135772673\\
4	2	0.00512830174865328\\
4	3	0.0170719820058953\\
4	4	1.00002602560134\\
4	5	0.0209437499331864\\
4	6	0.0184593257884122\\
4	7	0.0145749872939283\\
4	8	0.0192027496297948\\
};
 \addlegendentry{\scriptsize $t_1 = 4$}

\addplot3 [color=black, dotted, mark=+, mark options={solid, black} ]
 table[row sep=crcr] {%
5	1	0.00801243932292706\\
5	2	0.00323403447048174\\
5	3	0.0247666748369509\\
5	4	0.0209437499331864\\
5	5	1.00001548252851\\
5	6	0.0173174955718891\\
5	7	0.0128500951204616\\
5	8	0.0136844193205752\\
};
 \addlegendentry{\scriptsize $t_1 = 5$}

\addplot3 [color=black, dashed, mark=square, mark options={solid, black} ]
 table[row sep=crcr] {%
6	1	0.0148170171360081\\
6	2	0.0144297857531365\\
6	3	0.0181724644288342\\
6	4	0.0184593257884122\\
6	5	0.0173174955718891\\
6	6	0.999966245358332\\
6	7	0.0238480602272186\\
6	8	0.0130843258989047\\
};
 \addlegendentry{\scriptsize $t_1 = 6$}

\addplot3 [color=black, dashed, mark=triangle, mark options={solid, black} ]
 table[row sep=crcr] {%
7	1	0.00876082975635962\\
7	2	0.0103614054299998\\
7	3	0.0195912948315161\\
7	4	0.0145749872939283\\
7	5	0.0128500951204616\\
7	6	0.0238480602272186\\
7	7	1.00001289672055\\
7	8	0.0263138761883249\\
};
 \addlegendentry{\scriptsize $t_1 = 7$}

\addplot3 [color=black, dashed, mark=asterisk, mark options={solid, black} ]
 table[row sep=crcr] {%
8	1	0.0131446815555399\\
8	2	0.00511848250693002\\
8	3	0.00698404371587171\\
8	4	0.0192027496297948\\
8	5	0.0136844193205752\\
8	6	0.0130843258989047\\
8	7	0.0263138761883249\\
8	8	0.999915786364506\\
};
 \addlegendentry{\scriptsize $t_1 = 8$}

\end{axis}

\end{tikzpicture}%
	\caption{Correlation coefficient between \ac{frequency-domain} RCN from different layers \modi{for $k=256$}.}
	\label{fig:cor}
\end{figure}
%\begin{figure}[!htbp]
%	\centering
%	\includegraphics[width=0.45\textwidth, height=0.4\textwidth]{fig_RCNcor9L.pdf}
%	\caption{Correlation coefficient between RCN from different layers. The correlation coefficient between RCN from layer $t_1$ and $t_2$ {is} 1 if and only if $t_1=t_2$, otherwise it is close to 0. }
%	\label{fig:cor}
%\end{figure}

\subsection{RCN Power Estimation in an AWGN Channel} \label{sec:sim-rcn-est}
% 1. to show the realtion between estimation accuracy and rim number
% 2. to show the estimation accuracy of each layer under different SNR.

We examine the performance of the worst-case RCN power model in \eqref{eq:Prcnle_t}. 
%, \modi{i.e., $\mathcal{P}\{ \boldsymbol{\delta}_{t} \} \le \frac{1}{4}\mathcal{P}\{ \mathbf{e}_t \} =  \frac{1}{4N^2}\sum_{k=0}^{N-1}\mathcal{P}\{E_t(k)\}$.}
%\cc{of the estimation of $E_t(k)$ using $\mathsf{f}(d_\text{min}, \sigma^2, M)$ and $\mathcal{P}\{ \delta_{t}(n) \} \le \frac{1}{4}\mathcal{P}\{ e_t(n) \} =  \frac{1}{4N}\mathcal{P}\{\mathbf{E}_t\} = \frac{1}{4N^2}\sum_{k=0}^{N-1}\mathcal{P}\{E_t(k)\}$}. 
\cc{Since the accuracy of the model depends on $\mathsf{f}(\cdot, \cdot, \cdot)$} \ac{which is approximated by considering a specific} number of rims in the QAM constellation as discussed in Appendix \ref{Sec:Appen1}, we evaluate performance under different numbers of rims.

%{As discussed in Appendix \ref{Sec:Appen1}, the performance of \eqref{eq:prcn_le} is also related to how many rims are considered. Therefore, in the following, the number of rims used in estimating the RCN power is also compared through simulation. }
%\cc{Recalling that $\mathcal{P}\{ \delta_{t}(n) \} \le \frac{1}{4}\mathcal{P}\{ e_t(n) \}$, the RCN power $\mathcal{P}\{\boldsymbol{\delta}_{t}\}$ is estimated through the power of detection error $\mathcal{P}\{\mathbf{e}_t\} =  \frac{1}{N}\mathcal{P}\{\mathbf{E}_t\} = \frac{1}{N^2}\sum_{k=0}^{N-1}\mathcal{P}\{E_t(k)\}$, where the accuracy of estimating $\mathcal{P}\{E_t(k)\}$ is related to the number of rims as discussed in Appendix \ref{Sec:Appen1}, thus, in this simulation we also compare the effect of different numbers of rims used in the RCN power estimation process.}

Fig. \ref{fig:Prcn_cmp_l1} compares the \ac{simulated} and the estimated RCN power from layer 1 \cc{$\mathcal{P}\{{\boldsymbol{\delta}}_{1}\}$} \ac{using \eqref{eq:Prcnle_t}}, under a flat AWGN channel, \modi{where $\mathcal{P}\{E_t(k)\}$ is estimated using \eqref{eq:P_Et} with $1$, $2$, or $3$ rims.}
%{}The estimation is obtained using \eqref{eq:prcn_le} with $1$, $2$, or $3$ rims.} 
When only one rim is considered, $p_b, p_c$ in \eqref{eq:p_abc} should be set to $0$, and when only two rims are considered, $p_c$ in \eqref{eq:p_abc} should be set to $0$. The figure shows that the estimated RCN power is more accurate \ac{when more rims are considered at low effective SNR, and is accurate at moderate/high effective SNR for any number of rims}. Moreover, $3$ rims \ccc{are shown to be }enough to obtain a good estimation in the whole tested range of $\gamma_\text{eff}$. \cc{It can also be seen that \ac{using the power of detection error $\mathcal{P}\{\mathbf{e}_1\}$ to approximate} RCN power $\mathcal{P}\{{\boldsymbol{\delta}}_{1}\}$ (as suggested in \cite[(13)]{laco-perf3/papr-reduction}) results in an overestimation \ac{of RCN power}.} 

Fig. \ref{fig:Prcn_cmp} compares simulated and estimated $\mathcal{P}\{ {\boldsymbol{\delta}}_{t} \}$, $t=1,\dots,8$ using 3 rims under a flat AWGN channel with $\gamma_\text{eff}=0$, $10$dB, and $20$dB, respectively. It can be seen that the estimated  $\mathcal{P}\{ {\boldsymbol{\delta}}_{t}\}$ closely follows the simulation results. The estimation performance is generally better \ac{when $\gamma_\text{eff}$ is moderate/large.} When $\gamma_\text{eff}$ is small, accurate estimation of detection-error power $\mathcal{P}\{E_t(k)\}$ requires more than 3 rims. \cc{Similarly, in high layers such as layers $6,7,8$, RCN is accumulated leading to stronger noise, and thus more rims are also required for accurate estimation of $\mathcal{P}\{E_t(k)\}$. This leads to an underestimation of \ac{RCN power in some cases}. Another reason for the underestimation of RCN power in high layers is the model mismatch when \ac{the number of effective subcarriers is small}.}

%The underestimation shown in high layers is for the same reason, {since} RCN is accumulated in high layers so that more rims are required for accurate estimation. 

%In the following simulation, RCN power estimation {is} always supposed to use 3 rims unless otherwise specified. 

\begin{figure}[!t]
	\centering
%	\pgfplotsset{compat=newest} 
%	\pgfplotsset{plot coordinates/math parser=false} 
	\subfloat[RCN power from layer 1 versus $\gamma_\text{eff}$.]{\label{fig:Prcn_cmp_l1}
		% This file was created by matlab2tikz.
%
%The latest updates can be retrieved from
%  http://www.mathworks.com/matlabcentral/fileexchange/22022-matlab2tikz-matlab2tikz
%where you can also make suggestions and rate matlab2tikz.
%
\definecolor{mycolor1}{rgb}{0.00000,0.44700,0.74100}
\begin{tikzpicture}

\begin{axis}[%
width=2.4in,
height=2.3in,
at={(0,0)},
scale only axis,
xmin=0,
xmax=30,
xlabel style={font=\color{white!15!black}},
xlabel={\cc{$\gamma_\text{eff}$} (dB)},
ymode=log,
ymin=1e-3,
ymax=5,
yminorticks=true,
ylabel style={font=\color{white!15!black}},
ylabel={Power},
axis background/.style={fill=white},
xmajorgrids,
ymajorgrids,
yminorgrids,
legend style={at={(0,0)}, anchor=south west, legend cell align=left, align=left, draw=white!15!black}
]
\addplot [color=black, dashed, mark=triangle, mark options={solid, black}]
  table[row sep=crcr]{%
0	1.29532641369047\\
2	1.43972428729491\\
4	1.56387526125576\\
6	1.6844631215994\\
8	1.81460609391107\\
10	1.97055357142857\\
12	2.15694293758707\\
14	2.31857210029172\\
16	2.23612413781653\\
18	1.74533411331426\\
20	0.970424107142853\\
22	0.316318266328697\\
24	0.0457148378829794\\
26	0.0015699166695934\\
28	0\\
30	0\\
};
\addlegendentry{\scriptsize simulated \cc{$\mathcal{P}\{\mathbf{e}_1\}$}}

\addplot [color=red, dashed, mark=star, mark options={solid, red}]
table[row sep=crcr]{%
	0	0.323831603422618\\
	2	0.359931071823729\\
	4	0.39096881531394\\
	6	0.421115780399849\\
	8	0.453651523477767\\
	10	0.492638392857142\\
	12	0.539235734396767\\
	14	0.57964302507293\\
	16	0.559031034454132\\
	18	0.436333528328564\\
	20	0.242606026785713\\
	22	0.0790795665821743\\
	24	0.0114287094707448\\
	26	0.00039247916739835\\
};
\addlegendentry{\scriptsize \cc{simulated $\frac{\mathcal{P}\{\mathbf{e}_1\}}{4}$}}

\addplot [color=blue, dashed, mark=square, mark options={solid, blue}]
  table[row sep=crcr]{%
0	0.218356518908773\\
2	0.259644700759136\\
4	0.299671566005768\\
6	0.339126350329682\\
8	0.38078595570592\\
10	0.426956446798958\\
12	0.479627055512064\\
14	0.526921560014224\\
16	0.519148649889092\\
18	0.414481583595111\\
20	0.235393681769847\\
22	0.0779590326040945\\
24	0.011348364229233\\
26	0.000390170740600789\\
28	0\\
30	0\\
};
\addlegendentry{\scriptsize simulated \cc{$\mathcal{P}\{\boldsymbol{\delta}_{1}\}$}}

\addplot [color=black, dashed, mark=+, mark options={solid, black}]
  table[row sep=crcr]{%
0	0.0618095494377746\\
2	0.0933490415334737\\
4	0.138769381839635\\
6	0.201800426453213\\
8	0.284469021499682\\
10	0.383434766533561\\
12	0.484002249102574\\
14	0.554089139808581\\
16	0.547409403256365\\
18	0.432041842749165\\
20	0.241587559801272\\
22	0.0793255191476275\\
24	0.0113686741910936\\
26	0.000443437280757578\\
28	2.21890674101353e-06\\
30	4.31359088082235e-10\\
};
\addlegendentry{\scriptsize estimated RCN power, 1 rim}

\addplot [color=black, dashed, mark=x, mark options={solid, black}]
  table[row sep=crcr]{%
0	0.143771432130756\\
2	0.199897060686144\\
4	0.265714074661043\\
6	0.33432628300597\\
8	0.397516343053939\\
10	0.454115995117647\\
12	0.511337523388395\\
14	0.559091831729638\\
16	0.547692262788403\\
18	0.432044348917059\\
20	0.241587560986905\\
22	0.0793255191476326\\
24	0.0113686741910936\\
26	0.000443437280757578\\
28	2.21890674101353e-06\\
30	4.31359088082235e-10\\
};
\addlegendentry{\scriptsize estimated RCN power, 2 rims}

\addplot [color=black, dashed, mark=o, mark options={solid, black}]
  table[row sep=crcr]{%
0	0.192746037085559\\
2	0.250656828416745\\
4	0.307805788492625\\
6	0.359044302571859\\
8	0.406020295703253\\
10	0.455389144186364\\
12	0.511389357162504\\
14	0.559092103805996\\
16	0.547692262844573\\
18	0.432044348917059\\
20	0.241587560986905\\
22	0.0793255191476326\\
24	0.0113686741910936\\
26	0.000443437280757578\\
28	2.21890674101353e-06\\
30	4.31359088082235e-10\\
};
\addlegendentry{\scriptsize estimated RCN power, 3 rims}

\end{axis}

\end{tikzpicture}%
	}
%	\hspace{.2cm}
	\subfloat[RCN power from all layers.]{\label{fig:Prcn_cmp}
		% This file was created by matlab2tikz.
%
%The latest updates can be retrieved from
%  http://www.mathworks.com/matlabcentral/fileexchange/22022-matlab2tikz-matlab2tikz
%where you can also make suggestions and rate matlab2tikz.
%
\definecolor{mycolor1}{rgb}{1.00000,0.00000,1.00000}%
\begin{tikzpicture}

\begin{axis}[%
width=2.4in,
height=2.3in,
at={(0,0)},
scale only axis,
xmin=1,
xmax=8,
xtick = {1, 2,3,4,5,6,7,8},
xlabel style={font=\color{white!15!black}},
xlabel={Layer index},
ymode=log,
ymin=0.001,
ymax=2,
yminorticks=true,
ylabel style={font=\color{white!15!black}},
ylabel={Power},
axis background/.style={fill=white},
xmajorgrids,
ymajorgrids,
yminorgrids,
legend style={at={(0,0)}, anchor=south west, legend cell align=left, align=left, draw=white!15!black}
]
\addplot [color=blue, mark=+, mark options={solid, blue}]
  table[row sep=crcr]{%
1	0.218356518908773\\
2	0.133342301735317\\
3	0.0777821289766665\\
4	0.0436282154584948\\
5	0.0237497289859245\\
6	0.0126805059221739\\
7	0.00663086889187016\\
8	0.00351735491071428\\
};
\addlegendentry{\scriptsize 0dB-simulated}

\addplot [color=black, dashed, mark=+, mark options={solid, black}]
  table[row sep=crcr]{%
1	0.192746037085559\\
2	0.107426644974076\\
3	0.0577978785126185\\
4	0.0304318867557884\\
5	0.0158054791610902\\
6	0.00813521900421932\\
7	0.00416141204258343\\
8	0.00211932238885683\\
};
\addlegendentry{\scriptsize 0dB-estimated}

\addplot [color=blue, mark=o, mark options={solid, blue}]
  table[row sep=crcr]{%
1	0.426956446798958\\
2	0.33966780951125\\
3	0.256048241566272\\
4	0.180443138650768\\
5	0.119556437318473\\
6	0.0751402192625029\\
7	0.0445635554432215\\
8	0.0259267113095238\\
};
\addlegendentry{\scriptsize 10dB-simulated}

\addplot [color=black, dashed, mark=o, mark options={solid, black}]
  table[row sep=crcr]{%
1	0.455389144186364\\
2	0.369800800951995\\
3	0.276659011669458\\
4	0.186132591128151\\
5	0.113498313225328\\
6	0.064561090013791\\
7	0.0351677891412358\\
8	0.0186586212402996\\
};
\addlegendentry{\scriptsize 10dB-estimated}

\addplot [color=blue, mark=asterisk, mark options={solid, blue}]
  table[row sep=crcr]{%
1	0.235393681769847\\
2	0.285854585626088\\
3	0.339256233658507\\
4	0.331792298853642\\
5	0.278813080364893\\
6	0.221782983951934\\
7	0.164588460195928\\
8	0.118549107142857\\
};
\addlegendentry{\scriptsize 20dB-simulated}

\addplot [color=black, dashed, mark=asterisk, mark options={solid, black}]
  table[row sep=crcr]{%
1	0.241587560986905\\
2	0.302054432727171\\
3	0.373781407047876\\
4	0.372336957746851\\
5	0.318939258122056\\
6	0.254954629200642\\
7	0.186512023210142\\
8	0.122528909593023\\
};
\addlegendentry{\scriptsize 20dB-estimated}

\end{axis}

\end{tikzpicture}%
	}
	\caption{Comparison of simulated and estimated RCN power a flat AWGN channel.}
	\label{fig:Prcn_cmp_flat_major}
\end{figure}
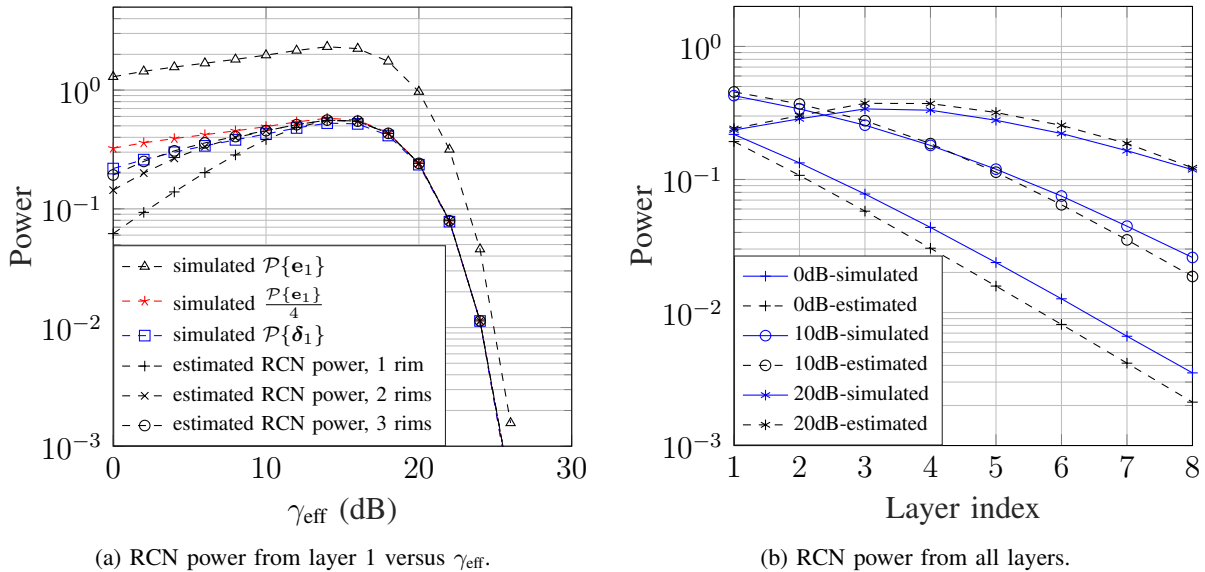

%\subsection{RCN Estimation in Frequency-selective Channel} \label{Sec:sim-fsChan}

We also run a similar simulation in a \cc{frequency-selective} channel with channel-inversion equalization at the receiver. The channel gain of the tested \ac{frequency-selective channel is the one shown in \cite[Fig. 4b]{laco-perf2-exp/bitLoading} converted to our $1024$-subcarrier LACO-OFDM system.} 
%shown in Fig. \ref{fig:fsChannel}, which is obtained from the experiment in \cite{laco-perf2-exp/bitLoading} \cc{where a point-to-point VLC link is tested and the frequency selectivity is a result of the devices' nonideal frequency response.}
%\begin{figure}[!htbp]
%	\centering
%	\includegraphics[width=0.45\textwidth, height=0.4\textwidth]{fig_channelRecords.pdf}
%	\caption{The \cc{frequency-selective} channel from the experiment in \cite{laco-perf2-exp/bitLoading}.}
%	\label{fig:fsChannel}
%\end{figure}
%
%The channel gain is shown in Fig. \ref{fig:fsChannel}, which is from the experiments in \cite{laco-perf2-exp/bitLoading}. 
Simulation results are given in Fig. \ref{fig:Prcn_cmp_fs_major}, which {shows similar trends as in} Fig. \ref{fig:Prcn_cmp_flat_major}, i.e., the estimated RCN power is close to the simulated value for each layer under each $\gamma_\text{eff}$.

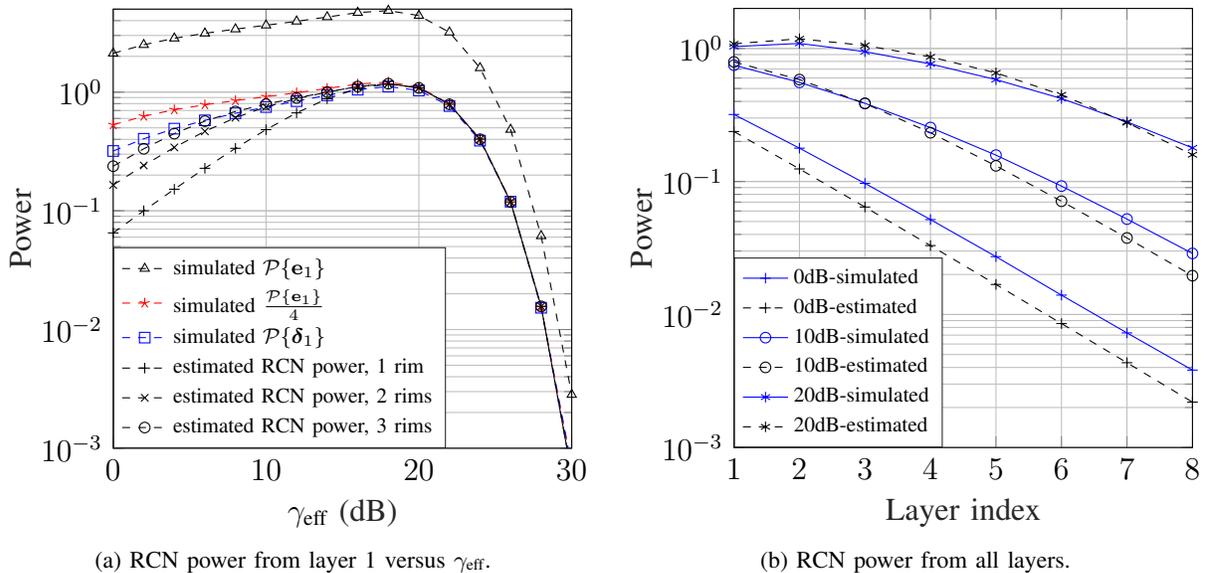
\begin{figure}[!htbp]
	\subfloat[RCN power from layer 1 versus $\gamma_\text{eff}$.]{
		\centering
		% This file was created by matlab2tikz.
%
%The latest updates can be retrieved from
%  http://www.mathworks.com/matlabcentral/fileexchange/22022-matlab2tikz-matlab2tikz
%where you can also make suggestions and rate matlab2tikz.
%
\definecolor{mycolor1}{rgb}{0.00000,0.44700,0.74100}

\begin{tikzpicture}

\begin{axis}[%
width=2.4in,
height=2.3in,
at={(0,0)},
scale only axis,
xmin=0,
xmax=30,
xlabel style={font=\color{white!15!black}},
xlabel={\cc{$\gamma_\text{eff}$} (dB)},
ymode=log,
ymin=1e-3,
ymax=5,
yminorticks=true,
ylabel style={font=\color{white!15!black}},
ylabel={Power},
axis background/.style={fill=white},
xmajorgrids,
ymajorgrids,
yminorgrids,
legend style={at={(0,0)}, anchor=south west, legend cell align=left, align=left, draw=white!15!black}
]
\addplot [color=black, dashed, mark=triangle, mark options={solid, black}]
  table[row sep=crcr]{%
0	2.12191636904763\\
2	2.5038601377307\\
4	2.8398597423248\\
6	3.12746209961971\\
8	3.39101968727081\\
10	3.65757068452382\\
12	3.94733215497724\\
14	4.29027399057158\\
16	4.67217272034537\\
18	4.86001936520517\\
20	4.41148809523808\\
22	3.19093008651509\\
24	1.59796346647373\\
26	0.483415849957819\\
28	0.0614056701324473\\
30	0.00282738095238095\\
};
\addlegendentry{\scriptsize simulated \cc{$\mathcal{P}\{\mathbf{e}_1\}$}}

\addplot [color=red, dashed, mark=star, mark options={solid}]
table[row sep=crcr]{%
	0	0.530479092261908\\
	2	0.625965034432675\\
	4	0.709964935581200\\
	6	0.781865524904927\\
	8	0.847754921817703\\
	10	0.914392671130955\\
	12	0.986833038744310\\
	14	1.07256849764290\\
	16	1.16804318008634\\
	18	1.21500484130129\\
	20	1.10287202380952\\
	22	0.797732521628773\\
	24	0.399490866618433\\
	26	0.120853962489455\\
	28	0.0153514175331118\\
	30	0.000706845238095238\\
};
\addlegendentry{\scriptsize \cc{simulated $\frac{\mathcal{P}\{\mathbf{e}_1\}}{4}$}}

\addplot [color=blue, dashed, mark=square, mark options={solid, blue}]
  table[row sep=crcr]{%
0	0.319046214818262\\
2	0.404447596537497\\
4	0.492243625860945\\
6	0.578219305052332\\
8	0.663123511949232\\
10	0.748535865040728\\
12	0.839084270030503\\
14	0.939083062062604\\
16	1.04773356199288\\
18	1.11360568038238\\
20	1.03291474978009\\
22	0.763501135526619\\
24	0.389745238013597\\
26	0.119443807288313\\
28	0.0152506196529412\\
30	0.00070240332951485\\
};
\addlegendentry{\scriptsize simulated \cc{$\mathcal{P}\{\boldsymbol{\delta}_{1}\}$}}

\addplot [color=black, dashed, mark=+, mark options={solid, black}]
  table[row sep=crcr]{%
0	0.0651537582501956\\
2	0.100033400487632\\
4	0.152088076805451\\
6	0.228176836391543\\
8	0.336129296833832\\
10	0.482677624136492\\
12	0.668508436214167\\
14	0.879024678015082\\
16	1.0717717780255\\
18	1.16977236305652\\
20	1.08428471131819\\
22	0.78908740984896\\
24	0.400781753427916\\
26	0.118864450817518\\
28	0.0156698771258528\\
30	0.000602597915485786\\
};
\addlegendentry{\scriptsize estimated RCN power, 1 rim}

\addplot [color=black, dashed, mark=x, mark options={solid, black}]
  table[row sep=crcr]{%
0	0.165075984933912\\
2	0.241019849268671\\
4	0.342048309155351\\
6	0.467403309463019\\
8	0.608803326669206\\
10	0.750759014484525\\
12	0.880102251513706\\
14	0.999445042405495\\
16	1.11378163017562\\
18	1.17678200896784\\
20	1.08466626957812\\
22	0.789091155743413\\
24	0.400781756067111\\
26	0.118864450817549\\
28	0.0156698771258528\\
30	0.000602597915485786\\
};
\addlegendentry{\scriptsize estimated RCN power, 2 rims}

\addplot [color=black, dashed, mark=o, mark options={solid, black}]
  table[row sep=crcr]{%
0	0.237487281828941\\
2	0.333014547027391\\
4	0.447010586246581\\
6	0.569313790561033\\
8	0.686528372856699\\
10	0.792086053793979\\
12	0.892902513743888\\
14	1.00120257376697\\
16	1.1138519785434\\
18	1.17678243817491\\
20	1.08466626972295\\
22	0.789091155743413\\
24	0.400781756067111\\
26	0.118864450817549\\
28	0.0156698771258528\\
30	0.000602597915485786\\
};
\addlegendentry{\scriptsize estimated RCN power, 3 rims}

\end{axis}

\end{tikzpicture}%
		\label{fig:Prcn_cmp_l1_fs}
	}
	\subfloat[RCN power from all layers.]{
		\centering
		% This file was created by matlab2tikz.
%
%The latest updates can be retrieved from
%  http://www.mathworks.com/matlabcentral/fileexchange/22022-matlab2tikz-matlab2tikz
%where you can also make suggestions and rate matlab2tikz.
%
\definecolor{mycolor1}{rgb}{1.00000,0.00000,1.00000}%
\begin{tikzpicture}

\begin{axis}[%
width=2.4in,
height=2.3in,
at={(0,0)},
scale only axis,
xmin=1,
xmax=8,
xtick = {1, 2,3,4,5,6,7,8},
xlabel style={font=\color{white!15!black}},
xlabel={Layer index},
ymode=log,
ymin=0.001,
ymax=2,
yminorticks=true,
ylabel style={font=\color{white!15!black}},
ylabel={Power},
axis background/.style={fill=white},
xmajorgrids,
ymajorgrids,
yminorgrids,
legend style={at={(0,0)}, anchor=south west, legend cell align=left, align=left, draw=white!15!black}
]
\addplot [color=blue, mark=+, mark options={solid, blue}]
  table[row sep=crcr]{%
1	0.319046214818262\\
2	0.178068220833952\\
3	0.0967224409100368\\
4	0.0517354495401605\\
5	0.0272194077783867\\
6	0.0139824439124803\\
7	0.00727236597684033\\
8	0.00382161458333334\\
};
\addlegendentry{\scriptsize 0dB-simulated}

\addplot [color=black, dashed, mark=+, mark options={solid, black}]
  table[row sep=crcr]{%
1	0.237487281828941\\
2	0.124293983910845\\
3	0.0642724242222026\\
4	0.0329811361614607\\
5	0.0168221799545646\\
6	0.0085493249321514\\
7	0.0043415690594577\\
8	0.00219520546581784\\
};
\addlegendentry{\scriptsize 0dB-estimated}

\addplot [color=blue, mark=o, mark options={solid, blue}]
  table[row sep=crcr]{%
1	0.748535865040728\\
2	0.55636287968582\\
3	0.386961802676769\\
4	0.254486355256617\\
5	0.157847308806794\\
6	0.0923905376517088\\
7	0.0521866601709869\\
8	0.0287723214285713\\
};
\addlegendentry{\scriptsize 10dB-simulated}

\addplot [color=black, dashed, mark=o, mark options={solid, black}]
  table[row sep=crcr]{%
1	0.792086053793979\\
2	0.583466182125666\\
3	0.386148368500489\\
4	0.232549475599582\\
5	0.131151757593468\\
6	0.0710800250911117\\
7	0.0376217765550282\\
8	0.019591800886742\\
};
\addlegendentry{\scriptsize 10dB-estimated}

\addplot [color=blue, mark=asterisk, mark options={solid, blue}]
  table[row sep=crcr]{%
1	1.03291474978009\\
2	1.09152886716618\\
3	0.946455938874549\\
4	0.764705450194442\\
5	0.583277510467361\\
6	0.420375453769025\\
7	0.280752479946969\\
8	0.179015997023809\\
};
\addlegendentry{\scriptsize 20dB-simulated}

\addplot [color=black, dashed, mark=asterisk, mark options={solid, black}]
  table[row sep=crcr]{%
1	1.08466626972295\\
2	1.18310136857598\\
3	1.05356130107178\\
4	0.864279099119369\\
5	0.655596224394413\\
6	0.448239111844311\\
7	0.276984955844419\\
8	0.158872452532526\\
};
\addlegendentry{\scriptsize 20dB-estimated}

\end{axis}

\end{tikzpicture}%
		\label{fig:Prcn_cmp_fs}
	}
	\caption{Comparison of simulated and estimated RCN power in \cc{a frequency-selective} AWGN channel.}
	\label{fig:Prcn_cmp_fs_major}
\end{figure}

\subsection{SER Evaluation} \label{sec:sim-serEst}

In this simulation, the SER for ADO-OFDM, HACO-OFDM and LACO-OFDM discussed in Sec. \ref{Sec:serEst} {is} examined. For a fair comparison, all schemes are tested under the same  electrical power $P_\text{elec}$ and \modi{a modulation order of $16$, i.e., $M_t(k) =16$, $\forall t,k\in \mathcal{B}_t$}. Each scheme equally distributes its effective power $P_\text{eff}$ in all effective subcarriers, \cc{where $P_\text{eff}$ is obtained from \cc{$P_\text{elec}$} using Table \ref{tab:power}.}
%As explained in Appendix \ref{Sec:Appen2}, the DC value in ADO-OFDM is chosen to be $d=\sqrt{\frac{3}{4} P_t}$ so that DCO-OFDM signal is positive with high probability. With the given $d$, the $P_t'=\frac{1}{6}P_t$ in the ADO-OFDM.

Fig. \ref{fig:ser_est_cmp_awgn} compares the simulated SER with evaluated SER using an RCN-aware and an RCN-unaware model, in a flat AWGN channel. In the RCN-aware evaluation, the total noise power is evaluated using \eqref{eq:Pzjle}, \ac{otherwise the total noise power is $\mathcal{P}\{V(k)\}$.} It can be seen the RCN-aware SER evaluation always matches the simulation results, while the RCN-unaware SER evaluation does not for most of the SNR in LACO-OFDM and ADO-OFDM. The figure shows that HACO-OFDM has the best tolerance to RCN among the three schemes, as the RCN-unaware SER is close to the simulation result. This is because PAM-DMT subcarriers only use imaginary symbols and thus are not affected by the the real part of RCN. 
%{\color{red} PS: I will comment on why HACO-OFDM has the best torlarance to RCN, which might be because that the 2nd layer of it only suffers from the imaginary part of RCN. }
%It can be seen that the theoretical SER always gives better performance than the one which ignores RCN. It also can be seen that the performance improvement in LACO-OFDM is larger than in ADO-OFDM and HACO-OFDM, and HACO-OFDM is shown to have the best tolerance to RCN among the three schemes. 
Similar results can be seen in Fig. \ref{fig:ser_est_cmp_fsChan} which compares the SER in the \cc{frequency-selective} AWGN channel in \cite[Fig. 4b]{laco-perf2-exp/bitLoading} with channel-inversion equalization at the receiver. Again, the RCN-aware SER evaluation closely matches the simulation results. 

\begin{figure}[!t]
	\subfloat[Flat AWGN channel.]{
		% This file was created by matlab2tikz.
%
%The latest updates can be retrieved from
%  http://www.mathworks.com/matlabcentral/fileexchange/22022-matlab2tikz-matlab2tikz
%where you can also make suggestions and rate matlab2tikz.
%
\definecolor{mycolor1}{rgb}{0.47059,0.67059,0.18824}%
\definecolor{mycolor2}{rgb}{0.85098,0.32941,0.10196}%
\definecolor{mycolor3}{rgb}{0.00000,0.45098,0.74118}%
\begin{tikzpicture}

\begin{axis}[%
width=2.4in,
height=2.3in,
at={(0,0)},
scale only axis,
xmin=5,
xmax=30,
xlabel style={font=\color{white!15!black}},
xlabel={$\gamma\text{ (dB)}$},
ymode=log,
ymin=6e-5,
ymax=1,
yminorticks=true,
ylabel style={font=\color{white!15!black}},
ylabel={SER},
axis background/.style={fill=white},
xmajorgrids,
ymajorgrids,
yminorgrids,
legend style={at={(0.03,0.03)}, anchor=south west, legend cell align=left, align=left, draw=white!15!black, font=\scriptsize},
]

\addlegendimage{only marks, mark=square, mycolor3},
\addlegendentry[text width=,text depth=]{ADO-OFDM},
\addlegendimage{only marks, mark=o, mycolor1},
\addlegendentry[text width=,text depth=]{HACO-OFDM},
\addlegendimage{only marks, mark=triangle, mycolor2},
\addlegendentry[text width=,text depth=]{LACO-OFDM},

\addlegendimage{solid}
\addlegendentry[text width=,text depth=]{ simulation results}
\addlegendimage{dotted}
\addlegendentry[text width=,text depth=]{ RCN-unaware evaluation}
\addlegendimage{only marks, mark = *}
\addlegendentry[text width=,text depth=]{ RCN-aware evaluation}

\addplot [color=mycolor3,  mark=square, mark options={solid, mycolor3}]
table[row sep=crcr]{%
0	0.884411546	\\
2	0.869171429	\\
4	0.848903718	\\
6	0.822277886	\\
8	0.787482192	\\
10	0.741152838	\\
12	0.680719961	\\
14	0.601822114	\\
16	0.505164188	\\
18	0.388219569	\\
20	0.256024266	\\
22	0.122969276	\\
24	0.032316634	\\
26	0.003691389	\\
28	0.000168689	\\
30	1.17E-06	\\
};
%\addlegendentry[text width=6em,text depth=]{\scriptsize ado-simulation}

\addplot [color=mycolor3, draw=none, only marks, mark=square*, mark options={solid, fill=mycolor3, mycolor3}]
table[row sep=crcr]{%
0	0.883573399	\\
2	0.867997968	\\
4	0.847701838	\\
6	0.821151733	\\
8	0.786204343	\\
10	0.739988714	\\
12	0.679152317	\\
14	0.60075755	\\
16	0.50357805	\\
18	0.389034375	\\
20	0.262353862	\\
22	0.135060351	\\
24	0.0365417	\\
26	0.00371437	\\
28	0.000162304	\\
30	1.59E-06	\\
};
%\addlegendentry[text width=6em,text depth=]{\scriptsize ado-evaluation with RCN}

\addplot [color=mycolor3, dotted, mark=square, mark options={solid, mycolor3}]
table[row sep=crcr]{%
0	0.882481512	\\
2	0.865790801	\\
4	0.843416648	\\
6	0.813255562	\\
8	0.772493776	\\
10	0.717570879	\\
12	0.644480844	\\
14	0.549875896	\\
16	0.433614116	\\
18	0.302880601	\\
20	0.175576636	\\
22	0.076503388	\\
24	0.021556351	\\
26	0.003112521	\\
28	0.000159169	\\
30	1.59E-06	\\
};
%\addlegendentry[text width=6em,text depth=]{\scriptsize ado-evaluation without RCN}

\addplot [color=mycolor1, mark=o, mark options={solid, mycolor1}]
table[row sep=crcr]{%
0	0.851172603	\\
2	0.826360078	\\
4	0.793692955	\\
6	0.751345205	\\
8	0.696921331	\\
10	0.628153816	\\
12	0.546684736	\\
14	0.45516047	\\
16	0.358772211	\\
18	0.263177495	\\
20	0.17721272	\\
22	0.110312133	\\
24	0.061579648	\\
26	0.027074755	\\
28	0.007905284	\\
30	0.00125225	\\
};
%\addlegendentry[text width=6em,text depth=]{\scriptsize haco-simulation}

\addplot [color=mycolor1, draw=none, only marks, mark=*, mark options={solid, fill=mycolor1, mycolor1}]
table[row sep=crcr]{%
0	0.84936324	\\
2	0.824270842	\\
4	0.791398742	\\
6	0.748359648	\\
8	0.692745283	\\
10	0.623015085	\\
12	0.539652345	\\
14	0.446291061	\\
16	0.349687465	\\
18	0.256667842	\\
20	0.174079368	\\
22	0.109892739	\\
24	0.061631833	\\
26	0.026986266	\\
28	0.007891589	\\
30	0.001229129	\\
};
%\addlegendentry[text width=6em,text depth=]{\scriptsize haco-evaluation with RCN}

\addplot [color=mycolor1, dotted, mark=o, mark options={solid, mycolor1}]
table[row sep=crcr]{%
0	0.846735169	\\
2	0.81982654	\\
4	0.784348906	\\
6	0.737792852	\\
8	0.677552675	\\
10	0.601844333	\\
12	0.511488746	\\
14	0.412156274	\\
16	0.314826092	\\
18	0.230938739	\\
20	0.163892958	\\
22	0.108460252	\\
24	0.06158832	\\
26	0.026986139	\\
28	0.007891589	\\
30	0.001229129	\\
};
%\addlegendentry[text width=6em,text depth=]{\scriptsize haco-evaluation without RCN}

\addplot [color=mycolor2, mark=triangle, mark options={solid, mycolor2}]
table[row sep=crcr]{%
0	0.867227202	\\
2	0.847174364	\\
4	0.821501761	\\
6	0.787766145	\\
8	0.742272603	\\
10	0.684118004	\\
12	0.60972818	\\
14	0.518281018	\\
16	0.411905088	\\
18	0.292145401	\\
20	0.16647182	\\
22	0.052384344	\\
24	0.003366145	\\
26	8.43E-05	\\
28	3.91E-07	\\
30	0.00E+00	\\
};
%\addlegendentry[text width=6em,text depth=]{\scriptsize laco-simulation}

\addplot [color=mycolor2, draw=none, only marks, mark=triangle*, mark options={solid, fill=mycolor2, mycolor2}]
table[row sep=crcr]{%
0	0.866131752	\\
2	0.8462545	\\
4	0.820661412	\\
6	0.78738785	\\
8	0.743798584	\\
10	0.68693029	\\
12	0.614383027	\\
14	0.525153138	\\
16	0.419716162	\\
18	0.300494165	\\
20	0.173386411	\\
22	0.056537203	\\
24	0.002603781	\\
26	7.10E-05	\\
28	4.40E-07	\\
30	1.62E-10	\\
};
%\addlegendentry[text width=6em,text depth=]{\scriptsize laco-evaluation with RCN}

\addplot [color=mycolor2, dotted, mark=triangle, mark options={solid, mycolor2}]
table[row sep=crcr]{%
0	0.861518209	\\
2	0.837668442	\\
4	0.805490039	\\
6	0.762004983	\\
8	0.703513001	\\
10	0.62601706	\\
12	0.526583894	\\
14	0.406263892	\\
16	0.274383104	\\
18	0.151051847	\\
20	0.060824681	\\
22	0.015191757	\\
24	0.001821071	\\
26	6.96E-05	\\
28	4.40E-07	\\
30	1.62E-10	\\
};
%\addlegendentry[text width=6em,text depth=]{\scriptsize laco-evaluation without RCN}

\end{axis}

\end{tikzpicture}%
		\label{fig:ser_est_cmp_awgn}
	}
	\subfloat[Frequency-selective AWGN channel.]{
		% This file was created by matlab2tikz.
%
%The latest updates can be retrieved from
%  http://www.mathworks.com/matlabcentral/fileexchange/22022-matlab2tikz-matlab2tikz
%where you can also make suggestions and rate matlab2tikz.
%

\definecolor{mycolor1}{rgb}{0.47059,0.67059,0.18824}%
\definecolor{mycolor2}{rgb}{0.85098,0.32941,0.10196}%
\definecolor{mycolor3}{rgb}{0.00000,0.45098,0.74118}%
\begin{tikzpicture}

\begin{axis}[%
width=2.4in,
height=2.3in,
at={(0,0)},
scale only axis,
xmin=5,
xmax=34,
xlabel style={font=\color{white!15!black}},
xlabel={$\gamma \text{ (dB)}$},
ymode=log,
ymin=6e-5,,
ymax=1,
yminorticks=true,
ylabel style={font=\color{white!15!black}},
ylabel={SER},
axis background/.style={fill=white},
xmajorgrids,
ymajorgrids,
yminorgrids,
legend style={at={(0.03,0.03)}, anchor=south west, legend cell align=left, align=left, draw=white!15!black, font=\scriptsize},
]

\addlegendimage{only marks, mark=square, mycolor3},
\addlegendentry[text width=,text depth=]{ADO-OFDM},
\addlegendimage{only marks, mark=o, mycolor1},
\addlegendentry[text width=,text depth=]{HACO-OFDM},
\addlegendimage{only marks, mark=triangle, mycolor2},
\addlegendentry[text width=,text depth=]{LACO-OFDM},

\addlegendimage{solid}
\addlegendentry[text width=,text depth=]{ simulation results}
\addlegendimage{dotted}
\addlegendentry[text width=,text depth=]{ RCN-unaware evaluation}
\addlegendimage{only marks, mark = *}
\addlegendentry[text width=,text depth=]{ RCN-aware evaluation}

\addplot [color=mycolor3, mark=square, mark options={solid, mycolor3}]
  table[row sep=crcr]{%
0	0.901529159	\\
2	0.890971037	\\
4	0.877746575	\\
6	0.860427984	\\
8	0.83758317	\\
10	0.807513503	\\
12	0.768034442	\\
14	0.715913112	\\
16	0.64813816	\\
18	0.562044618	\\
20	0.457770059	\\
22	0.336185127	\\
24	0.204023679	\\
26	8.46E-02	\\
28	1.87E-02	\\
30	1.97E-03	\\
32	8.36E-05	\\
34	1.17E-06	\\
};
%\addlegendentry{ado-simulation}

\addplot [color=mycolor3, draw=none, mark=square*, mark options={solid, fill=mycolor3, mycolor3}]
  table[row sep=crcr]{%
0	0.900763371	\\
2	0.890354057	\\
4	0.876824153	\\
6	0.859219432	\\
8	0.836258538	\\
10	0.806171003	\\
12	0.766532567	\\
14	0.714289949	\\
16	0.646275504	\\
18	0.560356438	\\
20	0.456725748	\\
22	0.338690677	\\
24	0.213415991	\\
26	9.54E-02	\\
28	2.03E-02	\\
30	1.97E-03	\\
32	8.65E-05	\\
34	9.03E-07	\\
};
%\addlegendentry{ado-theoretical value-considering RCN}

\addplot [color=mycolor3, dashdotted, mark=square, mark options={solid, mycolor3}]
  table[row sep=crcr]{%
0	0.900390534	\\
2	0.889574527	\\
4	0.875229079	\\
6	0.856065439	\\
8	0.830305894	\\
10	0.79554541	\\
12	0.748663466	\\
14	0.685942703	\\
16	0.603723345	\\
18	0.500111032	\\
20	0.378109671	\\
22	0.249262646	\\
24	0.133991949	\\
26	5.33E-02	\\
28	1.35E-02	\\
30	1.77E-03	\\
32	8.56E-05	\\
34	9.03E-07	\\
};
%\addlegendentry{ado-theoretical value-ignoring RCN}

\addplot [color=mycolor1, mark=o, mark options={solid, mycolor1}]
  table[row sep=crcr]{%
0	0.878347358	\\
2	0.861618591	\\
4	0.840409002	\\
6	0.81238728	\\
8	0.775686693	\\
10	0.727919374	\\
12	0.668130724	\\
14	0.593923875	\\
16	0.509220744	\\
18	0.416892172	\\
20	0.323137378	\\
22	0.23114501	\\
24	0.149925049	\\
26	8.99E-02	\\
28	4.69E-02	\\
30	1.89E-02	\\
32	0.004968885	\\
34	0.000710176	\\
};
%\addlegendentry{haco-simulation}

\addplot [color=mycolor1, draw=none, mark=*, mark options={solid, fill=mycolor1, mycolor1}]
  table[row sep=crcr]{%
0	0.877047577	\\
2	0.860287131	\\
4	0.8385304	\\
6	0.810130226	\\
8	0.772966344	\\
10	0.724657902	\\
12	0.663205793	\\
14	0.58805848	\\
16	0.50123938	\\
18	0.407812942	\\
20	0.314176307	\\
22	0.225204508	\\
24	0.147929163	\\
26	8.96E-02	\\
28	4.71E-02	\\
30	1.89E-02	\\
32	0.005007568	\\
34	0.000706609	\\
};
%\addlegendentry{haco-theoretical value-considering RCN}

\addplot [color=mycolor1, dotted, mark=o, mark options={solid, mycolor1}]
  table[row sep=crcr]{%
0	0.8759633	\\
2	0.858289658	\\
4	0.835049061	\\
6	0.804426421	\\
8	0.764166475	\\
10	0.711745023	\\
12	0.644952162	\\
14	0.563157571	\\
16	0.469218187	\\
18	0.370895963	\\
20	0.279212832	\\
22	0.202368403	\\
24	0.140359479	\\
26	8.87E-02	\\
28	4.71E-02	\\
30	1.89E-02	\\
32	0.005007568	\\
34	0.000706609	\\
};
%\addlegendentry{haco-theoretical value-ignoring RCN}

\addplot [color=mycolor2, mark=triangle, mark options={solid, mycolor2}]
  table[row sep=crcr]{%
0	0.889597456	\\
2	0.875869667	\\
4	0.858695108	\\
6	0.836363014	\\
8	0.806896673	\\
10	0.768148924	\\
12	0.718350098	\\
14	0.653223092	\\
16	0.572456556	\\
18	0.475378278	\\
20	0.365063796	\\
22	0.245422114	\\
24	0.124794129	\\
26	2.85E-02	\\
28	1.46E-03	\\
30	4.21E-05	\\
32	3.91E-07	\\
34	0	\\
};
%\addlegendentry{laco-simulation}

\addplot [color=mycolor2, draw=none, mark=triangle*, mark options={solid, fill=mycolor2, mycolor2}]
  table[row sep=crcr]{%
0	0.888522474	\\
2	0.874924233	\\
4	0.857534091	\\
6	0.835236287	\\
8	0.806436618	\\
10	0.768912678	\\
12	0.71990363	\\
14	0.65668793	\\
16	0.57758154	\\
18	0.482604155	\\
20	0.373450194	\\
22	0.253630667	\\
24	0.130982133	\\
26	3.09E-02	\\
28	1.20E-03	\\
30	3.78E-05	\\
32	2.59E-07	\\
34	1.39E-10	\\
};
%\addlegendentry{laco-theoretical value-considering RCN}

\addplot [color=mycolor2, dotted, mark=triangle, mark options={solid, mycolor2}]
  table[row sep=crcr]{%
0	0.886823344	\\
2	0.871564501	\\
4	0.851151002	\\
6	0.823681469	\\
8	0.786600575	\\
10	0.736637483	\\
12	0.67000475	\\
14	0.583231561	\\
16	0.475166656	\\
18	0.350375345	\\
20	0.222507398	\\
22	0.113161304	\\
24	0.041438705	\\
26	9.34E-03	\\
28	1.02E-03	\\
30	3.74E-05	\\
32	2.59E-07	\\
34	1.39E-10	\\
};
%\addlegendentry{laco-theoretical value-ignoring RCN}

\end{axis}

\end{tikzpicture}%
		\label{fig:ser_est_cmp_fsChan}
	}
	\caption{Simulated and evaluated SER comparison for ADO-OFDM, HACO-OFDM and LACO-OFDM in an AWGN channel, where \modi{$16$-QAM (or $16$-PAM for the PAM-DMT layer of HACO-OFDM)} is used in all subcarriers. The filled (empty) circle, triangle, and square markers represent the theoretical RCN-aware (RCN-unaware) SER evaluation of ADO-OFDM, HACO-OFDM and LACO-OFDM, respectively.}
	\label{fig:ser_est_cmp_major}
\end{figure}
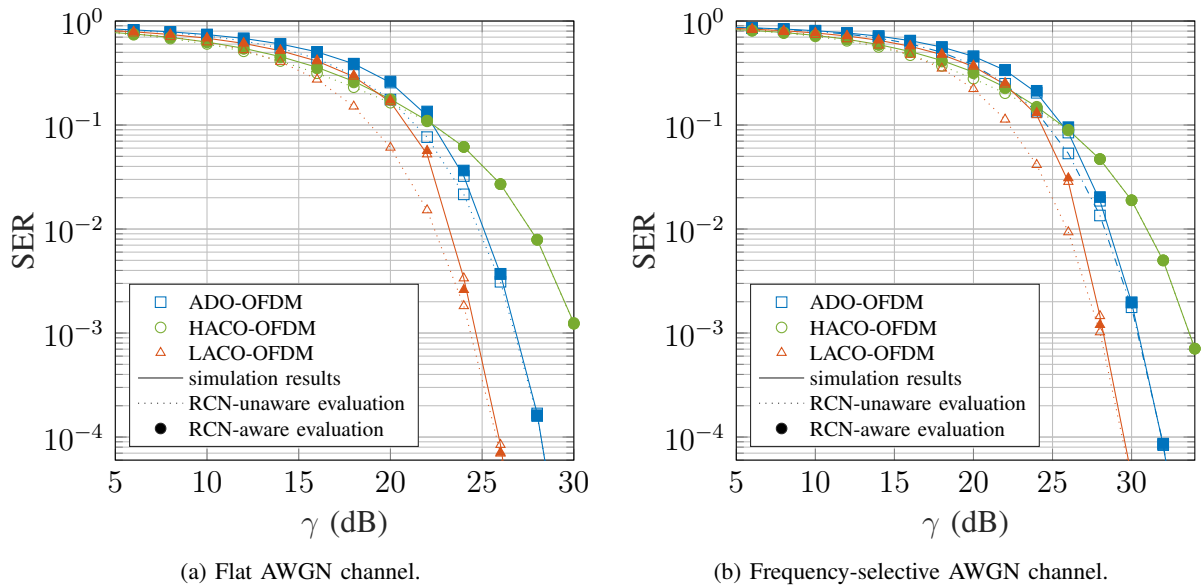

\ac{Note that a similar trend as in Fig. \ref{fig:ser_est_cmp_major} has been observed for smaller $N$, such as $N=64$ instead of $1024$, which indicates that the model is relevant even at relatively small $N$.}

\subsection{SER-controlled LACO-OFDM} \label{sec:sim-optimization}
Here, we examine the SER-controlled LACO-OFDM scheme proposed in Sec. \ref{sec:optimization} under the \cc{frequency-selective} channel in \cite[Fig. 4b]{laco-perf2-exp/bitLoading}. We use \eqref{eq:pe_laco_sys} to obtain the theoretical SER performance of the proposed resource allocation scheme. The theoretical SER are compared to simulated SER to show the accuracy of the proposed SER calculation. For resource allocation, the SER target is set at $p_{\rm e}=10^{-2}$ (as an example) which leads to a BER in the order of $10^{-3}$. The convergence criterion is set $\epsilon = 10^{-3} \frac{\| {\bf{P}}_z^{i} \|^2}{N}$.

Fig. \ref{fig:sim-optimizaton-ser} compares the theoretical and simulation SER performance of the resource allocation scheme. Both RCN-aware and {RCN-unaware} resource allocation are considered, where {RCN-unaware} resource allocation is realized by \ac{using $P_z^{(i+1)}=\mathcal{P}\{V(k)\}$ in the resource allocation algorithm instead of $P_z^{(i+1)}=\mathcal{P}\{V(k)\}+\sum_{t=1}^{j_k-1}P_t^{\rm w}$.} It can be seen that most theoretical and simulated SER values match for most $P_{\rm eff}$. Mismatch happens when {the average number} of bits per subcarrier is 1 or 3, i.e., when 2QAM and 8QAM are used, which makes \eqref{eq:pe_qam} inaccurate \cite{book-digitalComm5th}. It can also be seen that RCN-aware resource allocation controls the SER perfectly under the $10^{-2}$ constraint for almost all $\gamma_\text{eff}$, which guarantees the SER performance. 
%For gray-coding at high $\gamma_\text{eff}$, {\color{red} $\text{BER} \approx {\text{SER}} /(\text{avaraged number of bits per subcarrier})$, \cite{bitloading3}. }
{For gray-coding at high $\gamma_\text{eff}$, BER can be evaluated from SER through the relation $\text{BER} \approx \frac{\text{SER}}{\bar{n}_b}$, where $\bar{n}_b$ is the average number of bits per subcarrier among all the layers.} Take the $15-30$dB $\gamma_\text{eff}$ region as an example. In this region, the average number of bits per subcarrier is from 3 to 8, {thus the BER is in the range from $1.25\times 10^{-3}$ to $3.33\times 10^{-3}$. When the BER range} is known, coding schemes {can be chosen} to provide guaranteed coding gain \cite{snrGap,superFEC}, and leading to a more predictable coding performance (BER after decoding). Therefore, the proposed RCN power model and the proposed SER-controlled LACO-OFDM resource allocation scheme can guarantee reliable system performance. \ac{Fig. \ref{fig:sim-optimizaton-ser} also shows that 2 iterations of the resource allocation algorithm lead to an SER which is} almost the same as the SER after convergence.

Fig. \ref{fig:bit_number_perSc} shows the average number of bits per subcarrier allocated under each $\gamma_\text{eff}$. \ac{The figure shows that the RCN-aware scheme is more conservative in allocating bits than the RCN-unaware scheme, since it takes RCN into account. This is reflected as higher reliability in Fig. \ref{fig:sim-optimizaton-ser} since the target SER is more likely to be met than when RCN-unaware allocation is used.}

%{It is also worth to note that, from Fig. \ref{fig:optimization}(a), even $\gamma_\text{eff}$ is as high as more than 25dB, which means $\gamma \ge 28$ dB, RCN still causes SER {outage} as shown in the performance of RCN-unaware system, which suggests that the common assumption that \textit{RCN can be eliminated under a high SNR} \cite{cmp2/eaco, cmp5/powerAlloc} {is not practical}. }

\begin{figure}[!t]
	\subfloat[SER versus effective SNR.]{
		\input{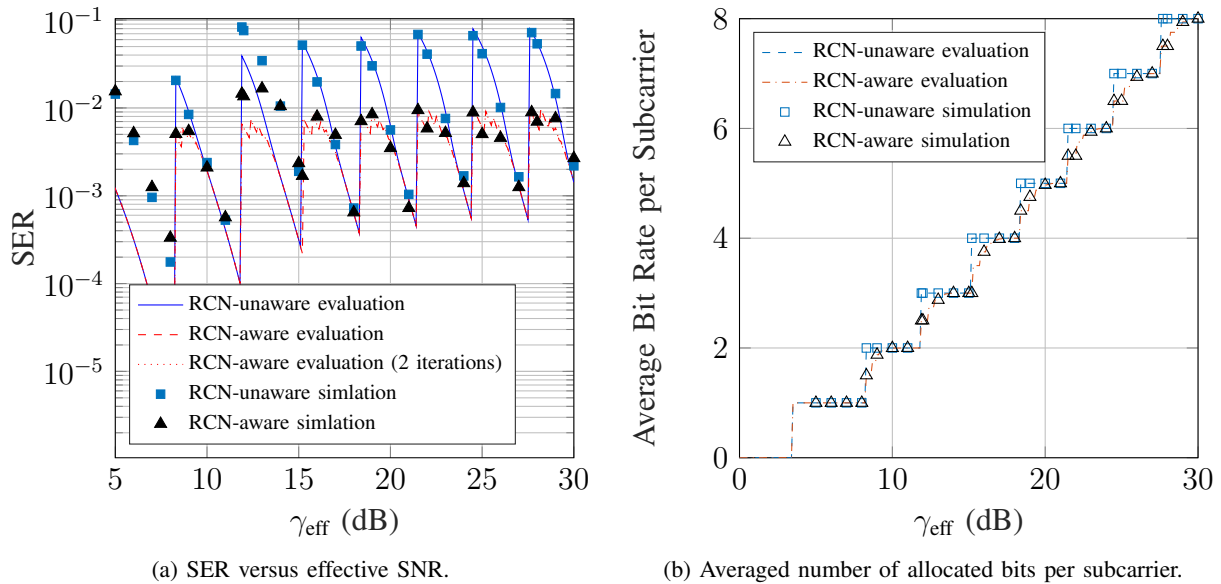}
		\label{fig:sim-optimizaton-ser}
	}
	\subfloat[Averaged number of allocated bits per subcarrier.]{
		% This file was created by matlab2tikz.
%
%The latest updates can be retrieved from
%  http://www.mathworks.com/matlabcentral/fileexchange/22022-matlab2tikz-matlab2tikz
%where you can also make suggestions and rate matlab2tikz.
%
\definecolor{mycolor1}{rgb}{0.00000,0.44700,0.74100}%
\definecolor{mycolor2}{rgb}{0.85000,0.32500,0.09800}%
\definecolor{mycolor3}{rgb}{0.00000,0.45098,0.74118}%
\begin{tikzpicture}

\begin{axis}[%
width=2.4in,
height=2.3in,
at={(0,0)},
scale only axis,
xmin=0,
xmax=30,
xlabel style={font=\color{white!15!black}},
xlabel={\cc{$\gamma_\text{eff}$} (dB)},
ymin=0,
ymax=8,
ylabel style={font=\color{white!15!black}},
ylabel={Average Bit Rate per Subcarrier},
axis background/.style={fill=white},
xmajorgrids,
ymajorgrids,
legend style={at={(0.03,0.97)}, anchor=north west, legend cell align=left, align=left, draw=white!15!black, font=\scriptsize},
]
\addplot [color=mycolor1, dashed]
  table[row sep=crcr]{%
0	0\\
3.4	0\\
3.5	1\\
8.2	1\\
8.3	2\\
11.8	2\\
11.9	3\\
15.1	3\\
15.2	4\\
18.3	4\\
18.4	5\\
21.4	5\\
21.5	6\\
24.4	6\\
24.5	7\\
27.5	7\\
27.6	8\\
30	8\\
};
\addlegendentry{RCN-unaware evaluation}

\addplot [color=mycolor2, dashdotted]
  table[row sep=crcr]{%
0	0\\
3.4	0\\
3.5	1\\
8.2	1\\
8.3	1.50097847358121\\
8.6	1.50097847358121\\
8.7	1.75146771037182\\
8.9	1.75146771037182\\
9	1.87671232876712\\
9.1	1.87671232876712\\
9.2	1.93933463796478\\
9.3	1.93933463796478\\
9.4	1.9706457925636\\
9.5	1.9706457925636\\
9.6	1.98630136986301\\
9.7	1.98630136986301\\
9.8	1.99412915851272\\
9.9	1.99804305283757\\
11.8	2\\
11.9	2.50097847358121\\
12.3	2.50097847358121\\
12.4	2.75146771037182\\
12.7	2.75146771037182\\
12.8	2.87671232876712\\
13	2.87671232876712\\
13.1	2.93933463796478\\
13.2	2.93933463796478\\
13.3	2.9706457925636\\
13.4	2.9706457925636\\
13.5	2.98630136986301\\
13.6	2.98630136986301\\
13.7	2.99412915851272\\
13.8	2.99804305283757\\
15.2	3\\
15.3	3.50097847358121\\
15.7	3.50097847358121\\
15.8	3.75146771037182\\
16.1	3.75146771037182\\
16.2	3.87671232876712\\
16.4	3.87671232876712\\
16.5	3.93933463796478\\
16.7	3.93933463796478\\
16.8	3.9706457925636\\
16.9	3.9706457925636\\
17	3.98630136986301\\
17.1	3.98630136986301\\
17.2	3.99412915851272\\
17.4	4\\
18.3	4\\
18.4	4.50097847358121\\
18.9	4.50097847358121\\
19	4.75146771037182\\
19.3	4.75146771037182\\
19.4	4.87671232876712\\
19.6	4.87671232876712\\
19.7	4.93933463796478\\
19.9	4.93933463796478\\
20	4.9706457925636\\
20.1	4.9706457925636\\
20.2	4.98630136986301\\
20.3	4.98630136986301\\
20.4	4.99412915851272\\
20.5	4.99804305283757\\
21.4	5\\
21.5	5.50097847358121\\
22	5.50097847358121\\
22.1	5.75146771037182\\
22.4	5.75146771037182\\
22.5	5.87671232876712\\
22.8	5.87671232876712\\
22.9	5.93933463796478\\
23	5.93933463796478\\
23.1	5.9706457925636\\
23.2	5.9706457925636\\
23.3	5.98630136986301\\
23.4	5.98630136986301\\
23.5	5.99412915851272\\
23.6	5.99412915851272\\
23.8	6\\
24.4	6\\
24.5	6.49902152641879\\
24.7	6.50097847358121\\
25.1	6.50097847358121\\
25.2	6.75146771037182\\
25.5	6.75146771037182\\
25.6	6.87671232876712\\
25.8	6.87671232876712\\
25.9	6.93933463796478\\
26.1	6.93933463796478\\
26.2	6.9706457925636\\
26.3	6.9706457925636\\
26.4	6.98630136986301\\
26.5	6.98630136986301\\
26.6	6.99412915851272\\
26.7	6.99804305283757\\
27.5	7\\
27.6	7.49902152641879\\
27.8	7.50097847358121\\
28.1	7.50097847358121\\
28.2	7.75146771037182\\
28.6	7.75146771037182\\
28.7	7.87671232876712\\
28.9	7.87671232876712\\
29	7.93933463796478\\
29.2	7.93933463796478\\
29.3	7.9706457925636\\
29.4	7.9706457925636\\
29.5	7.98630136986301\\
29.6	7.98630136986301\\
29.7	7.99412915851272\\
29.9	8\\
30	8\\
};
\addlegendentry{RCN-aware evaluation}

\addplot [color=mycolor3, draw=none, only marks, mark=square, mark options={solid, scale = 0.8, mycolor3}]
table[row sep=crcr]{%
	5	1\\
	6	1\\
	7	1\\
	8	1\\
	8.3	2\\
	9	2\\
	10	2\\
	11	2\\
	11.9	3\\
	12	3\\
	13	3\\
	14	3\\
	15	3\\
	15.2	4\\
	16	4\\
	17	4\\
	18	4\\
	18.4	5\\
	19	5\\
	20	5\\
	21	5\\
	21.5	6\\
	22	6\\
	23	6\\
	24	6\\
	24.5	7\\
	25	7\\
	26	7\\
	27	7\\
	27.7	8\\
	28	8\\
	29	8\\
	30	8\\
};
\addlegendentry{RCN-unaware  simulation}

\addplot [color=black, draw=none, only marks, mark=triangle, mark options={solid, scale = 1.3, black}]
  table[row sep=crcr]{%
5	1\\
6	1\\
7	1\\
8	1\\
8.3	1.50097847358121\\
9	1.87671232876712\\
10	1.99804305283757\\
11	2\\
11.9	2.50097847358121\\
12	2.50097847358121\\
13	2.87671232876712\\
14	3\\
15	3\\
15.2	3\\
16	3.75146771037182\\
17	3.98630136986301\\
18	4\\
18.4	4.50097847358121\\
19	4.75146771037182\\
20	4.9706457925636\\
21	5\\
21.5	5.50097847358121\\
22	5.50097847358121\\
23	5.93933463796478\\
24	6\\
24.5	6.49902152641879\\
25	6.50097847358121\\
26	6.93933463796478\\
27	7\\
27.7	7.50097847358121\\
28	7.50097847358121\\
29	7.93933463796478\\
30	8\\
};
\addlegendentry{RCN-aware simulation}

\end{axis}
\end{tikzpicture}%
		\label{fig:bit_number_perSc}
	}
	\caption{Performance of SER-controlled LACO-OFDM under a $10^{-2}$ target SER constraint. The dotted line and the dashed line are almost overlapping.}
	\label{fig:optimization}
\end{figure}

\section{Conclusion} \label{Sec:conclusion}
In this paper, we studied residual clipping noise in several enhanced ACO-OFDM schemes, including ADO-OFDM, HACO-OFDM, and LACO-OFDM. We proposed a worst-case RCN power model, which provides an accurate evaluation of total noise power and symbol error rate, and is the first  accurate RCN power model in the literature. Using the proposed RCN power model, we have proposed an RCN-aware resource allocation algorithm for enhanced ACO-OFDM, which is reliable in terms of achieving a target SER performance. \ac{The proposed model can be used in general system analysis and optimization research in enhanced ACO-OFDM, such as bit loading, power allocation, data rate maximization, error rate minimization, etc., and is hence of practical significance.}

\ac{It is worthwhile to note that, as future work, it would be interesting to extend this model to systems with a peak intensity constraint, where peak-clipping distortion should be accounted for. Moreover,} it is interesting to study the trade-off between complexity and \ac{error-rate performance when the RCN-aware design is combined with channel coding}, which will shed light on the design of practical eACO-OFDM systems.

\begin{appendices}

\section{Derivation of Power Relations} \label{Sec:Appen2}
This section proves the power relations listed in Table \ref{tab:power}. We first recall Lemma \ref{Lem1} in Sec. \ref{sec:statistics_analysis} which will be used in what follows. Lemma \ref{Lem1} states that if the number of loaded subcarriers in $S(k)$ is large, and $S(k)$ is zero-mean and bounded for all $k$ and Hermitian symmetric, then $s(n)$ converges in distribution for all $n$ to a Gaussian distribution $\mathcal{N} ( 0, \sigma^2 )$ for some $\sigma^2$. Consequently, clipping out negative $s(n)$ generates $x(n)=(s(n))^+ = \frac{s(n) + |s(n)|}{2}$ whose distribution (for all $n$) is $\mathcal{N}_c \big( 0, \sigma^2, [ 0, \infty ) \big)$ defined by the probability density function
\begin{equation}
	p_X(x) = \begin{cases}
	\frac{1}{\sqrt{2\pi}\sigma}{e^{ -\frac{x^2}{2\sigma^2} }} & x > 0,\\
	\frac{1}{2}\sdelta(x) & x = 0,
	\end{cases}
\end{equation}
with mean $\mathbb{E}\{x(n)\} = \frac{\sigma}{\sqrt{2\pi}}$ and second moment $\mathbb{E}\{x^2(n)\} = \frac{\sigma^2}{2}$, where $\sdelta(x)$ is the Dirac impulse.

An ACO-OFDM signal is given by $x_{\rm aco}(n) = (s_{\rm aco}(n))^+$. By Lemma \ref{Lem1}, $s_{\rm aco}(n) \sim \mathcal{N}(0,\sigma^2)$ for some $\sigma^2$. Hence, $x_{\rm aco}(n) \sim \mathcal{N}_c \big( 0, \sigma^2, [ 0, \infty ) \big)$ for all $n$. Thus, $P_\text{elec}=\frac{1}{N}\sum_{n=0}^{N-1}\mathbb{E}[x_{\rm aco}^2(n)]$ by the law of large numbers, leading to $P_\text{elec}= \frac{\sigma^2}{2}$. Similarly, we can write $P_\text{opt} = \frac{1}{N}\sum_{n=0}^{N-1}\mathbb{E}[x_{\rm aco}(n)] = \frac{\sigma}{\sqrt{2\pi}}$. We also have $P_\text{eff} = \mathcal{P}\{\frac{\mathbf{s}_{\rm aco}}{2}\} = \frac{1}{N}\sum_{n=0}^{N-1} \mathbb{E}[(\frac{s_{\rm aco}(n)}{2})^2] = \frac{\sigma^2}{4}$. Therefore, we have that $P_\text{elec}=2P_{\rm eff}$ and $P_\text{opt}=\sqrt{\frac{2P_{\rm eff}}{\pi}}$ as stated in Table \ref{tab:power}.

For DCO-OFDM, the signal $x_{\rm dco}(n)$ is given by $x_{\rm dco}(n) = (s_{\rm dco}(n) + d_{\rm dco})^+$. Recall that $s_{\rm dco}(n)\sim \mathcal{N} ( 0, \sigma^2 )$ for some $\sigma^2$ by Lemma \ref{Lem1}. By choosing $d_{\rm dco} = 3\sigma$, we can approximate $x_{\rm dco}(n)\approx s_{\rm dco}(n) + d_{\rm dco}$. Hence, we can approximate  $x_{\rm dco}(n)$ to be $\mathcal{N} ( d_{\rm dco}, \sigma^2 )$. Then, we have $P_\text{elec} = \frac{1}{N}\sum_{n=0}^{N-1}\mathbb{E}[x_{\rm dco}^2(n)] = 10\sigma^2$ and $P_\text{opt} = \frac{1}{N}\sum_{n=0}^{N-1} \mathbb{E}[x_{\rm dco}(n)] = 3\sigma$. We also have $P_\text{eff} = \mathcal{P}\{\mathbf{s}_{\rm dco}\} = \sigma^2$. We conclude that $P_\text{elec}=10P_{\rm eff}$ and $P_\text{opt}=3\sqrt{P_{\rm eff}}$ as shown in Table \ref{tab:power}.

The analysis of PAM-DMT omitted since it is similar to that of ACO-OFDM, leading to the relations $P_\text{elec}=2P_{\rm eff}$ and $P_\text{opt}=\sqrt{\frac{2P_{\rm eff}}{\pi}}$ as stated in Table \ref{tab:power}.

For an ADO-OFDM, the transmit signal is $x_{\rm ado}(n) = x_{1}(n) + x_{2}(n)$, where $x_{1}(n) = (s_{\rm aco}(n))^+ $ is the ACO-OFDM signal in layer 1, $x_{2}(n) = (s_{\rm dco}^{(2)}(n) + d_{\rm dco}^{(2)})^+$ is the DCO-OFDM signal in layer 2. Note that $s_{\rm aco}(n) \sim \mathcal{N}(0, \sigma_1^2)$ and $s_{\rm dco}^{(2)}(n)\sim\mathcal{N}(0,\sigma_2^2)$ for some $\sigma_1^2$ and $\sigma^2_2$ by Lemma \ref{Lem1}. Then $x_{1}(n) \sim \mathcal{N}_c \big( 0, \sigma_1^2, [ 0, \infty ) \big) $. Moreover, by choosing $d_{\rm dco}^{(2)} = 3\sigma_2$, $x_{2}(n) \sim \mathcal{N} ( 3\sigma_2, \sigma_2^2 )$. Thus, using the law of large numbers, we have 
$P_\text{elec} = \frac{1}{N}\sum_{n=0}^{N-1}\mathbb{E}[x_{\rm ado}^2(n)] = \frac{1}{N}\sum_{n=0}^{N-1}\mathbb{E}[(x_{1}(n) + x_{2}(n))^2] = \frac{\sigma_1^2}{2} + 10\sigma_2^2 + \frac{6}{\sqrt{2\pi}}\sigma_1\sigma_2$ and 
$P_\text{opt} = \frac{1}{N}\sum_{n=0}^{N-1} \mathbb{E}[x_{\rm ado}(n)] = \frac{1}{N}\sum_{n=0}^{N-1} \mathbb{E}[x_{1}(n) + x_{2}(n)] = \frac{\sigma_1}{\sqrt{2\pi}} + 3\sigma_2$. We also have $P_\text{eff} = \mathcal{P}\{\frac{\mathbf{s}_{\rm aco}}{2} + \mathbf{s}_{\rm dco}^{(2)} \} = \frac{\sigma_1^2}{4} + \sigma_2^2$. Letting $\sigma_2 = \alpha\sigma_1$ for some $\alpha$, then $P_\text{elec}$, $P_\text{opt}$, and $P_\text{eff}$ can be related through $\sigma_1$ for a general $\alpha$.  By setting $\alpha = \frac{1}{2}$ which equally distributes $P_\text{eff}$ in all effective subcarriers of ADO-OFDM, we obtain $P_{\rm elec}=(6+\frac{6}{\sqrt{2\pi}})P_{\rm eff}$ and $P_{\rm opt}=(\frac{1}{\sqrt{\pi}}+\frac{3}{\sqrt{2}})\sqrt{P_{\rm eff}}$ as shown in Table \ref{tab:power}.

%\cc{For a ADO-OFDM signal $x(n)=\ut{x}(n) + d = \ut{x}_1(n) + \ut{x}_2(n) + d$, where $x_1(n) = (s_1(n))^+ = \ut{x}_1(n) + d_1$ is the ACO-OFDM signal in layer 1, $x_2(n) = (s_2(n) + d_2)^+ = \ut{x}(n) + d_2$ is the DCO-OFDM signal in layer 2, and $d = d_1+d_2$. Note that $s_i(n) \sim \mathcal{N}(0, \sigma_i^2)$, $i=1,2$. Therefore $x_1(n) \sim \mathcal{N}_c \big( 0, \sigma_1^2, [ 0, \infty ) \big) $ with $\sigma_1^2 = \mathbb{E}[s_1^2(n)]$ and $d_1 = \mathbb{E}[x_1(n)] = \frac{\sigma_1}{\sqrt{2\pi}}$, and by choosing $d_2 = 3\sigma_2$, $x_2(n) \sim \mathcal{N} ( 3\sigma_2, \sigma_2^2 )$. 
%Then, we have $P_\text{AC} = \mathbb{E}[\ut{x}^2(n)] = \mathbb{E}[\ut{x}_1^2(n)]+\mathbb{E}[\ut{x}_2^2(n)] = (\frac{1}{2} - \frac{1}{2\pi})\sigma_1^2 + \sigma_2^2$, 
%$P_\text{DC} = P_\text{opt}^2 = d^2 = (\frac{\sigma_1}{\sqrt{2\pi}}+3\sigma_2)^2$, and 
%$P_\text{elec} = P_\text{AC} + P_\text{DC} = \frac{1}{2}\sigma_1^2 + 10\sigma_2^2 + \frac{6}{\sqrt{2\pi}}\sigma_1\sigma_2$.
%We also have $P_\text{eff} = \mathbb{E}[(\frac{s_1(n)}{2} + s_2(n))^2] = \frac{\sigma_1^2}{4} + \sigma_2^2$.
%For general analysis, let $\sigma_2 = \alpha\sigma_1 = \alpha \sigma$, then the relation among the powers can be built as they are all related to $\sigma^2$. Specifically, the results shown in Table \ref{tab:power} are obtained by setting $\alpha = \frac{1}{2}$ which equalizes the average symbol power in all layers. }

In HACO-OFDM, the transmit signal is $x_{\rm haco}(n) = x_{1}(n) + x_{2}(n)$, where $x_{1}(n) = (s_{\rm aco}(n))^+ $ is the ACO-OFDM signal in layer 1 and $x_{2}(n) = (s_{\rm pam}(n))^+$ is the PAM-DMT signal in layer 2. Using Lemma \ref{Lem1}, we have that $s_{\rm aco}(n) \sim \mathcal{N}(0, \sigma_1^2)$ and $s_{\rm pam}(n)\sim\mathcal{N}(0,\sigma_2^2)$ for some $\sigma_1^2$ and $\sigma_2^2$. Therefore, $x_{i}(n) \sim \mathcal{N}_c \big( 0, \sigma_i^2, [ 0, \infty ) \big) $. Then, using the law of large numbers, we have 
$P_\text{elec} = \frac{1}{N}\sum_{n=0}^{N-1}\mathbb{E}[x_{\rm haco}^2(n)] = \mathbb{E}[(x_{1}(n) + x_{2}(n))^2] = \frac{\sigma_1^2+\sigma_2^2}{2}+ \frac{\sigma_1\sigma_2}{\pi}$ and 
$P_\text{opt} = \frac{1}{N}\sum_{n=0}^{N-1} \mathbb{E}[x_{\rm haco}(n)] = \mathbb{E}[x_{1}(n) + x_{2}(n)] = \frac{\sigma_1+\sigma_2}{\sqrt{2\pi}}$. 
We also have $P_\text{eff} = \mathcal{P}\{\frac{\mathbf{s}_{\rm aco}}{2}+\frac{\mathbf{s}_{\rm pam}}{2}\} = \frac{\sigma_1^2 + \sigma_2^2}{4}$. In general, letting $\sigma_2 = \alpha\sigma_1$ for some $\alpha$, we can obtain relations among $P_{\rm elec}$, $P_{\rm opt}$ and $P_{\rm eff}$ through $\sigma_1$. By setting $\alpha = 1$ which equally distributes $P_\text{eff}$ in all effective subcarriers of HACO-OFDM, we obtain $P_{\rm elec}=(2+\frac{2}{\pi})P_{\rm eff}$ and $P_{\rm opt}=\frac{2}{\sqrt{\pi}}\sqrt{P_{\rm eff}}$ as shown in Table \ref{tab:power}.

%\cc{For an HACO-OFDM signal $x(n)=\ut{x}(n) + d = \ut{x}_1(n) + \ut{x}_2(n) + d$, where $x_i(n) = (s_i(n))^+ = \ut{x}_i(n) + d_i$, $i=1,2$ and $d = d_1+d_2$. Note that $s_i(n) \sim \mathcal{N}(0, \sigma_i^2)$, $i=1,2$. Therefore $x_i(n) \sim \mathcal{N}_c \big( 0, \sigma_i^2, [ 0, \infty ) \big) $ with $\sigma_i^2 = \mathbb{E}[s_i^2(n)]$ and $d_i = \frac{\sigma_i}{\sqrt{2\pi}}$.  
%Then, we have $P_\text{AC} = \mathbb{E}[\ut{x}^2(n)] = \mathbb{E}[\ut{x}_1^2(n)]+\mathbb{E}[\ut{x}_2^2(n)] = (\frac{1}{2} - \frac{1}{2\pi})(\sigma_1^2 + \sigma_2^2)$, 
%$P_\text{DC} = P_\text{opt}^2 = d^2 = ( \frac{\sigma_1}{\sqrt{2\pi}} + \frac{\sigma_1}{\sqrt{2\pi}} )^2$, and 
%$P_\text{elec} = P_\text{AC} + P_\text{DC} = \frac{1}{2}(\sigma_1^2 + \sigma_2^2 )+ \frac{\sigma_1\sigma_2}{\pi}$. 
%We also have $P_\text{eff} = \mathbb{E}[(\frac{s_1(n)}{2} + \frac{s_2(n)}{2})^2] = \frac{\sigma_1^2 + \sigma_2^2}{4}$, which turns out to be always a fixed portion of $P_\text{AC}$. 
%For general analysis, let $\sigma_2 = \alpha\sigma_1 = \alpha \sigma$, then the relation among the powers can be built as they are all related to $\sigma^2$. Specifically, the results shown in Table \ref{tab:power} are obtained by setting $\alpha = 1$ which equalizes the average symbol power in all layers. }

For LACO-OFDM, the transmit signal is $x_{\rm laco}(n) = \sum_{i=1}^J x_{i}(n)$, where $x_{i}(n) = (s_{\rm aco}^{(i)}(n))^+$ is the ACO-OFDM signal in layer $i$. Using Lemma \ref{Lem1}, we have that $s_{\rm aco}^{(i)}(n) \sim \mathcal{N}(0, \sigma_i^2)$ for some $\sigma_i^2$. Therefore, $x_{i}(n) \sim \mathcal{N}_c \big( 0, \sigma_i^2, [ 0, \infty ) \big) $. Then, using the law of large numbers, we have 
$P_\text{elec} = \frac{1}{N}\sum_{n=0}^{N-1}\mathbb{E}[x_{\rm laco}^2(n)] = \frac{1}{N}\sum_{n=0}^{N-1}\mathbb{E}[(\sum_{i=1}^J x_{i}(n))^2] = \sum_{i=1}^J (\frac{1}{2} - \frac{1}{2\pi})\sigma_i^2 + (\sum_i \frac{\sigma_i}{\sqrt{2\pi}})^2$ and 
$P_\text{opt} = \frac{1}{N}\sum_{n=0}^{N-1} \mathbb{E}[x_{\rm laco}(n)] = \frac{1}{N}\sum_{n=0}^{N-1} \mathbb{E}[\sum_{i=1}^J x_{i}(n)] = \sum_i \frac{\sigma_i}{\sqrt{2\pi}} $. We also have $P_\text{eff} = \mathcal{P}\{\sum_{i=1}^J\frac{\mathbf{s}_{\rm aco}^{(i)}}{2}\} = \frac{\sum_{i=1}^J\sigma_i^2}{4}$. Let $\sigma_i^2=\alpha_i\sigma^2$ for some $\alpha_i$ and $\sigma^2$. Then a relation between $P_\text{elec}$, $P_\text{opt}$, and $P_\text{eff}$ can be readily obtained through $\sigma^2$. By choosing $\alpha_i = 2^{J-i}$ for $i\in\{1,\ldots,J\}$, which means that $P_{\rm eff}$ is equally distributed in all effective subcarriers, then we obtain 
$P_\text{elec} = (2- \frac{2}{\pi} + \frac{2}{(3-2\sqrt{2})\pi} \frac{\sqrt{2}^J-1}{\sqrt{2}^J+1})P_\text{eff}$ and 
$P_\text{opt}=\sqrt{ \frac{2}{(3-2\sqrt{2})\pi} \frac{\sqrt{2}^J-1}{\sqrt{2}^J+1}P_\text{eff} }$ as shown in Table \ref{tab:power}.

%\cc{For a $J$-layer LACO-OFDM signal $x(n) = \ut{x} + d = \sum_{i=1}^J \ut{x}_i(n) + d$, where $x_i(n) = (s_i(n))^+ = \ut{x}_i(n) + d_i$ and $d = (\sum_{i=1}^{J} d_i)^2$.  Note that $s_i(n) \sim \mathcal{N}(0, \sigma_i^2)$, $i=1,\dots, J$. Therefore $x_i(n) \sim \mathcal{N}_c \big( 0, \sigma_i^2, [ 0, \infty ) \big) $ with $\sigma_i^2 = \mathbb{E}[s_i^2(n)]$ and $d_i = \frac{\sigma_i}{\sqrt{2\pi}}$. 
%Then, we have 
%$P_\text{AC} = \mathbb{E}[\ut{x}^2(n)] = \mathbb{E}[(\sum_{i=1}^J \ut{x}_i(n))^2] = (\frac{1}{2} - \frac{1}{2\pi}) \sum_{i=1}^J \sigma_i^2$, 
%$P_\text{DC} = P_\text{opt}^2 = d^2 = (\sum_i \frac{\sigma_i}{\sqrt{2\pi}})^2$, and 
%$P_\text{elec} = P_\text{AC} + P_\text{DC} = \sum_{i=1}^J (\frac{1}{2} - \frac{1}{2\pi})\sigma_i^2 + (\sum_i \frac{\sigma_i}{\sqrt{2\pi}})^2$. 
%We also have $P_\text{eff} = \mathbb{E}[(\sum_{i=1}^J \frac{s_i(n)}{2})^2] = \frac{\sum_{i=1}^J\sigma_i^2}{4}$, which is always a fixed portion of $P_\text{AC}$.
%For general analysis, let $\frac{\sigma_i}{\alpha_i} = \sigma$, then the relation among the powers can be built as they are all related to $\sigma^2$. Specifically, the results shown in Table \ref{tab:power} are obtained by setting $J = 9$ and $\alpha_i = 2^{i-1}$ which equalize the average symbol power in all layers. }

\section{Approximation of Detection-Error Power for QAM Constellations}\label{Sec:Appen1}

%Consider one point of an $M$-QAM constellation is received. Suppose the noise is circular Gaussian distributed with zero mean and variance of $2\sigma^2$, i.e., $\mathcal{CN}(0,2\sigma^2)$, and the minimum Euclidean distance of the constellation is $d_\text{min}$. For ML detection, the noise will confuse the detector between the actual position and its neighbors.  
%\modi{Here, we introduce the method to approximate $\mathsf{f}(d_\text{min}, \sigma^2, M)$.}
Consider an $M$-QAM symbol $x$ that is transmitted over a flat AWGN channel, and received at the receiver as $y=x+n$ where $n$ is $\mathcal{CN}(0,\sigma^2)$. The receiver adopts an ML detector to detect $\hat{x}$. Let the position of $x$ in the constellation be position $0$ as shown in Fig. \ref{fig:rims}. Noise will confuse the ML detector between the transmitted symbol and its neighbors, \ac{and an error will take place when $y$ falls outside position 0.}

\begin{figure}[!t]
	\centering
	\includegraphics[width=0.45\textwidth ]{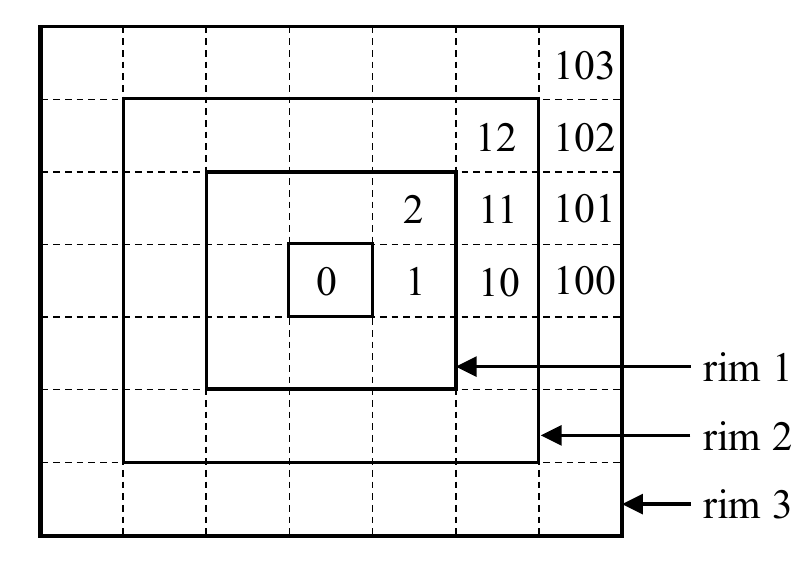}
	
	\caption{Geometry of 3 rims.}
	\label{fig:rims}
\end{figure}
We focus on neighbors of the transmitted symbol within the first three rims of position 0 and assume that \ccc{the probability of $y$ falling outside} the first three rims is negligible. We index the positions in the three rims using one-digit numbers starting from 1, two-digit numbers starting from 10, and three-digit numbers starting from 100, respectively. We also denote the Euclidean distance between position 0 and position $i$ by $d_i$, the minimum Euclidean distance of the used $M$-QAM constellation by $d_\text{min}$, and the error probability of detecting a symbol $\hat{x}$ at position $i$ by $p_i$ \ac{(i.e., the probability that $y$ falls in position $i$)}. 

Note that an error event in a QAM constellation can be considered as two independent error events in a PAM constellation \ac{where the noise is the real part or the imaginary part of $n$ \cite{book-digitalComm5th}. Denote the real part of $n$ by $n_{\rm r}$, and let $p_a$, $p_b$, and $p_c$ be the probabilities that $n_{\rm r}$ is larger than $\frac{d_{\rm min}}{2}$, $\frac{3d_{\rm min}}{2}$, and $\frac{5d_{\rm min}}{2}$, respectively, i.e.,}
\begin{align}\label{eq:p_abc}
p_a = \mathsf{Q}\left(  \frac{d_\text{min}}{\sqrt{2}\sigma} \right),\;\;
p_b = \mathsf{Q}\left(  \frac{3d_\text{min}}{\sqrt{2}\sigma} \right),\;\;
p_c = \mathsf{Q}\left(  \frac{5d_\text{min}}{\sqrt{2}\sigma} \right). 
\end{align}
Then, for the first three rims in a QAM constellation, we have
\begin{align}
	p_1 &= (p_a - p_b)(1 - 2p_a), \;\; \;
	p_2 = (p_a - p_b)^2; \\
	p_{10} &= (p_b - p_c) (1-2p_a), \;\;
	p_{11} = (p_b - p_c) (p_a - p_b), \;\;
	p_{12} = (p_b - p_c)^2; \\
	p_{101} &= p_c (p_a - p_b), \;\; \quad\quad\;\;
	p_{102} = p_c (p_b - p_c), \;\; \quad\quad\;\;
	p_{103} = p_c^2.
\end{align}

%	\begin{equation}
%	p_1 = (p_a - p_b)(1 - 2p_a), \;\; 
%	p_2 = (p_a - p_b)^2;
%	\end{equation}
%	\begin{equation}
%	p_{10} = (p_b - p_c) (1-2p_a), \;\;
%	p_{11} = (p_b - p_c) (p_a - p_b), \;\;
%	p_{12} = (p_b - p_c)^2;
%	\end{equation}
%	\begin{equation}
%	p_{101} = p_c (p_a - p_b), \;\; 
%	p_{102} = p_c (p_b - p_c), \;\;
%	p_{103} = p_c^2.
%	\end{equation}

%\begin{align}
%&p_1 = (p_a - p_b)(1 - 2p_a), \;\; 
%p_2 = (p_a - p_b)^2; \\ 
%&p_{10} = (p_b - p_c) (1-2p_a), \;\;
%p_{11} = (p_b - p_c) (p_a - p_b),  \;\; p_{12} = (p_b - p_c)^2; \\
%&p_{100} = p_c (1-2p_a), \;\;
%p_{101} = p_c (p_a - p_b),  p_{102} = p_c (p_b - p_c), \;\;
%p_{103} = p_c^2.
%\end{align}

For the first three rims of position 0, the power of detection error can be expressed as 
\begin{equation}\label{eq:pe_0}
P_0 = \ac{\mathbb{E}\{|x-\hat{x}|^2\} =} \sum_{i\in \mathcal{A}} {d_i^2 p_i n_i},
\end{equation}
where $\mathcal{A} = \{ 1,\; 2,\; 10,\; 11,\; 12,\; 100,\; 101,\; 102,\; 103 \}$ and $n_i$ is the number of neighboring points at a distance $d_i$ \ac{from $x$}.%,  which can be obtained from $d_\text{min}$, e.g., $d_1 = d_\text{min}$, $d_{10} = 2d_\text{min}$, $d_{100} = 3d_\text{min}$, and so on.

\ac{The former discussion applies when $x$ is in position $0$.} To estimate the power of detection errors when \ac{all points of the $M$-QAM constellation are considered, we assume that all points in the constellation are transmitted with the same probability. In this case,} $n_i$ in \eqref{eq:pe_0} should be replaced with the average number of neighbors at a distance $d_i$ from $x$, which is denoted by $\bar{n}_i^{M-\text{QAM} }$. \cc{Then the power of detection errors for a $M$-QAM constellation\cc{, $P_{\rm e}^{ M-\text{QAM} }$,} can be obtained by modifying \eqref{eq:pe_0} to}
\begin{equation}\label{eq:Pe_qam}
P_{\rm e}^{ M-\text{QAM} } = \sum_{i\in \mathcal{A}} { d_i^2 p_i \bar{n}_i^{ M-\text{QAM} } }.
\end{equation}

The calculation of \eqref{eq:Pe_qam} only requires \modi{the minimum Euclidean distance $d_\text{min}$, noise power $\sigma^2$ and constellation size $M$}. Therefore we can write 
\begin{equation}\label{eq:Pe_qam-func}
P_{\rm e}^{ M-\text{QAM} } = \cc{\mathsf{f}}(d_\text{min}, \modi{\sigma^2}, M),
\end{equation}
as given in Definition \ref{definition}.

\end{appendices}

\bibliographystyle{IEEEtran}
\bibliography{references}

\end{document}